\documentclass[english,aps,notitlepage,nofootinbib]{revtex4-1}
\usepackage{ae,aecompl}
\usepackage[T1]{fontenc}
\usepackage[utf8]{inputenc}
\usepackage{babel}
\usepackage{units}
\usepackage{amsmath}
\usepackage{array}
\usepackage{ragged2e}
\usepackage{amsbsy}
\usepackage{amstext}
\usepackage{amssymb}
\usepackage{multirow}
\usepackage{graphicx}
\usepackage{esint}
\usepackage[T1]{fontenc}
\usepackage{stmaryrd}
\usepackage{babel}
\usepackage[dvipsnames]{xcolor}
\usepackage[colorlinks,citecolor=orange,linkcolor=orange,urlcolor=orange]{hyperref}
\newcommand{\lyxmathsym}[1]{\ifmmode\begingroup\def\b@ld{bold}
  \text{\ifx\math@version\b@ld\bfseries\fi#1}\endgroup\else#1\fi}

\providecommand{\tabularnewline}{\\}
\renewcommand{\textcolor}[2]{#2}
\newenvironment{lyxlist}[1]
	{\begin{list}{}
		{\settowidth{\labelwidth}{#1}
		 \setlength{\leftmargin}{\labelwidth}
		 \addtolength{\leftmargin}{\labelsep}
		 }}
	{\end{list}}

\usepackage{babel}

\makeatother

\begin{document}
\preprint{APS/123-QED}

\title{Generalized Fourier Transforms for Momentum-Space Construction on Riemannian Manifolds}
\author{Seramika Ariwahjoedi$^{1}$}
\email[Corresponding author: ]{sera001@brin.go.id}

\author{Muhammad Farchani Rosyid$^{2}$}
\email{farchani@ugm.ac.id}

\author{Andika Kusuma Wijaya$^{1}$}
\email{andika.kusuma.wijaya@mail.ugm.ac.id}

\address{$^{1}$Research Center for Quantum Physics,\\ National Research and
Innovation Agency (BRIN), South Tangerang 15314, Banten, Indonesia,
}

\address{$^{2}$Department of Physics, Universitas Gadjah Mada,\\ Jl. Bulaksumur 55281, Yogyakarta, Indonesia,}

%\address{$^{3}$Program Studi Pendidikan Fisika, ISBI Singkawang,\\ Kode Pos 79151 Singkawang, Indonesia.~\\}

\date{\today}% It is always \today, today,

\begin{abstract}

We extend Fourier analysis to curved spaces by defining a Generalized
Fourier Transform (GFT) on any Riemannian manifold $\Sigma$ via spectral
decomposition. Under minimal requirements that the transform is an
isometric isomorphism and has a kernel diagonalizing the Laplace-Beltrami
operator, we prove that the GFT satisfies a generalized Parseval-Plancherel
theorem. To resolve the spectral degeneracy that obscures \textquotedblleft momentum
space\textquoteright\textquoteright{} in such settings, we require
the degenerate sector to be resolved by a local, symmetry-adapted
maximal Abelian commuting set (a fiberwise MASA), constructed from
geometric differential operators, most notably from Killing data when
such symmetries are available. We provide a constructive algorithm
for generating these commuting operators and show that the resulting
momentum label spaces $\mathcal{F}$ (discrete, continuous, or mixed)
reflect geometric symmetry constraints. We introduce a dual classification:
(i) by MASA completeness and St$\ddot{\textrm{a}}$ckel separability,
and (ii) by the topology of $\mathcal{F}.$ Finally, we distinguish
unitary changes induced by true isometries (which preserve the GFT
structure) from changes of coordinate-adapted degeneracy resolution/separation
schemes, which may induce inequivalent $k$-space labelings (e.g.
Cartesian vs spherical constructions in $\mathbb{R}^{3}$) while remaining
unitarily equivalent on $\mathcal{L}^{2}\left[\Sigma\right]$. This
symmetry-adapted harmonic analysis is intended as a foundation for
curved-space mode decompositions; dynamical applications are developed
in\textcolor{blue}{{} the subsequent work.}

\end{abstract}

\maketitle

\section{Introduction}

Fourier transform is one of the most powerful mathematical tools in
modern analysis, it plays a crucial role in fields ranging from signal
processing and quantum mechanics to differential geometry and mathematical
physics. At its core, it provides a way to decompose a function into
a sum of oscillatory modes and offers a dual perspective between spatial
(or temporal) and momentum (or frequency) domains. In Euclidean space,
the standard Fourier transform is well understood, where \textcolor{blue}{its
construction utilizes} the classical translation group as a relevant
and important example of Pontryagin duality \cite{Spectral1,Folland,Grigoryan,Spectral2}.
However, when dealing with more general spaces, particularly curved
Riemannian manifolds, the notion of Fourier analysis becomes less
straightforward. This motivates the need for a Generalized Fourier
Transform (GFT) that extends these principles beyond flat space while
preserving the core properties of invertibility (or isomorphism),
orthonormality (or isometricity), and spectral decomposition.

Several attempts have been made to generalize the Fourier transform
to non-Euclidean spaces. Approaches based on Pontryagin duality have
laid the groundwork for understanding Fourier analysis on (locally
compact Abelian) groups and homogeneous spaces \cite{Pontryagin1,Pontryagin2,Pontryagin3}.
Another common route is the spectral theory approach, where Fourier-type
transforms are constructed from the spectrum of the Laplace-Beltrami
operator \cite{Spectral1,Grigoryan,Spectral2,Plancherel1,Plancherel2};
the Helgason-Fourier transform could be considered as a generalization
to the Fourier transform using this approach, where it generalize
Fourier transform in $\mathbb{R}^{n}$ into the Riemannian symmetric
space of non-compact types \cite{Helgason1,Helgason2,Helgason3,Helgason4,Helgason5}.
The other approaches may consist of the Gelfand transform \cite{Gelfand1,Gelfand2},
noncommutative harmonic analysis \cite{NoncommuteHarmonic1,NoncommuteHarmonic2,NoncommuteHarmonic3},
and Fourier Integral Operators \cite{FIO1,FIO2}. All these methods
have provided valuable insights in advancing the field.

Building on classical harmonic analysis and separation-of-variables
theory, we develop a systematic framework applicable to Riemannian
manifolds by introducing a Generalized Fourier Transform (GFT), whose
kernel is defined through the spectral decomposition of the Laplace--Beltrami
operator. The present paper focuses on the mathematical framework:
construction of the transform, resolution of spectral degeneracy,
and a systematic classification of the resulting GFTs and their momentum
spaces. Physical applications (generalized momentum re-definition,
curved $k$-space \cite{curvedmomemtum1,curvedmomentum2,curvedmomentum3},
examples such as Hopf fibration \cite{Hopf,CNYang}, Unruh effect
\cite{Unruh,Fulling,Davies}, and Aharonov--Bohm effect \cite{ABeffect})
will \textcolor{blue}{be addressed in the subsequent work.}

\paragraph{Main results. }

The main results in this paper are the listed as follows:
\begin{enumerate}
\item \textbf{Local-MASA principle.} We show that a \textit{physical} GFT
basis (or kernel) on a Riemannian manifold $\Sigma$ shall arise from
a maximal Abelian set (MASA) of \textbf{local} commuting operators
that commutes with the Laplace-Beltrami operator on that manifold;
non-local MASAs, although allowed algebraically, lead to unphysical
label spaces.
\item \textbf{Algorithmic construction of local MASAs}. A rank-by-rank procedure,
adapting Benenti-Eisenhart-Kalnins-Miller-style separation \cite{Sep1,Sep2,Sep3,Sep4,Sep5,Sep6,Sep7}
and Killing-tensor machinery to a GFT-specific goal: constructing
local commuting algebras that resolve spectral degeneracy and determine
the $k$-space structure.
\item \textbf{Dual GFT classification grid.} We introduce a two-level taxonomy
on GFT: (i) Type I--III by MASA completeness and St$\ddot{\textrm{a}}$ckel
separability; (ii) discrete/continuous/semi-discrete (mixed) by the
topology of the joint spectrum. We provide two corresponding flowcharts
for curved-space Fourier analysis. 
\item \textbf{Gauge-freedom analysis}: We distinguish unitary transformations
induced by true isometries from changes in coordinate-adapted degeneracy
resolution/separation schemes, which can induce inequivalent momentum
label spaces (e.g. $\mathbb{R}^{3}$ \textrightarrow{} $\mathbb{R}^{+}\times\mathbb{Z}^{2}$
arising from Cartesian vs spherical constructions).
\item \textbf{Worked spectral-topology examples}: Cartesian versus spherical
MASAs in $\mathbb{R}^{3}$ and rational vs irrational flows on $\mathbb{T}^{2}$
explicitly demonstrate how different local operator choices alter
$k$-space while remaining unitarily equivalent in the Hilbert space
$\mathcal{L}^{2}\left[\Sigma\right]$.
\item \textbf{Gauge-fixing criterion in degenerate sectors}. We make explicit
that in the presence of spectral degeneracy, the $x-k$ duality is
unique \textit{only} after one local MASA is fixed; this resolves
an ambiguity left implicit in earlier treatments.
\end{enumerate}
As a physical application of our work, we propose a generalized definition
on momentum, which is context-dependent in physics terminology. In
curved backgrounds, \textquotedblleft momentum\textquoteright\textquoteright{}
labels are ambiguous unless tied to symmetry. Here we adopt a spectral
notion: we construct a local, symmetry-adapted MASA of geometric differential
operators (built from Killing vectors/tensors) that commute with the
Laplace-Beltrami operator, and take $k$-space to be their joint spectrum.
This provide the definition of momentum that (i) reduce to canonical
momenta in flat, translationally invariant cases and (ii) remain meaningful
on curved manifolds where no global translational momentum exists.

Furthermore, isometries preserve the topology of $k$-space, whereas
generic changes in coordinate-induced MASA choice need not. This symmetry-adapted
(MASA) labeling clarifies several physics-facing issues. It explains
discrete vs continuous mode spectra within one scheme and provides
an operational interpretation (\textquotedblleft one commuting set
per measurement context\textquotedblright ) of the generalized Fourier
Transform. The full proposal on the generalized momentum definition
will be developed in part II of this work.

Finally, the point of this framework is not merely to restate spectral
expansions, but to make the non-uniqueness of Fourier-like transforms
on curved spaces explicit, controllable, and classifiable. In particular,
by tying degeneracy resolution to local (symmetry-adapted) commuting
operators, the induced $k$-space $\mathcal{F}$ becomes an interpretable
object rather than an arbitrary labeling convention, it clarifies
when \textquotedblleft momentum labels\textquotedblright{} are meaningful
and how they behave under symmetry and coordinate-adapted choices.

\paragraph{Organization. }

Section II reviews the spectral theory foundations and defines the
GFT and its mathematical structures; this includes the properties
in the $x$ and $k$ spaces, and the proof for the generalized Parseval-Plancherel
theorem. Section III discussed basis freedom in degenerate spectra,
contrasting local vs. non-local operators, and emphasizing the physical
importance of locality. Section IV gives the MASA-construction algorithm:
a procedure to build a commuting set of local, symmetry-adapted operators
using Killing vectors/tensors. Section V studies coordinate/gauge
freedom, especially the impact of isometries and (St$\ddot{\textrm{a}}$ckel)
coordinates on the GFT kernel structure.

Section VI introduces the dual classification for GFTs: one based
on MASA/St$\ddot{\textrm{a}}$ckel completeness, and one based on
the topology of the Fourier space (discrete, continuous, semi-discrete).
Section VII highlight some subtleties around the GFT framework and
provides 2 illustrative examples on how coordinate-adapted transformation
and isometry affect the degenerate sector of the GFT. We conclude
with a discussion of open directions (a brief proposal on the generalized
momentum definition is given here) to pave the way\textcolor{blue}{{}
for the development of the physical applications.}

\section{Generalized Fourier Transform on Riemannian Manifold}

In this section we construct a unitary, norm-preserving direct-integral
spectral transform on a (region of a) Riemannian manifold that diagonalizes
the Laplace--Beltrami operator, hence generalizing the Euclidean
Fourier transform in $\mathbb{R}^{n}$.

\subsection{Spectral Decomposition Theorem}

Let the \textit{spatial domain} (or $x$-space), labeled by $\Sigma$,
be an $n$-dimensional connected, oriented, smooth manifold possibly
with boundary (appropriate boundary conditions are imposed so that
the relevant operator are self-adjoint). Notice that $\Sigma$ could
be compact or non-compact. Let $\mathbf{x}=x^{i}$ be a local coordinate
patch on $\Sigma$ and let us equip $\Sigma$ with a Riemannian (positive
definite) metric $\mathbf{q}_{\left[\mathbf{x}\right]}$ that is regular
on $\Sigma.$ This geometrization of $\Sigma$ allows us to define
geometrical quantities on $\Sigma$ such as length, angle, and curvatures.
One could define a measure $d\mu_{\Sigma}$ on $\Sigma$ as the volume
form:
\begin{equation}
d\mu_{\Sigma}=\,^{n}\textrm{vol}_{\Sigma}=\sqrt{q_{\left[\mathbf{x}\right]}}\textrm{d}^{n}\mathbf{x},\label{eq:measureinX}
\end{equation}
hence allowing us to do integration on $\Sigma$. One could construct
a space of square-integrable functions over $\Sigma$, namely, $\mathcal{L}^{2}\left[\Sigma\right]$,
and define a standard inner product as follows:
\begin{equation}
\left\langle \psi_{1},\psi_{2}\right\rangle _{\mathcal{H}_{x}}=\intop_{\Sigma}d\mu_{\Sigma}\:\psi_{1\left[\mathbf{x}\right]}^{*}\psi_{2\left[\mathbf{x}\right]},\label{eq:QMinnerproductxreps}
\end{equation}
with $\psi_{1},\psi_{2}\in\mathcal{L}^{2}\left[\Sigma\right].$ Equipping
$\mathcal{L}^{2}\left[\Sigma\right]$ with an inner product lifted
the function space $\mathcal{L}^{2}\left[\Sigma\right]$ to be a Hilbert
space over the $x$-space, $\mathcal{H}_{x}\sim\mathcal{L}^{2}\left[\Sigma\right]$.

Let $\hat{\mathcal{O}}:\mathrm{dom}_{\left[\hat{\mathcal{O}}\right]}\subset\mathcal{H}_{x}\rightarrow\mathcal{H}_{x}$
be a densely defined linear operator in $\mathcal{H}_{x}$ and let
$\hat{\mathcal{O}}^{*}:\mathrm{dom}_{\left[\hat{\mathcal{O}}^{*}\right]}\subset\mathcal{H}_{x}\rightarrow\mathcal{H}_{x}$
be the adjoint of $\hat{\mathcal{O}}^{*},$ i.e., $\left\langle \hat{\mathcal{O}}\psi_{1},\psi_{2}\right\rangle _{\mathcal{H}_{x}}=\left\langle \psi_{1},\hat{\mathcal{O}}^{*}\psi_{2}\right\rangle _{\mathcal{H}_{x}}$
for all $\psi_{1}\in\mathrm{dom}_{\left[\hat{\mathcal{O}}\right]}$
and $\psi_{2}\in\mathrm{dom}_{\left[\hat{\mathcal{O}}^{*}\right]}$.
$\hat{\mathcal{O}}$ is self-adjoint if $\hat{\mathcal{O}}=\hat{\mathcal{O}}^{*}$
and $\mathrm{dom}_{\left[\hat{\mathcal{O}}\right]}=\mathrm{dom}_{\left[\hat{\mathcal{O}}^{*}\right]}$.

Every $\mathcal{L}^{2}\left[\Sigma\right]$ admits plenty of self-adjoint
operators (identity, multiplication, projections). As primary work,
we assume the operators of interest are geometric differential, in
particular the Laplace-Beltrami operator $\triangle$. For such operators,
essential self-adjointness is automatic on complete Riemannian manifolds,
while for incomplete manifolds one must supply appropriate boundary
conditions or select a suitable self-adjoint extension.

By the spectral theorem, any self-adjoint operator $\hat{\mathcal{O}}$
in $\mathcal{L}^{2}\left[\Sigma\right]$ admits spectral decomposition.
Conversely, given such a spectral decomposition, one recovers a unique
self-adjoint operator. To fix notation, let $\mathcal{\hat{P}}_{\left[\cdot\right]}$
denote the projection-valued measure (PVM) associated with the (densely-defined)
self-adjoint operator \textit{$\hat{\mathcal{O}}$ }acting on $\mathcal{H}_{x}$.
\begin{lyxlist}{00.00.0000}
\item [{\textbf{Proposition.}}] (Spectral Decomposition): Let\textit{ $\mathcal{S}\subset\mathbb{R}$
be the spectrum of $\hat{\mathcal{O}}.$ Then $\hat{\mathcal{O}}$
admits a spectral representation:
\begin{equation}
\hat{\mathcal{O}}=\intop_{\mathcal{S}}\lambda\,d\mathcal{\hat{P}}_{\left[\lambda\right]},\label{eq:spectraltheorem}
\end{equation}
where $\lambda\in\mathcal{S}$ is the spectral parameter and $d\mathcal{\hat{P}}_{\left[\lambda\right]}$
is the increment of the PVM associated with $\hat{\mathcal{O}}$.}
\end{lyxlist}
The spectral theorem (\ref{eq:spectraltheorem}) may be viewed a diagonalization
of the self-adjoint operator \textit{$\hat{\mathcal{O}}$} on $\Sigma$.
As a consequence to this theorem, the self-adjointness of an operator
in $\mathcal{H}_{x}$ guarantees the existence of a set of generalized
eigenfunctions/spectral modes that are: (i) orthonormal with respect
to inner product (\ref{eq:QMinnerproductxreps}), (ii) yields a resolution
of identity/ spectral decomposition in $\mathcal{H}_{x}$, (iii) admits
real eigenvalues \textit{$\lambda\in\mathcal{S}$, }with degeneracy
labeled by an index $\alpha$ (possibly, but not necessarily dependent
on $\lambda$). Such degeneracies typically arise from geometric symmetries
of $\Sigma$, and it has an important role in the GFT construction.
This theorem allows us to generalized the standard Fourier transform
on $\mathbb{R}^{n}$ to general Riemannian manifold admitting self-adjoint
geometric operator.

\subsection{Generalized Fourier Transform Construction}

To construct the generalized Fourier transform on $\Sigma$, we require
a map satisfying three minimal conditions: (a)\label{a} the direct-integral
spectral transform is an isomorphism between Hilbert spaces, \textcolor{blue}{hence
invertible}, (b)\label{b} it and its inverse are isometric (i.e.,
preserve the norm of inner products in $\mathcal{L}^{2}\left[\Sigma\right]$,
\textcolor{blue}{hence unitary}), (c)\label{c} its kernel diagonalizes
the Laplace-Beltrami operator $\triangle$. \textcolor{blue}{These
requirements are realized by the spectral theorem applied to the self-adjoint
operator $\hat{\mathcal{O}}:=-\triangle$, which yield the corresponding
unitary spectral transform. We refer to this transform as the }\textit{\textcolor{blue}{Generalized
Fourier Transform}}\textcolor{blue}{{} (GFT) on $\Sigma$.}

The Laplace-Beltrami operator $\triangle$ is defined by:
\begin{equation}
\triangle=\frac{1}{\sqrt{q_{\left[\mathbf{x}\right]}}}\frac{\partial}{\partial x^{i}}\left(\sqrt{q_{\left[\mathbf{x}\right]}}q^{ij}\frac{\partial}{\partial x^{j}}\right),\label{eq:LaplaceBeltrami}
\end{equation}
with $q_{\left[\mathbf{x}\right]}=\det q_{ij}$ and $q^{ij}$ and
are, respectively, the determinant and the metric component (written
in coordinate $\mathbf{x}=x^{i}$) of the metric $\mathbf{q}_{\left[\mathbf{x}\right]}$.
On a Riemannian manifold $\Sigma$, $\triangle$ is essentially self-adjoint
on $C^{\infty}\left[\Sigma\right]$ (smooth functions of compact support
on $\Sigma$) and admits a unique self-adjoint extension on $\mathcal{L}^{2}\left[\Sigma\right]$
when $\Sigma$ is geodesically complete \cite{Spectral1,Grigoryan,Spectral2}.

The diagonalization of $\triangle$ is the Helmholtz equation in $\Sigma$:
\begin{equation}
\left(\triangle+\lambda\right)\hat{f}_{\left[\mathbf{x};\lambda,\alpha\right]}=0,\label{eq:Helmholtz}
\end{equation}
where $\hat{f}_{\left[\mathbf{x};\lambda,\alpha\right]}$ is a set
of generalized eigenfunction (not necessarily in $\in\mathcal{L}^{2}\left[\Sigma\right]$
for continuous spectrum, normalized distributionally with respect
to a measure) that solve (\ref{eq:Helmholtz}), labeled by its eigenvalues
and degeneracy $\left(\lambda,\alpha\right),$ where now $\lambda\in\mathcal{S}\subseteq\mathbb{R}^{+},$
 as a consequence of the ellipticity of $\triangle$ in $\Sigma$
\cite{Spectral1,Spectral2}. The generalized eigenfunction satisfies
an orthonormality condition with respect to the inner product (\ref{eq:QMinnerproductxreps}). 

\paragraph*{Remarks.}

$\alpha$ is a degeneracy label; for the case where the multiplicity
is finite/countable, $\alpha$ serve as an index, not a coordinate.
In symmetry-adapted constructions it may be replaced by eigenvalues
of a commuting set (then continuous labels can appear).

The self-adjoint operator considered in the spectral decomposition
is $\hat{\mathcal{O}}:=-\triangle$, which is positive semi-definite.
By restricting $\hat{\mathcal{O}}:=-\triangle$, the requirement (c)\nameref{c}
of the GFT construction is satisfied, while requirement (a)\nameref{a}
and (b)\nameref{b} could be shown as consequences of the spectral
theorem (\ref{eq:spectraltheorem}).

The collection of $\left(\lambda,\alpha\right)$ construct an abstract,
topological set of real spectral parameters and its degeneracy:
\begin{equation}
\mathcal{F}:=\bigcup_{\lambda\in\mathcal{S}}\left\{ \lambda\right\} \times\left.\mathcal{F}\right|_{\lambda}\cong\left\{ \left(\lambda,\alpha\right)|\lambda\in\mathcal{S},\alpha\in\left.\mathcal{F}\right|_{\lambda}\right\} \label{eq:kspace}
\end{equation}
with $\mathcal{F}$ is the disjoint union of all the degeneracy fiber
$\left.\mathcal{F}\right|_{\lambda}$. In the most general setting,
$\mathcal{F}$ is not geometrical, i.e., the introduction of a metric
in $\mathcal{F}$ is not necessary. Let us call this space as the
$k$-space or the \textit{momentum domain}.

The topology of $\mathcal{F}$ depends on the structure of the spectrum
of the Laplace-Beltrami operator (\ref{eq:LaplaceBeltrami}); $\mathcal{F}$
could be continuous, semi-continuous, or even discrete (under additional
conditions) if $\Sigma$ is non-compact, and typically discrete\footnote{Throughout this paper, we use the term \textquotedbl discrete''
spectrum as adopted by physicists, i.e., to describe ``a countable
pure-point set with no continuous component''. It needs to be kept
in mind that in a rigorous manner, discreteness and countability are
distinct; all discrete sets are countable, but not all countable sets
are topologically discrete.} if $\Sigma$ is compact. For the case where $\mathcal{F}$ is continuous
and smooth, it could be considered as a (differentiable) manifold.
We will only consider the continuous-spectrum case for clarity; the
purely discrete $\mathcal{F}$ (compact $\Sigma$) or mixed cases
follow by replacing integrals by sums (or sums + integrals).

We equip $\mathcal{F}$ with a reference measure to define $\mathcal{L}^{2}\left[\mathcal{F}\right].$
A product measure $d\lambda d\alpha$ on $\mathcal{F}$ yields:
\begin{equation}
d\mu_{\mathcal{F}}=d\mu_{\mathcal{S}}\otimes d\mu_{\left.\mathcal{F}\right|_{\lambda}}=\rho_{\left[\lambda,\alpha\right]}d\lambda d\alpha,\label{eq:measurek}
\end{equation}
with $\rho_{\left[\lambda,\alpha\right]}$ is the density weight relative
to the Lebesgue/counting product. For discrete $\left(\lambda,\alpha\right),$
$d\lambda d\alpha$ acts as a counting measure. At this stage, $\rho_{\left[\lambda,\alpha\right]}$
is not fixed canonically by $\left(\Sigma,\boldsymbol{q}\right)$
alone: its precise form is tied to the normalization convention for
the generalized eigenfunctions $\hat{f}_{\left[\mathbf{x};\lambda,\alpha\right]}$
(equivalently, to the choice of orthonormal basis within degenerate
fibers). Once a specific orthonormal family $\hat{f}_{\left[\mathbf{x};\lambda,\alpha\right]}$
is chosen, the pair $\left(\hat{f}_{\left[\mathbf{x};\lambda,\alpha\right]},d\mu_{\mathcal{F}}\right)$
must satisfy the orthonormality/completeness relations, which fixes
$d\mu_{\mathcal{F}}$ up to unitary rotations inside degenerate fibers.
This freedom corresponds to gauge choices in the degenerate sectors
(detailed in Section V); in symmetry-adapted cases, $\rho$ can be
fixed via the Liouville measure induced by the local MASA (e.g. from
Killing symmetries).

With this measure, we define the momentum-space Hilbert space over
$\mathcal{F}$, namely $\mathcal{H}_{k}\sim\mathcal{L}^{2}\left[\mathcal{F}\right]$
with a standard inner product:
\begin{equation}
\left\langle \phi_{1},\phi_{2}\right\rangle _{\mathcal{H}_{k}}=\intop_{\mathcal{F}}d\mu_{\mathcal{F}}\phi_{1\left[\lambda,\alpha\right]}^{*}\phi_{2\left[\lambda,\alpha\right]}.\label{eq:explicitinnerprod}
\end{equation}
We assume $\left(\mathcal{F},d\mu_{\mathcal{F}}\right)$ is a $\sigma$-finite\footnote{A measure space is called $\sigma$-finite if it can be covered by
a countable union of measurable pieces each having finite measure.} measure space so that $\mathcal{H}_{k}\sim\mathcal{L}^{2}\left[\mathcal{F}\right]$
is separable. Since $\mathcal{H}_{x}\sim\mathcal{L}^{2}\left[\Sigma\right]$
and $\mathcal{H}_{k}\sim\mathcal{L}^{2}\left[\mathcal{F}\right]$
are separable Hilbert spaces, an abstract unitary identification exists.
However, such an identification is non-canonical and carries no information
about the operator $\hat{\mathcal{O}}=-\triangle.$ In this work we
construct the specific spectral transform $\mathcal{U}$ associated
with $\hat{\mathcal{O}},$ characterized by the property that it diagonalizes
$\hat{\mathcal{O}}$ (mapping it to multiplication by the spectral
parameter). The measure on the spectral space and the unitarity of
$\mathcal{U}$ are fixed accordingly via the Plancherel/Parseval relation.
\begin{lyxlist}{00.00.0000}
\item [{\textbf{Definition.}}] (Generalized Fourier Transform): \textit{Given
$\psi_{\left[\mathbf{x}\right]}\in\mathcal{H}_{x}\sim\mathcal{L}^{2}\left[\Sigma\right]$,
where $\Sigma$ is an (region of an) $n$-dimensional manifold equipped
with a Riemannian metric $\mathbf{q},$ the generalized Fourier transform
is defined as the map:}
\begin{equation}
\begin{array}{cccc}
\mathcal{U}: & \mathcal{H}_{x} & \rightarrow & \mathcal{H}_{k},\\
 & \psi_{\left[\mathbf{x}\right]} & \mapsto & \phi_{\left[\lambda,\alpha\right]},
\end{array}\label{eq:map-1}
\end{equation}
\textit{specified by:
\begin{equation}
\phi_{\left[\lambda,\alpha\right]}=\left(\mathcal{U}\psi\right)_{\left[\lambda,\alpha\right]}:=\left\langle \hat{f}_{\left[\mathbf{x};\lambda,\alpha\right]},\psi_{\left[\mathbf{x}\right]}\right\rangle _{\mathcal{H}_{x}}\label{eq:GFT}
\end{equation}
with the Fourier kernel $\hat{f}_{\left[\mathbf{x,}\lambda,\alpha\right]}$
is the generalized eigensolution to the Helmholtz equation in $\Sigma$,
satisfying the orthonormality condition:}
\begin{equation}
\left\langle \hat{f}_{\left[\mathbf{x};\lambda',\alpha'\right]},\hat{f}_{\left[\mathbf{\mathbf{x};\lambda,\alpha}\right]}\right\rangle _{\mathcal{H}_{x}}=\intop_{\Sigma}d\mu_{\Sigma}\:\hat{f}_{\left[\mathbf{x};\lambda',\alpha'\right]}^{*}\hat{f}_{\left[\mathbf{\mathbf{x};\lambda,\alpha}\right]}=\delta_{\rho\left[\lambda,\alpha;\lambda',\alpha'\right]}.\label{eq:orthonormalq-1}
\end{equation}
\end{lyxlist}
Here $\delta_{\rho\left[\lambda,\alpha;\lambda',\alpha'\right]}$
is the Dirac-delta distribution in $\mathcal{F}:$
\begin{equation}
\intop_{\mathcal{F}}d\mu_{\mathcal{F}}\delta_{\rho\left[\lambda,\alpha;\lambda',\alpha'\right]}\phi_{\left[\lambda,\alpha\right]}=\phi_{\left[\lambda',\alpha'\right]}.\label{eq:ddeltaf}
\end{equation}
defined with respect to the measure $d\mu_{\mathcal{F}}$ (\ref{eq:measurek})
as:
\begin{equation}
\delta_{\rho\left[\lambda,\alpha;\lambda',\alpha'\right]}=\frac{\delta_{\left[\lambda,\alpha;\lambda',\alpha'\right]}}{\rho_{\left[\lambda,\alpha\right]}},\label{eq:ddeltak}
\end{equation}
where $\rho_{\left[\lambda,\alpha\right]}$ is the weight of the measure
(\ref{eq:measurek}) (its normalization will be fixed by the Parseval-Plancherel
identity). By this definition, the requirement (c), where the kernel
of the transform needs to diagonalized the Laplace-Beltrami operator,
is satisfied. As a consequence of (\ref{eq:GFT}), one could prove
the following statement:

\textbf{Proposition.} (Inverse Transform):\textit{ From completeness
(spectral theorem), equation (\ref{eq:GFT}) admits the inverse:
\begin{equation}
\psi_{\left[\mathbf{x}\right]}=\intop_{\mathcal{F}}d\mu_{\mathcal{F}}\phi_{\left[\lambda,\alpha\right]}\hat{f}_{\left[\mathbf{\mathbf{x};\lambda,\alpha}\right]},\label{eq:inverseGFT}
\end{equation}
where $\phi_{\left[\lambda,\alpha\right]}\in\mathcal{L}^{2}\left[\mathcal{F},d\mu_{\mathcal{F}}\right]$
is the spectral coefficient function, $\mathcal{F}$ is the spectral
parameter space (including degeneracy labels), and $d\mu_{\mathcal{F}}$
is the spectral measure associated with the Laplace-Beltrami operator.
(For purely discrete spectra, replace integrals by sums).}

The unitarity (isometric isomorphism, requirements (a)--(b)) follows
from the generalized Parseval-Plancherel theorem (next subsection).

\subsection{Parseval-Plancherel Theorem}

The proof of the Parseval--Plancherel theorem relies on the spectral
resolution of the identity. Let $\left.\mathcal{H}_{x}\right|_{\lambda}\subset\mathcal{H}_{x}$
be the Hilbert fiber defined by fixing $\lambda$, the space  is spanned
by $\hat{f}_{\left[\mathbf{\mathbf{x};\lambda,\alpha}\right]}$ across
degeneracy label $\alpha$. We define the spectral projection density
(in the position representation) as the integral kernel obtained by
summing/integrating over the degeneracy fiber $\left.\mathcal{F}\right|_{\lambda}$:
\begin{equation}
\mathcal{K}_{\left[\mathbf{x},\mathbf{x}',\lambda\right]}=\intop_{\left.\mathcal{F}\right|_{\lambda}}d\mu_{\left.\mathcal{F}\right|_{\lambda}}\hat{f}_{\left[\mathbf{\mathbf{x}';\lambda,\alpha}\right]}^{*}\hat{f}_{\left[\mathbf{\mathbf{x};\lambda,\alpha}\right]}.\label{eq:disc}
\end{equation}
(For discrete fibers, the integral is replaced by a sum). 

The completeness of the orthonormal basis (or the resolution of the
identity $\mathbb{I}_{\mathcal{H}_{x}}$) is guaranteed by the spectral
theorem. In terms of integral kernels, this completeness relation
reads:
\begin{align}
\left\langle \hat{f}_{\left[\mathbf{\mathbf{x}';\lambda,\alpha}\right]},\mathbb{I}_{\mathcal{H}_{x}}\hat{f}_{\left[\mathbf{\mathbf{x};\lambda,\alpha}\right]}\right\rangle _{\mathcal{H}_{x}} & =\intop_{\mathcal{S}}d\mu_{\mathcal{S}}\:\mathcal{K}_{\left[\mathbf{x},\mathbf{x}',\lambda\right]},\nonumber \\
 & =\intop_{\mathcal{F}}d\mu_{\mathcal{F}}\:\hat{f}_{\left[\mathbf{\mathbf{x}';\lambda,\alpha}\right]}^{*}\hat{f}_{\left[\mathbf{\mathbf{x};\lambda,\alpha}\right]}=\delta_{q\left[\mathbf{x},\mathbf{x}'\right]},\label{eq:completeness}
\end{align}
with $\delta_{q\left[\mathbf{x},\mathbf{x}'\right]}$ denoting the
geometric Dirac-delta distribution in $\Sigma,$ defined with respect
to the volume measure (\ref{eq:measureinX}). To ensure coordinate
invariance, it is related to the standard Euclidean delta $\delta_{\left[\mathbf{x},\mathbf{x}'\right]}$
by:
\begin{equation}
\delta_{q\left[\mathbf{x},\mathbf{x}'\right]}=\frac{\delta_{\left[\mathbf{x},\mathbf{x}'\right]}}{\sqrt{q_{\left[\mathbf{x}\right]}}},\label{eq:ddeltaq}
\end{equation}
such that:
\[
\intop_{\Sigma}d\mu_{\Sigma}\delta_{q\left[\mathbf{x},\mathbf{x}'\right]}\psi_{\left[\mathbf{x}\right]}=\psi_{\left[\mathbf{x}'\right]}.
\]
Using this resolution of the identity, we ensure the isometry. Expanding
the inner product (\ref{eq:QMinnerproductxreps}) in $\mathcal{H}_{x}$
and substituting the inverse GFT (\ref{eq:inverseGFT}) to $\psi_{2\left[\mathbf{x}\right]}$:
\[
\left\langle \psi_{1},\psi_{2}\right\rangle _{\mathcal{H}_{x}}=\intop_{\Sigma}d\mu_{\Sigma}\:\psi_{1\left[\mathbf{x}\right]}^{*}\left(\intop_{\mathcal{F}}d\mu_{\mathcal{F}}\phi_{2\left[\lambda,\alpha\right]}\hat{f}_{\left[\mathbf{\mathbf{x};\lambda,\alpha}\right]}\right).
\]
By rearranging the order of integration (Fubini\textquoteright s theorem)
and recognizing $\left(\mathcal{U}\psi_{1}\right)_{\left[\lambda,\alpha\right]}^{*}:$

\begin{align}
\left\langle \psi_{1},\psi_{2}\right\rangle _{\mathcal{H}_{x}} & =\intop_{F}d\mu_{\mathcal{F}}\phi_{1\left[\lambda,\alpha\right]}^{*}\phi_{2\left[\lambda,\alpha\right]},\nonumber \\
 & =\left\langle \phi_{1},\phi_{2}\right\rangle _{\mathcal{H}_{k}},\label{eq:PlancherelGeneralized}
\end{align}
where the last equality is obtained using the inner product (\ref{eq:explicitinnerprod}).

Relation (\ref{eq:PlancherelGeneralized}) is the statement of isometricity
of the GFT (\ref{eq:GFT}).
\begin{lyxlist}{00.00.0000}
\item [{\textbf{Theorem.}}] (Generalized Plancherel-Parseval): \textit{Let
$\Sigma$ be a Riemannian manifold and let $\mathcal{U}:\mathcal{H}_{x}\rightarrow\mathcal{H}_{k}$
be the generalized Fourier transform (\ref{eq:GFT}) associated with
the self-adjoint operator }$\hat{\mathcal{O}}=-\triangle.$\textit{
Then for any }$\psi_{1},\psi_{2}\in\mathcal{H}_{x}\sim\mathcal{L}^{2}\left[\Sigma\right]:$\textit{
\[
\left\langle \psi_{1},\psi_{2}\right\rangle _{\mathcal{H}_{x}}=\intop_{F}d\mu_{\mathcal{F}}\phi_{1\left[\lambda,\alpha\right]}^{*}\phi_{2\left[\lambda,\alpha\right]},
\]
where $\phi_{i}=\mathcal{U}\psi_{i}.$ Equivalently:
\[
\left\Vert \psi\right\Vert _{\mathcal{H}_{x}}^{2}=\left\Vert \phi\right\Vert _{\mathcal{H}_{k}}^{2}.
\]
Consequently, the generalized Fourier transform $\mathcal{U}$ is
a unitary isomorphism.}
\end{lyxlist}

\paragraph{Proof sketch.}

The spectral theorem for unbounded sself-adjoint operators on a Hilbert
space guarantees the existence of a unique direct-integral decomposition:
\[
\mathcal{H}_{x}\simeq\intop_{\mathcal{S}}^{\oplus}d\mu_{\mathcal{S}}\left.\mathcal{H}_{x}\right|_{\lambda},
\]
where $\left.\mathcal{H}_{x}\right|_{\lambda}$ denotes the fiber
Hilbert space associated with the spectral value $\lambda$, with
multiplicity determined by the structure of the degeneracy fiber $\left.\mathcal{F}\right|_{\lambda}.$
The corresponding projection-valued measure yields a resolution of
the identity:
\[
\mathbb{I}_{\mathcal{H}_{x}}=\intop_{\mathcal{S}}d\mu_{\mathcal{S}}\:\mathcal{K}_{\left[\cdot,\cdot,\lambda\right]}=\intop_{\mathcal{F}}d\mu_{\mathcal{F}}\:\hat{f}_{\left[\mathbf{\cdot;\lambda,\alpha}\right]}^{*}\hat{f}_{\left[\mathbf{\cdot;\lambda,\alpha}\right]},
\]
where $\hat{f}_{\left[\mathbf{x}\mathbf{;\lambda,\alpha}\right]}$
denotes a generalized eigenfunction of $\hat{\mathcal{O}}.$ For arbitrary
$\psi_{1},\psi_{2}\in\mathcal{H}_{x},$ we insert the identity to
obtain:
\[
\left\langle \psi_{1},\psi_{2}\right\rangle _{\mathcal{H}_{x}}=\left\langle \psi_{1},\mathbb{I}_{\mathcal{H}_{x}}\psi_{2}\right\rangle _{\mathcal{H}_{x}}=\intop_{\mathcal{F}}d\mu_{\mathcal{F}}\:\left\langle \psi_{1},\hat{f}_{\left[\mathbf{\cdot;\lambda,\alpha}\right]}\right\rangle _{\mathcal{H}_{x}}^{*}\left\langle \hat{f}_{\left[\cdot\mathbf{;\lambda,\alpha}\right]},\psi_{2}\right\rangle _{\mathcal{H}_{x}}.
\]
By the definition of the generalized Fourier transform coefficients
and the weighted orthonormality relations of the eigenfunctions, the
right-hand side becomes:
\[
\intop_{\mathcal{F}}d\mu_{\mathcal{F}}\:\left(\mathcal{U}\psi_{1}\right)_{\left[\lambda,\alpha\right]}^{*}\left(\mathcal{U}\psi_{2}\right)_{\left[\lambda,\alpha\right]}=\left\langle \mathcal{U}\psi_{1},\mathcal{U}\psi_{2}\right\rangle _{\mathcal{H}_{k}}.
\]
This establishes the presservation of the inner product. $\blacksquare$

Surjectivity of $\mathcal{U}$ follows from completeness of the generalized
eigenfunctions, which guarantees reconstruction of any $\psi\in\mathcal{H}_{x}$
from its spectral coefficients. Boundedness and invertibility are
guaranteed by the $\sigma$-finiteness of the spectral measure and
the structure of the direct-integral decomposition. The residual freedom
in choosing an orthonormal basis within each degenerate fiber $\left.\mathcal{H}_{x}\right|_{\lambda}$
corresponds to unitary transformations acting fiberwise, leaving the
overall unitarity of $\mathcal{U}$ invariant. Such gauge freedom
is addressed in later sections through symmetry-adapted choices of
local maximal Abelian sets of commuting operators (MASA - maximal
Abelian sub-algebra) .

This confirms that the GFT is a unitary isomorphism mapping the Laplace-Beltrami
operator to multiplication by the spectral parameter $\lambda$. For
readers interested in a more detailed  treatment, one could consult
\cite{Spectral1,Grigoryan,Spectral2}.

\subsection{Degeneracies and Non-Uniqueness of the Generalized Fourier Transform}

A general self-adjoint operator $\hat{\mathcal{O}}$ could admit degeneracies:
its spectrum $\lambda$ could have multiplicity greater than 1, labeled
by $\alpha$. For the Laplace-Beltrami operator (\ref{eq:LaplaceBeltrami}),
such degeneracies principally arise from the underlying symmetries
of the Riemannian manifold $\left(\Sigma,\boldsymbol{q}\right)$.
The presence or absence of this degeneracy provides a natural classification
for GFTs.

\subsubsection{The Non-Degenerate Case}

Let us consider the case where $\triangle$ contains no degeneracy,
i.e., where the spectrum is simple. As a consequence, the multiplicity
$\alpha$ of every spectral value $\lambda\in\mathcal{S}$ is equal
to 1. The generalized eigenfunction is labeled completely by $\lambda$
alone, hence we write it as $\hat{f}_{\left[\mathbf{x};\lambda\right]}.$
Each Hilbert fiber $\left.\mathcal{H}_{x}\right|_{\lambda}$ has dimension
1. The topological set of real eigenvalues $\mathcal{F}$ (\ref{eq:kspace})
is isomorphic to the spectrum itself:
\[
\mathcal{F}:=\bigcup_{\lambda\in\mathcal{S}}\left\{ \lambda\right\} \times1\cong\mathcal{S},
\]
where $\mathcal{S}$ is a subset of $\mathbb{R}^{+}.$

A manifold that does not admit a symmetry in general will typically
posses non-degenerate Laplace-Beltrami operator. This is supported
by the theorem on generic metrics \cite{uhlenbeck}: for a generic
Riemannian metric on any compact manifold (dimension $\ge2$) all
eigenvalues of $\triangle$ are simple. Physically, a generic $C^{\infty}$
metric perturbation on $\boldsymbol{q}$ breaks the geometric symmetries
and lifts the degeneracies.

A simple example is the non-compact half-line $\mathbb{R}^{+}$ (with
Dirichlet or Neumann boundary conditions), where the spectrum is continuous
and simple. (\textit{Remark}: the full line R remains degenerate with
multiplicity 2 due to reflection symmetry.)

Degeneracies arise prominently only in systems with symmetric geometries;
only in this case it is relevant to discuss the freedoms arising due
to the degeneracy in the operator.

\subsubsection{The Degenerate Case}

In the presence of spectral degeneracy, the definition of GFT in (\ref{eq:GFT})
imposes necessary but insufficient conditions to uniquely determine
the kernel $\hat{f}_{\left[\mathbf{x}\mathbf{;\lambda,\alpha}\right]}.$
While the kernel must diagonalize the Laplace-Beltrami operator $\triangle$
and satisfy the orthonormality condition (\ref{eq:orthonormalq-1}),
these constraints leave a residual degree of freedom. Given a fixed
eigenvalue $\lambda$, there exists a unitary freedom to choose an
orthonormal basis in the Hilbert fiber $\left.\mathcal{H}_{x}\right|_{\lambda}$.
This freedom corresponds to arbitrary unitary rotations mixing the
degenerate states labeled by $\alpha$. Consequently, the GFT kernel
(\ref{eq:GFT}) is not unique; distinct choices of basis yield distinct
Fourier transforms $\mathcal{U}$. As we will discuss in the following
sections, this ambiguity is physically resolved by selecting a maximal
commuting set of operators choice in $\mathcal{L}^{2}\left[\Sigma\right]$. 

Furthermore, distinct from the basis mixing freedom, there is a normalization
freedom. The Dirac-delta distribution $\delta_{\rho\left[\lambda,\alpha;\lambda',\alpha'\right]}$
(\ref{eq:ddeltak}) is defined with respect to the spectral measure
weight $\rho_{\left[\lambda,\alpha\right]}$. One is free to rescale
the eigenfunctions $\hat{f}_{\left[\mathbf{x};\lambda,\alpha\right]}$
and the weight $\rho_{\left[\lambda,\alpha\right]}$ inversely, such
that the completeness relation (\ref{eq:completeness}) remains invariant.
Nevertheless, once a normalization convention is fixed (e.g., to match
a specific asymptotic amplitude or group representation), the weight
$\rho_{\left[\lambda,\alpha\right]}$ becomes uniquely determined
by the Parseval identity. Therefore, within this framework, given
a fixed Riemannian manifold $\left(\Sigma,\boldsymbol{q}\right)$,
the generalized Fourier transform is not a single map, but a family
of unitary isomorphisms parameterized by (i) the choice of MASA used
to resolve degeneracy, and (ii) the normalization convention.

\subsubsection{Degrees of Freedom in the Generalized Fourier Transform.}

Let us return to the spectral theorem (\ref{eq:spectraltheorem}).
Given any Borel set $\mathcal{B}\subset\mathbb{R}$, there exists
a unique projection-valued measure $\mathcal{\hat{P}}_{\left[\cdot\right]}$
on $\mathcal{B}\subset\mathbb{R}$ such that the spectral decomposition
(\ref{eq:spectraltheorem}) is satisfied, with $\mathcal{\hat{P}}_{\left[\mathcal{B}\right]}\mathcal{\hat{P}}_{\left[\mathcal{B}'\right]}=\mathcal{\hat{P}}_{\left[\mathcal{B}\cap\mathcal{B}'\right]},$
$\mathcal{\hat{P}}_{\left[\mathbb{R}\right]}=\mathbb{I}$. In this
sense, the \emph{invariant} spectral content is carried by the pair
$\left(\hat{\mathcal{O}},\mathcal{\hat{P}}\right)$: once $\hat{\mathcal{O}}$
is fixed, the associated PVM $\mathcal{\hat{P}}$ is fixed.

A generalized Fourier transform (GFT) is then a choice of a unitary
spectral representation $\mathcal{U}:\mathcal{H}_{x}\rightarrow\mathcal{H}_{k}$
in which $\hat{\mathcal{O}}$ acts as multiplication by the spectral
parameter. For a fixed operator $\hat{\mathcal{O}}$, the non-uniqueness
of the GFT can be organized into four independent freedoms:
\begin{enumerate}
\item \emph{Operator choice.} Choosing a different self-adjoint operator
$\hat{\mathcal{O}}$ (hence a different $\hat{\mathcal{P}}$) changes
the spectral decomposition itself. This is the only freedom that changes
the invariant spectral data $\left(\hat{\mathcal{O}},\mathcal{\hat{P}}\right)$.
Since we fix the operator $\hat{\mathcal{O}}:=-\triangle$ (the
Laplace-Beltrami operator), the invariant spectral data will only
change if we change $\left(\Sigma,\boldsymbol{q}\right)$. We will
not consider this freedom further.
\item \emph{Basis rotation inside degenerate fibers.} For fixed $\hat{\mathcal{O}}$,
the spectrum may have multiplicity. In a direct-integral form $\mathcal{H}_{x}\simeq\int_{\mathcal{S}}^{\oplus}d\mu_{\mathcal{S}}\left.\mathcal{H}_{x}\right|_{\lambda}$,
one may choose a measurable family of unitary maps $\mathcal{W}_{\left[\lambda\right]}:\left.\mathcal{H}_{x}\right|_{\lambda}\to\left.\mathcal{H}_{x}\right|_{\lambda}$.
This corresponds to a $\lambda$-dependent unitary change of orthonormal
basis in each degenerate eigenspace (rotation in the degeneracy fiber/mixing
the modes labeled by $\alpha$). This changes the kernel/eigenfunctions
but leaves the PVM $\left(\hat{\mathcal{O}},\mathcal{\hat{P}}\right)$
invariant.
\item \emph{Spectral measure/normalization convention.} Within a fixed
spectral representation, one may rescale generalized eigenmodes by
a positive measurable weight $\hat{f}_{\left[\mathbf{x};\lambda,\alpha\right]}\mapsto\hat{f}'_{\left[\mathbf{x};\lambda,\alpha\right]}=\sqrt{w_{\left[\lambda,\alpha\right]}}\hat{f}_{\left[\mathbf{x};\lambda,\alpha\right]}$.
To preserve completeness/Plancherel, the spectral measure must transform
inversely, $d\mu_{\mathcal{F}}\mapsto d\mu'_{\mathcal{F}}=w_{\left[\lambda,\alpha\right]}^{-1}d\mu_{\mathcal{F}}$.
Thus the density appearing in $d\mu_{\mathcal{F}}$ is not an independent
choice; it is tied to the normalization convention of the kernel.
\item \emph{Coordinate (label) freedom on $\mathcal{F}$.} One may reparameterize
the spectral labels by a measurable bijection $\left(\lambda,\alpha\right)\mapsto\left(\lambda',\alpha'\right)$.
This induces a pushforward measure $d\mu'_{\mathcal{F}}=\left(\phi^{*}\right)d\mu_{\mathcal{F}}$
and correspondingly changes the coordinate form of the Dirac delta
by the appropriate Jacobian factors. This freedom changes only the
labeling/coordinate description of the same spectral decomposition
and does not affect the underlying PVM $\mathcal{\hat{P}}_{\left[\cdot\right]}$. 
\end{enumerate}
Therefore, for fixed $\left(\Sigma,\boldsymbol{q}\right)$, the GFT
is unique only up to (2) unitary rotations in Hilbert fibers, (3)
equivalent choices of spectral measure/normalization, and (4), choice
of coordinate/ label on \emph{$\mathcal{F}$.} These change the representation
but not the underlying spectral projector $\mathcal{\hat{P}}_{\left[\cdot\right]}$.
This paper focuses exclusively on resolving freedom (2) via the construction
of local set of maximally commuting operators. The later freedoms,
which relate to the geometrization of momentum space, will be discussed
in Part II of this work.

\section{Spectral Degeneracy and Operator Freedom}

For the next three sections, we focus on the degenerate case of GFT,
i.e., where the manifold admits sufficient symmetries resulting in
spectral degeneracy. This degeneracy grants us the freedom to construct
a complete set of commuting operators. Particularly in this section,
we will discuss a special set of operators that describe physical
observables: the\textit{ local differential operators.}

\subsection{Degeneracy and Orthonormal Basis Freedom}

The Helmholtz equation (\ref{eq:Helmholtz}) \textcolor{blue}{represents}
an eigenvalue problem (EVP) i.e., the diagonalization of operator
$\triangle$. For the case where $\Sigma$ has symmetries, the spectrum
of $\triangle$ may contains degeneracy. This is reflected by the
multiplicity of the eigenvalue $\lambda\in\mathcal{S}\subseteq\mathbb{R}^{+},$
labeled by $\alpha.$ 

Recall from Subsection II C that each distinct eigenvalue $\lambda\in\mathcal{S}$
is associated with the Hilbert fiber $\left.\mathcal{H}_{x}\right|_{\lambda}$,
spanned by the orthonormal basis $\hat{f}_{\left[\mathbf{x};\lambda,\alpha\right]}:$
\begin{equation}
\mathcal{H}_{x}\cong\mathcal{L}^{2}\left[\Sigma\right]=\intop_{\mathcal{S}}^{\oplus}d\mu_{\mathcal{S}}\left.\mathcal{H}_{x}\right|_{\lambda},\qquad\left(\mathcal{U}\psi\right)_{\left[\lambda\right]}\in\left.\mathcal{H}_{x}\right|_{\lambda}\forall\;\lambda\in\mathcal{S}.\label{eq:hilbertdecompose}
\end{equation}
Correspondingly, the Hilbert fiber on the momentum side is:
\[
\mathcal{H}_{k}\cong\mathcal{L}^{2}\left[\mathcal{F}\right]=\intop_{\mathcal{S}}^{\oplus}d\mu_{\mathcal{S}}\mathcal{L}^{2}\left[\left.\mathcal{F}\right|_{\lambda}\right],
\]
(or $\mathcal{\ell}^{2}\left[\left.\mathcal{F}\right|_{\lambda}\right]$
if $\alpha$ is discrete), where the dimension of the Hilbert fiber
is the cardinality of the degeneracy fiber $\left.\mathcal{F}\right|_{\lambda}$
of the fiber-bundle $\mathcal{F}$ in (\ref{eq:kspace}). Under the
GFT (\ref{eq:GFT}), these fibers are unitarily identified:
\begin{equation}
\left.\mathcal{H}_{x}\right|_{\lambda}\equiv\mathcal{L}^{2}\left[\left.\mathcal{F}\right|_{\lambda}\right],\label{eq:partition}
\end{equation}
hence they are isomorphic to each other.

The Helmholtz equation (\ref{eq:Helmholtz}) fixes the eigenvalue
$\lambda$, but leaves the basis vectors within the fiber $\left.\mathcal{H}_{x}\right|_{\lambda}$
undefined up to a unitary rotation. To resolve this ambiguity, we
look for auxiliary operators that commute with $\triangle$.
\begin{lyxlist}{00.00.0000}
\item [{\textbf{Theorem\label{theorem}.}}] (Commutant Invariant Eigenspace):
\textit{Let $\left(\Sigma,\boldsymbol{q}\right)$ be a Riemannian
manifold and let $\triangle$ be the self-adjoint Laplace--Beltrami
operator acting on a dense domain in $\mathcal{L}^{2}\left[\Sigma\right]$.
Let $\hat{\mathcal{O}}$ be a linear operator acting on a dense domain
$D\left[\hat{\mathcal{O}}\right]\subset\mathcal{L}^{2}\left[\Sigma\right],$
such that:
\begin{equation}
\left(\hat{\mathcal{O}}\triangle-\triangle\hat{\mathcal{O}}\right)\psi:=\left[\hat{\mathcal{O}},\triangle\right]\psi=0,\qquad\forall\;\psi\;\in\mathcal{D}\subset D\left[\triangle\right]\cap D\left[\hat{\mathcal{O}}\right],\label{eq:hypotheses}
\end{equation}
$\mathcal{D}$ is the common invariant core for both operators. Then
for every eigenvalue $\lambda\in\mathcal{S}$, the corresponding eigenspace
$E_{\lambda}:=\ker(\Delta-\lambda I)$ is invariant under $\hat{\mathcal{O}}$,
that is:
\[
f\in E_{\lambda}\Longrightarrow\hat{\mathcal{O}}f\in E_{\lambda}.
\]
}
\end{lyxlist}

\paragraph{Proof sketch. }

Let $f\in E_{\lambda}\cap\mathcal{D}$ be an arbitrary eigenfunction
of $\triangle$ corresponding to the eigenvalue $\lambda$. By definition
of the eigenspace, $f$ satisfies the Helmholtz equation: $\triangle f=\lambda f$.
We wish to determine the action of $\triangle$ on the vector $\hat{\mathcal{O}}f.$
Applying $\triangle$ from the left and announcing the commutation
hypothesis $\left[\hat{\mathcal{O}},\triangle\right]f=0$, we may
interchange the operators:
\[
\triangle\left(\hat{\mathcal{O}}f\right)=\hat{\mathcal{O}}\left(\triangle f\right).
\]
Substituting $\triangle f=\lambda f$ into the right-hand side gives:
\[
\hat{\mathcal{O}}\left(\triangle f\right)=\hat{\mathcal{O}}\left(\lambda f\right).
\]
By the linearity of $\hat{\mathcal{O}},$ the scalar eigenvalue $\lambda$
factors out: $\hat{\mathcal{O}}\left(\lambda f\right)=\lambda\left(\hat{\mathcal{O}}f\right).$
Combining these steps, we arrive at:
\[
\triangle\left(\hat{\mathcal{O}}f\right)=\lambda\left(\hat{\mathcal{O}}f\right).
\]
Therefore, vector $\hat{\mathcal{O}}f$ is an eigenvector of $\triangle$
with the same eigenvalue $\lambda$. Therefore, $\hat{\mathcal{O}}f\in E_{\lambda}$.
Since $f$ was arbitrary, the entire subspace $E_{\lambda}$ is invariant
under the action of $\hat{\mathcal{O}}$. $\blacksquare$
\begin{lyxlist}{00.00.0000}
\item [{\textbf{Corollary.}}] (Joint Diagonalization). \textit{Since $\hat{\mathcal{O}}$
maps $E_{\lambda}$ to itself, we can restrict the operator to this
subspace, denoted as $\left.\hat{\mathcal{O}}\right|_{E_{\lambda}}.$
If $\hat{\mathcal{O}}$ is also self-adjoint, the spectral theorem
for finite-dimensional (or compact) operators guarantee that $E_{\lambda}$
admits a basis of common eigenvectors for both $\triangle$ and $\hat{\mathcal{O}}$.
This provides the mechanism to resolve the degeneracy labeled by $\alpha$.}
\item [{\textbf{Corollary.}}] (Commutation-forced Degeneracy): \textit{Assume
the hypotheses (\ref{eq:hypotheses}). Fix an eigenvalue $\lambda$
and restrict $\hat{\mathcal{O}}$ to the invariant eigensspace $E_{\lambda}$.
If the restricted operator $\hat{\mathcal{O}}|_{E_{\lambda}}$ acts
non-trivially (i.e., it is not a scalar multiple of the identity),
then $\dim E_{\lambda}\ge2$. In other words, the existence of a non-scalar
commuting operator implies that the eigenvalue $\lambda$ is degenerate.
Conversely, if $\lambda$ is a simple eigenvalue, then necessarily
$\hat{\mathcal{O}}|_{E_{\lambda}}=\alpha_{\lambda}\mathbb{I}$ for
some $\alpha_{\lambda}\in\mathbb{C}$. Moreover, if $\hat{\mathcal{O}}$
is self-adjoint, then $\alpha_{\lambda}\in\mathbb{R}$.}
\end{lyxlist}

\paragraph{Proof sketch.}

Proceeding by contraposition. We assume that $\dim E_{\lambda}=1$.
Let $\hat{f}$ be a normalized basis vector for $E_{\lambda}$. Since
$E_{\lambda}$ is invariant under $\hat{\mathcal{O}},$ the vector
$\hat{\mathcal{O}}\hat{f}$ must be proportional to $\hat{f}$. That
is, $\hat{\mathcal{O}}\hat{f}=\alpha_{\lambda}\hat{f}$ for some scalar
$\alpha_{\lambda}$. Since any vector in $E_{\lambda}$ is a multiple
of $\hat{f}$, $\hat{\mathcal{O}}$ acts as multiplication by $\alpha_{\lambda}$
on the entire subspace. Thus, $\hat{\mathcal{O}}|_{E_{\lambda}}$
is a scalar operator. Logically, this implies that if $\hat{\mathcal{O}}|_{E_{\lambda}}$
is not scalar, the dimension cannot be 1. $\blacksquare$

The above statements are purely algebraic: they rely solely on the
existence of nontrivial operators commuting with $\triangle$. In
the following subsections, we will show that the geometric isometries
of $\left(\Sigma,\boldsymbol{q}\right)$ provide the canonical generators
for such operators.

\paragraph*{Remark {[}The Ground State Exception{]}.}

The presence of symmetries on $\Sigma$ does not guarantee that every
eigenvalue of $\triangle$ is degenerate. A trivial representation
of the symmetry group always exists. For example, on any compact connected
Riemannian manifold, the ground state (lowest eigenvalue $\lambda=0$)
corresponds to constant functions. This eigenspace is always 1-dimensional
(simple), regardless of the manifold\textquoteright s symmetry. In
this case, any symmetry operator acts as the identity (scalar) on
the constant function, consistent with the corollaries. Thus, the
refined physical statement is: Degeneracy is required only if the
symmetry breaks the \textquotedblleft shape\textquotedblright{} of
the eigenfunction (i.e., the operator acts non-scalarly).

\subsection{Local vs. Non-Local Operators}

We have established that degeneracy can be resolved by a commuting
operator. But which one? We now distinguish between synthetic operators
(mathematically constructed to force uniqueness) and local operators
(physically motivated by the geometry).

\subsubsection{Arbitrary (Synthetic) Operator Construction}

Let us demonstrate that a commuting operator always exists. We can
construct a self-adjoint operator $\hat{\mathcal{O}}$ explicitly
by defining its action on the basis states of each fiber $\left.\mathcal{H}_{x}\right|_{\lambda}$.
For a fixed eigenvalue $\lambda$, let the fiber be spanned by an
arbitrary orthonormal basis labeled by $\alpha$. We define the fiber-operator
$\left.\hat{\mathcal{O}}\right|_{\lambda}$ such that these basis
vectors are its eigenvectors:
\begin{equation}
\left.\hat{\mathcal{O}}\right|_{\lambda}\hat{f}_{\left[\mathbf{x};\lambda,\alpha\right]}:=\alpha\hat{f}_{\left[\mathbf{x};\lambda,\alpha\right]}.\label{eq:operators}
\end{equation}
Originally, $\alpha$ denotes an abstract degeneracy label; we fix
a preferred labeling by diagonalizing an auxiliary (commuting) self-adjoint
operator on each fiber, so that $\alpha$ is identified with its spectral
parameter. The self-adjointness of $\left.\hat{\mathcal{O}}\right|_{\lambda}$
guarantees that the degeneracy label are real: $\alpha\in\left.\mathcal{F}\right|_{\lambda}\subseteq\mathbb{R}$.
If the spectrum of $\left.\hat{\mathcal{O}}\right|_{\lambda}$ is
still degenerate, one may further refine the labeling by diagonalizing
an additional commuting operator, yielding joint spectral labels. 

As consequences of the definitions, (a) $\left.\hat{f}_{\left[\mathbf{x};\alpha\right]}\right|_{\lambda}$
diagonalize $\left.\hat{\mathcal{O}}\right|_{\lambda}$ with spectrum
$\alpha$, (b) $\left.\hat{f}_{\left[\mathbf{x};\alpha\right]}\right|_{\lambda}$
is complete and orthonormal on $\left.\mathcal{H}_{x}\right|_{\lambda}$
, and (c) $\left.\hat{\mathcal{O}}\right|_{\lambda}$ is linear, densely-defined,
and symmetric. Since $\alpha\in\left.\mathcal{F}\right|_{\lambda}\subseteq\mathbb{R}$,
then $\left.\hat{\mathcal{O}}\right|_{\lambda}$ is guaranteed to
be self-adjoint, furthermore, $\left.\hat{\mathcal{O}}\right|_{\lambda}$
is bounded if $\sup\left|\alpha\right|<\infty,$ otherwise, it is
unbounded.

One could extend $\left.\hat{\mathcal{O}}\right|_{\lambda}$ to the
entire full Hilbert space $\mathcal{H}_{x}=\mathcal{L}^{2}\left[\Sigma\right]$
by the direct sum (or direct integral, for continuous case):
\begin{align}
\hat{\mathcal{O}}=\bigoplus_{\lambda}\left.\hat{\mathcal{O}}\right|_{\lambda},\quad\mathrm{(discrete)},\qquad\hat{\mathcal{O}} & =\intop_{\mathcal{S}}^{\oplus}d\mu_{\mathcal{S}}\left.\hat{\mathcal{O}}\right|_{\lambda},\quad\mathrm{(continuous)}.\label{eq:generaloperators}
\end{align}
Since $\triangle$ acts as the scalar $\lambda$ on each Hilbert fiber
$\left.\mathcal{H}_{x}\right|_{\lambda}$, it trivially commutes with
any operator $\left.\hat{\mathcal{O}}\right|_{\lambda}$ that acts
purely within that fiber, i.e., $\left[\hat{\mathcal{O}},\hat{\triangle}\right]\psi_{\left[\mathbf{x}\right]}=0,$
for every $\psi_{\left[\mathbf{x}\right]}\in\mathcal{H}_{x}.$

Thus, one can always artificially construct a commuting operator $\hat{\mathcal{O}}$
to \textquotedblleft label\textquotedblright{} the degeneracy. Even
so, this construction is non-unique and generally lacks physical meaning.
The resulting operator $\hat{\mathcal{O}}$ is typically an integral
operator with no clear geometric origin.

\subsubsection{The Kernel Criterion for Locality}

To select a physically meaningful basis, we impose the constraint
of locality. An operator $\hat{\mathcal{O}}$ on $\mathcal{L}^{2}\left[\Sigma\right]$
is \textit{local in $\Sigma$} if the value of $\hat{\mathcal{O}}\psi_{\left[\mathbf{x}\right]}$
at a point $\mathbf{x}\in\Sigma$ depends only on the value of $\psi$
and its (finitely many) derivatives at $\mathbf{x}.$

Mathematically, this is characterized by the integral kernel $\mathcal{\hat{K}}_{\mathcal{O}\left[\mathbf{x},\mathbf{x}'\right]}$
in the representation:
\begin{equation}
\hat{\mathcal{O}}\psi_{\left[\mathbf{x}\right]}=\intop_{\Sigma}d\mu_{\Sigma\left[\mathbf{x}'\right]}\mathcal{\hat{K}}_{\mathcal{O}\left[\mathbf{x},\mathbf{x}'\right]}\psi_{\left[\mathbf{x}'\right]}.\label{eq:kerneldefs}
\end{equation}

\begin{lyxlist}{00.00.0000}
\item [{\textbf{Definition.}}] (Locality of an Operator): \textit{The operator
$\hat{\mathcal{O}}$ is local}\footnote{\textit{In this work, by \textquoteleft local\textquoteright{} we
mean finite-order differential operators, equivalently kernels given
by finite sums of $\delta$ and its derivatives supported on the diagonal.}}\textit{ for every $\psi_{\left[\mathbf{x}\right]}\in\mathcal{H}_{x}$
iff its kernel $\mathcal{\hat{K}}_{\mathcal{O}\left[\mathbf{x},\mathbf{x}'\right]}$
is a distribution supported solely on the diagonal $\mathbf{x}=\mathbf{x}'$. }
\end{lyxlist}
By Peetre's Theorem \cite{Peetre}, any such local linear operator
is necessarily a differential operator of finite order. Its kernel
takes the form:

\begin{equation}
\mathcal{\hat{K}}_{\mathcal{O}\left[\mathbf{x},\mathbf{x}'\right]}=\sum_{|i|\le r}c_{i\left[\mathbf{x}\right]}\,\partial_{x}^{i}\delta_{q\left[\mathbf{x},\mathbf{x}'\right]},\label{eq:localitycondition}
\end{equation}
which is a finite linear combination of delta functions and its derivatives.
$i$ is the order of the differential $\partial_{x}^{i}=\frac{\partial^{i}}{\partial x^{i}}$,
$r$ is a finite number, $c_{i\left[\mathbf{x}\right]}$ is the weight
coefficient, and $\delta_{q\left[\mathbf{x},\mathbf{x}'\right]}$
is the delta function of $\Sigma$ satisfying (\ref{eq:ddeltaq}).
Otherwise, $\mathcal{\hat{K}}_{\mathcal{O}\left[\mathbf{x},\mathbf{x}'\right]}$
has non-zero weight for $\mathbf{x}\neq\mathbf{x}'$, hence $\hat{\mathcal{O}}$
is non-local.

Inserting (\ref{eq:localitycondition}) to (\ref{eq:kerneldefs})
implies the explicit action of operator $\hat{\mathcal{O}}$ at $\psi$:
\begin{equation}
\hat{\mathcal{O}}\psi_{\left[\mathbf{x}\right]}=\sum_{|i|\le r}c_{i\left[\mathbf{x}\right]}\partial_{x}^{i}\psi_{\left[\mathbf{x}\right]};\label{eq:diffop}
\end{equation}
any such operator written in this form is automatically local. Hence,
a local operator acts on a function over a manifold as a finite order
derivative.

\paragraph{Remarks {[}The Problem with Synthetic Operators{]}.}

The synthetic operator constructed in (\ref{eq:generaloperators})
generally admits a kernel $\mathcal{\hat{K}}_{\mathcal{O}\left[\mathbf{x},\mathbf{x}'\right]}$
that is non-zero for $\mathbf{x}\neq\mathbf{x}'$. It is generally
\textit{non-local} \textit{in $\Sigma$}. As a consequence, they do
not inherit symmetry related to $\Sigma$ (in the sense of the isometry/Killing
field in $\Sigma$) and has no inherent algebraic relation, eventhough
it commutes with $\triangle$. Therefore, we cannot rely on arbitrary
mathematical construction, and search for commuting operators that
are intrinsically local differential operators. As we will see, such
operators arise from the geometric symmetries (isometries) of the
manifold $\Sigma$.

\subsection{Set of Commuting Local Operators}

From a physical perspective, the locality of operators is paramount.
Locality preserves causality, enables local conservation laws, construct
well-defined dynamics, and ensures that measurements are physically
realizable. Non-local operators are generally considered unphysical
unless introduced deliberately under strict control (e.g., Wilson
loops in gauge theory).

The Laplace-Beltrami operator $\triangle$ is the prototypical example
of a local operator. However, as noted, the spectrum of $\triangle$
is often degenerate due to the symmetries of $\Sigma$. Moreover,
if $\Sigma$ admits a sufficiently large commutative algebra of local
(differential) operators commuting with $\triangle$, for a set of
eigenfunction $\hat{f}_{\left[\mathbf{x};\lambda,\alpha\right]}$
that diagonalize $\triangle$, there exists a set of operators $\left\{ \hat{\mathcal{O}}_{i}\right\} $
sharing  the same eigenfunction $\hat{f}_{\left[\mathbf{x};\lambda,\alpha\right]}$.
To resolve this, we seek a complete set of commuting operators (CSCO)
that shares the same eigenfunctions.

Our specific task is to collect a set of operators $\left\{ \hat{\mathcal{O}}_{1},...,\hat{\mathcal{O}}_{i}\right\} ,$
such that each operator $\hat{\mathcal{O}}_{i}$:
\begin{description}
\item [{(i)}] Commutes with ``Hamiltonian'': $\left[\hat{\mathcal{O}}_{i},\triangle\right]=0$,
\item [{(ii)}] Commutes with each others (mutual commutation), namely $\left[\hat{\mathcal{O}}_{i},\hat{\mathcal{O}}_{j}\right]=0$
for all $i,j$ in $\left\{ \hat{\mathcal{O}}_{i}\right\} $, and,
\item [{(iii)}] Local: each $\hat{\mathcal{O}}_{i}$ is a local differential
operator (satisfying the kernel criterion of (\ref{eq:localitycondition})).
\end{description}
\begin{lyxlist}{00.00.0000}
\item [{\textbf{Definition.}}] (Local Basis): \textit{If a choice of orthonormal
basis $\hat{f}_{\left[\mathbf{x};\lambda,\alpha\right]}$ diagonalizes
such a set $\left\{ \triangle,\hat{\mathcal{O}}_{1},...,\hat{\mathcal{O}}_{i}\right\} ,$
then the choice of $\hat{f}_{\left[\mathbf{x};\lambda,\alpha\right]}$
is \textquotedbl local'', in the sense that it is an eigenfunction
that diagonalizes a set of commuting, local operators.}
\end{lyxlist}
This distinguishes it from arbitrary bases generated by synthetic
(non-local) operators.

As summary, for the degenerate case, the GFT admits freedoms: one
of them is the freedom to choose orthonormal basis in the Hilbert
fiber $\left.\mathcal{H}_{x}\right|_{\lambda}.$ Selecting an orthonormal
basis is equivalent to choosing a set of self-commuting operators
$\left\{ \hat{\mathcal{O}}_{i}\right\} $ that commute with $\triangle$.
While arbitrary mathematical choices exist (synthetic operators),
they are physically undesirable due to non-locality. A physically
meaningful local basis is related to a special cases of operators
satisfying the locality condition (\ref{eq:localitycondition}). Therefore,
the physical GFT problem reduces to a geometric one: Does the manifold
$\Sigma$ admit enough local differential operators to form a CSCO?
A systematic procedure to construct these using Killing vectors is
the subject of the next section.

\section{Geometric Operators and Symmetry-Adapted Bases}

In this section, we discuss a general method to obtain \textit{geometric
operators}: the local operators that respect the symmetry related
to $\Sigma$. The procedure adopts the concept of Liouville (or quantum)
integrability in symplectic geometry, where the existence of these
local operators are related to the existence of the Killing fields
on the manifold.

\paragraph*{The Problem of \textquotedbl Lost Geometry''.}

From a mathematical perspective, it is necessary to obtain operators
(and its corresponding bases) that encode the geometry of the underlying
manifold $\Sigma$. As discussed in Section II B, the structure of
the Hilbert space $\mathcal{H}_{x}=\mathcal{L}^{2}\left[\Sigma\right]$
is determined (up to isomorphism) solely by its dimension (or by the
cardinality of its orthonormal basis). As a consequence, the Hilbert
space ``forgets'' the geometry: the information on the topological
and geometrical structure of $\Sigma$ is lost at $\mathcal{H}_{x}.$
One could replace $\Sigma$ with an entirely different set $\Sigma'$
and their Hilbert space would remain isomorphic (asssuming they are
infinite dimensional and separable).

However, the geometric information of $\Sigma$ could be carried by
some specific sets of orthonormal bases. These symmetry-adapted bases
carry the information of the geometrical structure of $\Sigma$ because
they are eigenfunctions of geometric operators constructed from Killing
fields.

\paragraph*{Criteria for Geometric Operators. }

To obtain the basis that recover the geometry of $\Sigma$, the local
geometric operators we need to construct must satisfy two conditions:
(a) locality: they must be finite-order differentials operators (by
Peetre's theorem \cite{Peetre}), and (b) compatibility: they must
strictly commutes with the Laplace-Beltrami operator in $\Sigma$
(carrying the symmetry information of $\Sigma$). 

Candidates for such operators are constructed from Killing vectors
data\footnote{By \textquoteleft constructed from Killing data\textquoteright{} we
include operators in the algebra generated by Killing generators (e.g.
symmetrized products / elements of enveloping algebra), not only first-order
ones.} (isometry) on $\Sigma$. For a more general case where $\Sigma$
lacks sufficient Killing vectors (e.g. triaxial ellipsoids), one must
consider Killing tensors/higher-order symmetry data. Note that for
Killing tensors, the construction of a commuting operator is not automatic
due to ordering ambiguities and curvature obstruction; specific conditions
are required to guarantee strict commutativity with $\triangle$ \cite{prolongation,prolongation1a,prolongation2,obstruct1,obstruct2}. 

We use this fact to construct a set with maximal number of commuting
local operators (the geometric operators) for the degenerate case
of our GFT. 

\subsection{Killing Fields and Maximal Abelian Sub-Algebra (MASA)}

\subsubsection{Killing Vector Fields}

A vector field $\boldsymbol{K}=K^{i}\partial_{i}$ on $\left(\Sigma,\mathbf{q}\right)$
is a Killing vector field if it generates an isometry. Geometrically,
it must satisfies the Killing vector condition $\mathcal{L}_{\boldsymbol{K}}\mathbf{q}=0,$
which takes the local covariant form:
\begin{equation}
\nabla_{\left(i\right.}K_{\left.j\right)}=\frac{1}{2}\left(\nabla_{i}K_{j}+\nabla_{j}K_{i}\right)=0,\qquad i,j=1,...,n.\label{eq:Killingvector}
\end{equation}
By the Picard-Lindel$\ddot{\textrm{o}}$f theorem, the existence of
such fields generates a (local) one-parameter group of isometries
(global if the field is complete), where the corresponding integral
curve $\phi_{t}:\Sigma\rightarrow\Sigma$ satisfies \cite{ODE}:
\begin{equation}
\frac{d}{dt}\phi_{t\left[\mathbf{x}\right]}=\boldsymbol{K}\left(\phi_{t\left[\mathbf{x}\right]}\right).\label{eq:flow}
\end{equation}
One could consider the (local) flow $\phi_{t\left[\mathbf{x}\right]}$
as a diffeomorphism that preserve metric, i.e., a (Lie) group of isometries
in $\Sigma$, say $\mathcal{G}$.

The solution to (\ref{eq:Killingvector}) construct a (sub)space of
solution $\mathcal{K}\subseteq\oplus^{n}\mathcal{L}^{2}\left[\Sigma\right]$
with dimension $R\leq\frac{1}{2}n\left(n+1\right).$ The Killing vector
could be written as a linear combination of the basis in $\mathcal{K},$
namely $\boldsymbol{K}=\alpha^{a}\hat{\sigma}_{a}$, where $\alpha^{a}$
scalar coefficients and $\hat{\sigma}_{a}\in\mathfrak{g},$ with $1\leq a\leq R$,
are the generators of the Lie algebra $\mathfrak{g}$ of $\mathcal{G}$.
Notice that each generator $\hat{\sigma}_{a}$ could be written as
a linear combination of the coordinate basis, namely: $\hat{\sigma}_{a}=\sigma_{a}^{i}\partial_{i}$,
so that $\boldsymbol{K}$ could be written as: $\boldsymbol{K}=\alpha^{a}\sigma_{a}^{i}\partial_{i},$
where $\alpha^{a}\sigma_{a}^{i}=K^{i}$.

For $A,B\in T_{\mathbf{x}}\Sigma$, the Lie bracket operator $\left[..,..\right]$
is defined in $\mathfrak{g}$ as:
\begin{equation}
[A,B]^{\,i}=A^{j}\nabla_{j}B^{i}-B^{j}\nabla_{j}A^{i}.\label{eq:Liebracket}
\end{equation}
For the case of maximally symmetric space, where the rank $R$ is
maximal ($\frac{1}{2}n\left(n+1\right)$), the Laplace-Beltrami operator
$\triangle$ coincides with the quadratic Casimir element $\hat{\mathcal{C}}=\hat{\sigma}^{a}\hat{\sigma}_{a}$,
where the inner product in $\mathfrak{g}$ is defined by a bilinear
map.

\subsubsection{Maximal Abelian Sub-Algebra and Complete Set of Commuting (Local)
Operators}

To construct our CSCO, we cannot use the entire algebra $\mathfrak{g}$
because the generators do not necessarily commute ($\left[\hat{\sigma}_{a},\hat{\sigma}_{b}\right]\neq0$).
One could construct a sub-algebra $\mathfrak{h}\subset\mathfrak{g},$
defined by elements of $\mathfrak{g}$ that commutes with each other.
The basis that spans the sub-algebra $\mathfrak{h}$, say $\left\{ \hat{\mathcal{O}}_{a}\right\} $,
construct the maximal commuting set of local operators (or maximal
Abelian sub-algebra/MASA\footnote{Throughout this paper, the term \textquotedbl MASA'' will describe
a maximal commuting set of local operators arising from the Killing
fields. In general, MASA is only a maximal set of commuting elements
in a Lie algebra; we make sure to state clearly when we intent to
use the later. }) in $\mathfrak{g}$ which, in special case like $\mathbb{R}^{n}$,
is exactly the complete set of commuting operators (CSCO) in quantum
mechanics. The rank of $\mathfrak{h}$ is $r\leq R$, where $r$ is
the number of basis in the subalgebra, namely, the cardinal of $\left\{ \hat{\mathcal{O}}_{a}\right\} $.
In addition to the Laplace--Beltrami operator $\triangle$, the MASA
is \textit{complete} if its rank $r=n-1,$ namely: $\left\{ \hat{\mathcal{O}}_{1},...,\hat{\mathcal{O}}_{n-1}\right\} $
with functionally independent principal symbols. 

\subsubsection{Killing Tensor Fields and Hidden Symmetries}

However, for general curved Riemannian manifold, the generators $\left\{ \hat{\mathcal{C}}_{a}\right\} $
of $\mathfrak{h}$ is not sufficient to describe the degrees of freedom
of the degeneracy in $\triangle$. The degrees of freedom of a system
is normally defined as the topological dimension of $\Sigma$, which
is $n,$ while $\mathfrak{h}$ has dimension $r\leq R$. In addition
to $\triangle,$ we need $\left(n-1-r\right)$ more operators
to describe the remaining degrees of freedom.

To recover these missing operators, we look at the \textquotedbl hidden
symmetries'' in $\Sigma$. The object that could capture these hidden
symmetries are the Killing tensor fields $\overline{\boldsymbol{K}}$
(with its components $K_{ij}$), defined by solving:
\begin{equation}
\nabla_{\left(i\right.}K_{\left.jk\right)}=0,\qquad\overline{\boldsymbol{K}}=K^{ij}\partial_{i}\otimes\partial_{j},\;1\leq i,j\leq n.\label{eq:Killingtensor}
\end{equation}
In a similar way, we could expand the Killing tensor as $\overline{\boldsymbol{K}}=\beta^{a}\hat{\kappa}_{a}$,
with $\hat{\kappa}_{a}$ are the basis of the space of solution to
(\ref{eq:Killingtensor}). However, in constrast with the Killing
vectors, the collection of $\left\{ \hat{\kappa}_{a}\right\} $, generally,
does not span a Lie algebra; they span a linear tensor space $\mathfrak{i}$
that is not related to the isometry group in $\Sigma$. In physical
terms, unlike Killing vectors, Killing tensors do not generate geometric
isometries but rather correspond to conservation laws quadratic (or
higher) in momentum (like the Runge--Lenz vector).

Using the Schouten-Nijenhuis bracket $[...,...]_{{\rm SN}}$ \cite{Nijenhuis,Schouten,Operatorbook},
defined as a generalization of the Lie bracket (\ref{eq:Liebracket}):
\[
[A,B]_{\textrm{SN}}^{\,i_{1}\dots i_{r+s-1}}\;=\;r\,A^{j(i_{1}\dots i_{r-1}}\nabla_{j}B^{i_{r}\dots i_{r+s-1})}\;-\;s\,B^{j(i_{1}\dots i_{s-1}}\nabla_{j}A^{i_{s}\dots i_{r+s-1})},
\]
for $A,B$ are, respectively, symmetric contravariant tensors of rank
$r$ and $s,$ we could define the commutator inside $\mathfrak{i}$
and construct the maximally Abelian sub-algebra $\mathfrak{j}$ inside
$\mathfrak{i}$, where it's component $\left\{ \hat{\kappa}_{a}\right\} $
satisfies $\left[\hat{\kappa}_{a},\hat{\kappa}_{\beta}\right]_{\textrm{SN}}=0$.
These objects are ready to fill the missing operator parts.

Let us formalize out result by proposing a principle as follows. Throughout,
$\triangle$ denotes the (essentially) self-adjoint Laplace--Beltrami
operator on $\mathcal{H}:=\mathcal{L}^{2}\left(\Sigma,d\mu_{\Sigma}\right)$.
\begin{lyxlist}{00.00.0000}
\item [{\textbf{Definition.}}] (Fiberwise MASA/Degeneracy Resolution).
\textit{Let (\ref{eq:hilbertdecompose}) be the spectral decomposition
of (the Hilbert space of) $\triangle$, where $\left.\mathcal{H}_{x}\right|_{\lambda}$
is the multiplicity (degeneracy) fiber at spectral value $\lambda$.
A }\textbf{\textit{\emph{fiberwise MASA}}}\textit{ is a maximal abelian
$*$-subalgebra $\mathcal{\mathcal{D}}_{\lambda}\subset\mathcal{O}_{\left[\mathcal{H}\right]}$,
with $\mathcal{O}_{\left[\mathcal{H}\right]}$ is the set of operator
on $\mathcal{H}$. }
\end{lyxlist}
Equivalently, choosing $\mathcal{\mathcal{D}}_{\lambda}$ is the same
as choosing an orthonormal basis in each $\mathcal{H}_{\lambda}$
(up to phase), hence fixing the degeneracy labels $\alpha$. In this
work we are primarily interested in \emph{local} degeneracy resolutions,
i.e.\ those arising from commuting families of (essentially) self-adjoint
\emph{local differential operators} on $\Sigma$ that commute with
$\triangle$ (see Section 3 and 4).
\begin{lyxlist}{00.00.0000}
\item [{\textbf{Proposition.}}] (Resolving basis freedom via a commuting
MASA). \textit{Let $\triangle$ be (essentially) self-adjoint on $\mathcal{H}=\mathcal{L}^{2}\left(\Sigma\right)$
and assume we work in the discrete (point) spectrum so that each eigenspace
$E_{\lambda}:=\ker(\triangle-\lambda I)$ is finite-dimensional. Whenever
$\dim E_{\lambda}>1$, an eigenbasis of $\triangle$ is defined only
up to a unitary rotation. If a fiberwise MASA $\mathcal{\mathcal{D}}_{\lambda}\subset\mathcal{O}_{\left[\mathcal{H}\right]}$
is self-adjoint operators and commute with $\triangle$, then the
joint spectral (\ref{eq:kspace}) yields a joint spectral resolution
and hence a joint eigenbasis that fixes this freedom (up to phases)
by simultaneously diagonalizing $\mathcal{\mathcal{D}}_{\lambda}$
on each $E_{\lambda}$. }
\end{lyxlist}
Different choices of $\mathcal{\mathcal{D}}_{\lambda}$ may lead to
different joint spectra and therefore to different kernels.

\paragraph{Remark {[}Geometric Selection Principle{]}.}

To make the above mathematical choice physically meaningful, we restrict
our choice of MASA to the algebra generated by geometric symmetry
operators (Killing vectors and tensors), i.e., to commuting \emph{geometric
symmetry operators} (finite-order differential operators) that commute
with $\triangle$. This principle ensures that the resulting basis
functions are not just random mathematical constructs, but are symmetry-adapted
bases labeled by conserved quantities (e.g. local ``momentum'' labels/quantum
numbers).

In the next subsections, we provide a constructive scheme to obtain
such operators from Killing vectors/tensors (and their higher-order
generalizations), hence producing a geometrically motivated MASA that
resolves the basis freedom in physically local terms.

\subsection{The Operator Construction Algorithm}

We provide the algorithm to construct a set of commuting, local, geometric
operators in $\Sigma$ that simultaneously shares a same eigenfunction
with $\triangle$. This procedure builds the fiberwise MASA \textit{$\mathcal{\mathcal{D}}_{\lambda}$
}discussed in the previous section.
\begin{description}
\item [{Step$\;$1}] \textbf{Solve the Killing--vector equation}. Pick
any local chart $\left(x^{1},\dots,x^{n}\right)$ on an open patch
$U\subset\Sigma$. Solve the Killing vector condition (\ref{eq:Killingvector})
to obtain $\boldsymbol{K}=\alpha^{a}\hat{\sigma}_{a}$, expanded in
its internal space basis. The generator $\hat{\sigma}_{a}\in\mathfrak{g}$
is a function of first derivative operators.
\item [{Step$\;$2}] \textbf{Extract a commuting subset of vectors (maximal
Abelian sub-algebra)}. Select a maximal Abelian sub-algebra $\mathfrak{h}\subset\mathfrak{g}$
(rank $r=\dim\mathfrak{h}$). In practice: pick $r$ linearly independent
Killing vectors that pairwise commute:
\[
\left[\hat{\sigma}_{a},\hat{\sigma}_{b}\right]=0,\qquad a,b=1,\dots,r.
\]
Record $\left\{ \hat{\sigma}_{1},...,\hat{\sigma}_{r}\right\} $ as
the first block of commuting local operators:
\[
\hat{\mathcal{O}}_{a}^{\left(1\right)}:=\hat{\sigma}_{a}\in\mathcal{D}^{\left(1\right)},\qquad a=1,\dots,r,
\]
where $\mathcal{D}^{\left(1\right)}$ is the family of first-order
differential operators. Notice that $\hat{\sigma}_{a}$ could be written
as $\hat{\sigma}_{a}=\sigma_{a}^{i}\nabla_{i},$but since it acts
on functions, it is safe to write $\hat{\sigma}_{a}=\sigma_{a}^{i}\partial_{i}.$
\item [{Step$\;$3}] \textbf{Construct second-order symmetry operators
from rank-2 Killing tensors.} If $r<n-1,$ the vector algebra is insufficient,
hence we need $\left(n-1-r\right)$ more operators and this could
be obtained from the Killing tensors $\overline{\boldsymbol{K}}$: 
\begin{enumerate}
\item Solve the rank-2 Killing-tensor equation (\ref{eq:Killingtensor})
to obtain the symmetric tensors $\overline{\boldsymbol{K}}=\beta^{p}\hat{\kappa}_{p}$.
Expand each generator $\hat{\kappa}_{p}$ as $\hat{\kappa}_{p}=\kappa_{p}^{ij}\partial_{i}\otimes\partial_{j}$,
$1\leq i,j\leq n$.
\item Build second-order operators using the purely differential-geometric
map:
\begin{equation}
\begin{array}{cccc}
D: & \mathrm{Sym}\left[T_{\mathbf{x}}U\times T_{\mathbf{x}}U\right] & \rightarrow\mathcal{D}^{\left(2\right)},\\
 & \kappa_{p\left[\mathbf{x}\right]}^{ij} & \mapsto D_{\kappa^{\left(p\right)}} & :=-\nabla_{i}\bigl(\kappa_{p}^{ij}\nabla_{j}\bigr),
\end{array}\label{eq:diffmap}
\end{equation}
where $\mathcal{D}^{\left(2\right)}$ is the family of second-order
differential operators. If $\kappa_{p}^{ij}$ is constant, it reduces
to $D_{\kappa^{\left(p\right)}}=\kappa_{p}^{ij}\partial_{i}\partial_{j}$,
otherwise a first-derivative term appears. In curved space, this divergence
(symmetric) ordering makes $D_{\kappa^{\left(p\right)}}$ formally
self-adjoint. For Killing tensors $\overline{\boldsymbol{K}},$ $D_{\kappa^{\left(p\right)}}$
provides a natural candidate for a Laplace--Beltrami symmetry operator;
any remaining curvature/quantum-ordering conditions for exact commutation
could occur. The map (\ref{eq:diffmap}) is chosen because the form
$\nabla_{i}\!\bigl(\kappa_{p}^{ij}\,\nabla_{j}\bigr)$ guarantees
formally self-adjointness with respect to the inner-product (\ref{eq:QMinnerproductxreps}). 
\item Select $\left(n-1-r\right)$ linearly-independent tensors whose operators
commute with each other and with $\left\{ \hat{\mathcal{O}}_{a}^{\left(1\right)}\right\} $:
\[
\left[D_{\kappa^{\left(p\right)}},D_{\kappa^{\left(q\right)}}\right]=0,\qquad\left[D_{\kappa^{\left(p\right)}},\hat{\mathcal{O}}_{a}^{\left(1\right)}\right]=0,\qquad\left[D_{\kappa^{\left(p\right)}},\triangle\right]=0.
\]
Poisson-commutativity of principal symbols is the classical compatibility
condition. Promoting it to exact commutativity of differential operators
may require lower-order correction terms (ordering/curvature-dependent),
as discussed in the symmetry-operator literature \cite{prolongation2}:
\item Add them to the family of MASA: $\mathcal{D}:=\mathcal{D}^{\left(1\right)}\cup\mathcal{D}^{\left(2\right)}\cup...$,
by defining:
\[
\hat{\mathcal{O}}_{p}^{\left(2\right)}:=D_{\kappa^{\left(p\right)}}\in\mathcal{D}^{\left(2\right)},\qquad p=1,...,s,\quad s:=n-1-r.,
\]
Normally $s\le n-1-r$; if fewer tensors exist, go to Step 4.
\end{enumerate}
\item [{Step$\;$4}] \textbf{Higher order iteration (If necessary).} If
after Step 3 the family has fewer than $n-1$ independent commuting
operators, repeat the same procedure with rank-$m;$ $(m\ge3)$ Killing
tensors:
\begin{enumerate}
\item Solve the $m$-order Killing equation for $K_{j_{1}\dots j_{m}}:$
\[
\nabla_{\left(i\right.}K_{j_{1}\dots\left.j_{m}\right)}=0.
\]
\item Build the $m^{\text{th}}$-order operator:
\[
\hat{\mathcal{O}}_{p}^{\left(m\right)}:=D_{\kappa}^{(m)}=(-1)^{m}\,\nabla_{\left(i_{1}\right.}\kappa^{i_{1}...i_{m}}\nabla_{i_{2}}\cdots\nabla_{\left.i_{m}\right)}\in\mathcal{D}^{\left(m\right)},
\]
where $\mathcal{D}^{\left(m\right)}$ is the family of $m^{\mathrm{th}}$-order
differential operators. The ordering terms chosen so that $D_{\kappa}^{(m)}$
is formally self-adjoint. Add any independent solution to the commuting
family. Keep only those that: (i) commute with every operator already
in the set and (ii) raise the rank of the algebra.
\item Iterate Step 4 for rank $m+1$, $m+2$, ... as needed. If at some
finite rank the set reaches $\left(n-1\right)$ commuting operators
(with functionally independent symbols), the construction terminates.
No general bound on $m$ is known, and success is not guaranteed in
general.
\end{enumerate}
\item [{Step$\;$5}] \textbf{Verify functional independence.} Check that
the principal symbols of $\left\{ \hat{\mathcal{O}}_{1},...,\hat{\mathcal{O}}_{n-1}\right\} $
together with that of $\triangle$ are generically functionally
independent on $T_{\mathbf{x}}^{*}U$ (e.g. on an open dense subset).
If this is the case, their gradients are linearly independent almost
everywhere. If functional dependence occurs, discard the redundant
operator and seek an additional commuting operator within the chosen
class; if no such operator exists, the resulting commuting family
is maximal within that class.
\end{description}
The procedure constructs a hierarchy of commuting local differential
operators generated by Killing data. At each finite rank $m$, the
construction is algorithmic and yields a well-defined commuting family
$\left\{ \hat{\mathcal{O}}^{\left(i\right)}\right\} _{i\leq m}$.
If $\left\{ \hat{\mathcal{O}}^{\left(i\right)}\right\} _{i\leq m}$
provides $\left(n-1\right)$ independent labels (equivalently, resolves
the targeted eigenspace multiplicities), completeness is certified
and the algorithm terminates. Otherwise, the outcome at rank $m$
is a partial symmetry-adapted labeling, and the existence of additional
Killing-generated commuting operators at higher rank remains inconclusive
in full generality.

\paragraph{Remarks {[}The Semi-Algorithm{]}.}

The iterative construction is a semi-algorithm: it certifies a complete
local CSCO when it succeeds, but it does not provide a general guarantee
of success even in the infinite-rank limit. Nonetheless, for the physically
relevant cases of symmetric spaces (spheres, tori), this procedure
is guaranteed to terminate. 

\subsubsection*{Example in $\Sigma=\mathbb{R}^{3}$}

Let us give an illustrative examples of the maximal set of commuting
geometric operators in $\mathbb{R}^{3}$. In $\mathbb{R}^{3}$, there
exists 6 independent Killing vectors, related to their isometry group,
the Euclidean group $\mathrm{ISO}(3)=\mathrm{SO(3)}\ltimes\mathbb{R}^{3}.$
The generators consist of 3 translations $\hat{p}_{i}=-\mathbf{i}\partial_{i}$
and 3 rotations $\hat{L}_{i}=-\mathbf{i}\left(x_{j}\partial_{k}-x_{k}\partial_{j}\right).$
In $\mathbb{R}^{3}$, the Killing vectors are already sufficient to
construct a complete MASA so the Killing tensor is not considered.
\begin{center}
\begin{table}
\begin{centering}
\label{Table 1}%
\begin{tabular}{|c|c|c|c|}
\hline 
Sets & Cartesian set & Cylindrical set & Spherical set\tabularnewline
\hline 
\hline 
Laplacian & $\triangle=-\sum_{i=1}^{3}\hat{p}_{i}\hat{p}^{i}$ & $\triangle=\frac{1}{r}\frac{\partial}{\partial r}\left(r\frac{\partial}{\partial r}\right)-\frac{1}{r^{2}}\hat{L}_{z}-\hat{p}_{z}^{2}$ & $\triangle_{\left[r,\theta,\phi\right]}=\frac{1}{r^{2}}\frac{\partial}{\partial r}\left(r^{2}\frac{\partial}{\partial r}\right)+\left|\hat{L}\right|^{2}$,\tabularnewline
\hline 
\multirow{2}{*}{Local operators} & \multirow{2}{*}{$\hat{p}_{i}=-\mathbf{i}\partial_{i}$, $i=x,y,z$} & $\hat{p}_{z}=-\mathbf{i}\frac{\partial}{\partial z}$, & $\left|\hat{L}\right|^{2}=-\frac{1}{\sin\theta}\frac{\partial}{\partial\theta}\left(\sin\theta\frac{\partial}{\partial\theta}\right)+\frac{1}{\sin^{2}\theta}\hat{L}_{z}^{2}$\tabularnewline
 &  & $\hat{L}_{z}=-\mathbf{i}\partial_{\phi}$, & $\hat{L}_{z}=-\mathbf{i}\partial_{\phi}$,\tabularnewline
\hline 
\multirow{3}{*}{Involution } & $\left[\hat{p}_{i},\hat{p}_{j}\right]=0,$ & $\left[\hat{p}_{z},\hat{L}_{z}\right]=0,$ & $\left[\left|\hat{L}\right|^{2}\hat{L}_{z}\right]=0,$\tabularnewline
 & $\left[\triangle,\hat{p}_{i}\right]=0.$ & $\left[\triangle,\hat{p}_{z}\right]=0,$ & $\left[\triangle,\left|\hat{L}\right|^{2}\right]=0,$\tabularnewline
 &  & $\left[\triangle,\hat{L}_{z}\right]=0,$ & $\left[\triangle,\hat{L}_{z}\right]=0,$\tabularnewline
\hline 
MASA & $\left\{ \hat{p}_{x},\hat{p}_{y},\hat{p}_{z}\right\} $\footnote{There is a subtlety here: $\triangle$, which is a polynomial function
of $\hat{p}_{i}$, is not included in the Cartesian set of MASA. To
address this subtlety, we need to generalize the definition of MASA
as in Section VI B.} & $\left\{ \triangle,\hat{p}_{z},\hat{L}_{z}\right\} $ & $\left\{ \triangle,\left|\hat{L}\right|^{2},\hat{L}_{z}\right\} $\tabularnewline
\hline 
\end{tabular}
\par\end{centering}
\caption{Maximal set of commuting geometric operators in $\mathbb{R}^{3}.$}
\end{table}
\par\end{center}

These 3 sets describe 3 distinct, geometric, MASA in $\mathbb{R}^{3}.$
In fact, each sets are naturally related to 3 different (St$\ddot{\textrm{a}}$ckel)
local coordinates in $\mathbb{R}^{3},$ the Cartesian $\left(x,y,z\right)$,
cylindrical $\left(r,\phi,z\right)$, and spherical $\left(r,\theta,\phi\right)$
coordinates. This subject will be discussed in the next section.

Let us summarize this section in one  line: \textbf{The sets of commuting,
local, geometric operators are constructed from the MASA of the Killing
fields on $\Sigma$.}

\section{Gauge and Coordinate Freedom in GFT}

The choice of maximal set of geometric operators in $\mathcal{L}^{2}\left[\Sigma\right]$
would, in some cases, canonically leads to a natural coordinate chart
in $\Sigma$ (and vice versa). In this section, we discuss how the
choice of local coordinate chart and its transformation in $\Sigma$
affect the degenerate sector of the GFT. We also discuss a special
case of coordinate chart in $\Sigma$ where the Helmholtz equation
admits separation of variables.

\subsection{Diffeomorphism, Isometries and Coordinate Transformation}

Let $U_{a}\subset\Sigma$ be a local chart/patch on $\Sigma$ such
that $\cup_{a}U_{a}=\Sigma$ is an open cover of $\Sigma$. By the
definition of a (differentiable) manifold, there always exist a smooth
bijective map $\varphi_{a}:U_{a}\rightarrow U_{a}\left(\varphi_{a}\right)\subset\mathbb{R}^{n};$
the function $\varphi_{a}$ is the coordinate system of $\Sigma$
at patch $U_{a}$.

We consider two differentiable manifolds $\Sigma$ and $\Sigma'$
of dimension $n$, with local patches $U_{a}\subset\Sigma$ and $U_{b}\subset\Sigma'$.
Let us define a map $\begin{array}{ccc}
\phi: & U_{a} & \rightarrow U_{b}\end{array}$; if $\phi$ is a smooth bijection with a smooth inverse, then $\phi$
is a (local) \textit{diffeomorphism}. The diffeomorphism $\phi$ induces
a map between two copies of $\mathbb{R}^{n}$, which is the local
coordinate representation of $\phi$:
\begin{equation}
\begin{array}{ccccc}
\varphi_{b}\circ\phi\circ\varphi_{a}^{-1}: & \varphi_{a}\left(U_{a}\right) & \rightarrow & \varphi_{b}\left(U_{b}\right),\\
 & \mathbf{x} & \mapsto & \mathbf{x}' & =\varphi_{b}\circ\phi\circ\varphi_{a}^{-1}\left(\mathbf{x}\right).
\end{array}\label{eq:diffreps}
\end{equation}
 see FIG. 1. 
\begin{figure}[h]
\begin{centering}
\includegraphics{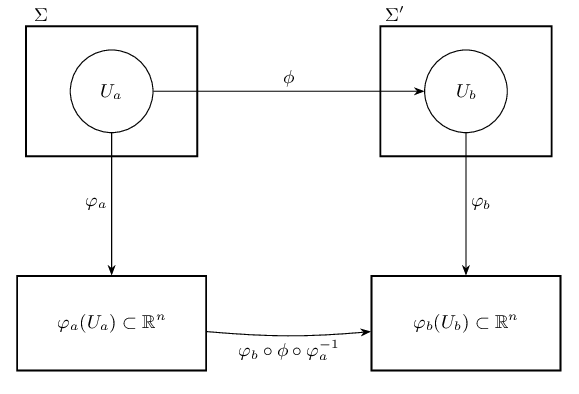}$\quad$\includegraphics[scale=0.75]{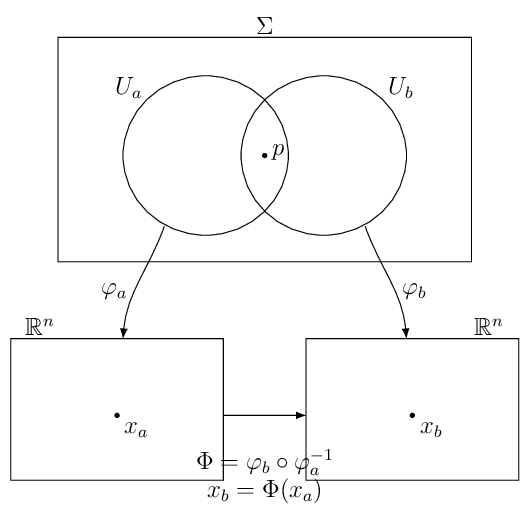}\caption{Commutative diagram representing the local coordinate expression of
a diffeomorphism $\phi$. The map $\phi$ transforms points from a
local patch $U_{a}\subset\Sigma$ to $U_{b}\subset\Sigma'$. The functions
$\varphi_{a}:U_{a}\rightarrow\mathbb{R}^{n};$ and $\varphi_{b}:U_{b}\rightarrow\mathbb{R}^{n}$
are the coordinate charts mapping these patches to the numerical domain
$\mathbb{R}^{n}$. The lower horizontal arrow, given by the composition
$\varphi_{b}\circ\phi\circ\varphi_{a}^{-1}$, represents the realization
of $\phi$ in local coordinates. \textit{Left panel}: If $\phi$ is
a non-trivial metric-preserving map, this represents an active isometry
(symmetry). \textit{Right panel}: If $\phi=\mathbb{I}$ (identity)
and $\Sigma=\Sigma'$, this diagram represents a passive coordinate
transformation (gauge freedom) between two overlapping charts.}
\par\end{centering}
\end{figure}

The map (\ref{eq:diffreps}) does not require a geometrical structure
on $\Sigma$ and $\Sigma'$. To introduce geometry, let us equip $\Sigma$
with a metric $\mathbf{q}$ and $\Sigma'$ with $\mathbf{q}'$. The
diffeomorphism $\phi$ is \textit{metric-preserving} if its pull-back
$\phi^{*}$ preserves the metric:
\begin{equation}
\phi^{*}\left[\mathbf{q}'\right]=\mathbf{q}.\label{eq:metricpreserving}
\end{equation}

For $\Sigma'=\Sigma$, the diffeomorphism map in (\ref{eq:diffreps})
admits two distinct but equivalent interpretations: the active and
the passive points of view. In the active interpretation, the diffeomorphism
acts on the field while the points are held fixed. In the passive
interpretation, the field is left unchanged, while the points of $\Sigma$
are relabeled by a change of coordinates.

Furthermore, there are two distinct cases where $\phi$ becomes metric-preserving:

\paragraph*{Case 1: coordinate (gauge) transformation (passive view):}

This is the trivial case where the diffeomorphism is the identity
map on the manifold, $\phi=\mathbb{I}.$ This condition forces the
domain and codomain of $\phi$ to coincide, namely, $U_{a},U_{b}$
are two distinct overlapping charts on the same manifold ($U_{a}\cap U_{b}\neq\left\{ \oslash\right\} $).
In this case, (\ref{eq:diffreps}) becomes the transition function
between charts:
\begin{equation}
\begin{array}{ccccc}
\varphi_{b}\circ\mathbb{I}\circ\varphi_{a}^{-1}: & \varphi_{a}\left(U_{a}\cap U_{b}\right) & \rightarrow & \varphi_{b}\left(U_{a}\cap U_{b}\right),\\
 & \mathbf{x} & \mapsto & \mathbf{x}' & =\varphi_{b}\circ\varphi_{a}^{-1}\left(\mathbf{x}\right).
\end{array}\label{eq:coordtransf}
\end{equation}
$\varphi_{b}\circ\varphi_{a}^{-1}$ in (\ref{eq:coordtransf}) is
a \textit{(gauge}\footnote{\textit{Here \textquoteleft gauge\textquoteright{} is used only in
the weak sense of descriptive redundancy under coordinate relabeling,
not in the sense of an internal gauge symmetry.}}\textit{) coordinate transformation} or local charts transition in
$\Sigma$. We classify this type of diffeomorphism as $\mathrm{id}\left(\Sigma\right)$:
the set of all gauge /coordinate transformations in $\Sigma.$ It
represents a redundancy in our description; the physical point has
not moved, only its label has changed. If interpreted in the active
point of view, the diffeomorphism $\phi$ act trivially as $\phi=\mathbb{I}.$ 

\paragraph*{Case 2: (non-trivial) isometry (active view):}

The second case is where $\phi$ is non-trivial, but still satisfy
the metric-preserving condition (\ref{eq:metricpreserving}). In this
case, $\phi$ is an \textit{isometry} on $\Sigma$ (and $\varphi_{b}\circ\phi\circ\varphi_{a}^{-1}$
is the coordinate representation of $\phi$). These transformations
actively move points to new locations while preserving distances.
The set of all such maps forms the \textit{isometry group} on $\Sigma$,
I$\left(\Sigma\right).$ \textcolor{blue}{In a set-theoretic sense,
a coordinate transformation can be considered as a trivial isometry.}

\subsection{Metric-Preserving Diffeomorphism and Unitary Transformation}

The Helmholtz equation (\ref{eq:Helmholtz}) is geometric, and therefore
coordinate invariant. Similarly the spectral decomposition (\ref{eq:spectraltheorem})
is independent of the choice of coordinate. GFT is fundamentally a
spectral decomposition of an operator $\hat{\mathcal{O}}=-\triangle,$
hence, the (abstract) GFT is coordinate invariant. We can view GFT
as a map that partitions the Hilbert space $\mathcal{H}_{x}=\mathcal{L}^{2}\left[\Sigma\right]$
into a Hilbert spectral subspaces $\left.\mathcal{H}_{x}\right|_{\lambda}$
as in (\ref{eq:partition}). This structure/partition inside $\mathcal{H}_{x}=\mathcal{L}^{2}\left[\Sigma\right]$
is intrinsic to the geometry and exists independently of any coordinate
system.

However, coordinate charts are practically unavoidable. To obtain
concrete solutions to the Helmholtz equation, one must specify a local
chart in $\Sigma$. Solving the PDE in different coordinate systems
produces solution sets that look different functionally. In this subsection,
we show that a metric-preserving diffeomorphism on $\Sigma$ induces
a unitary transformation o n $\mathcal{H}_{x}=\mathcal{L}^{2}\left[\Sigma\right]$
that allow us to rotate these solutions into one another.

Let $\phi$ (not to be confused with $\phi_{\left[\lambda,\alpha\right]}$,
the Fourier pair via the GFT) be a (local) metric-preserving diffeomorphism
satisfying (\ref{eq:metricpreserving}):
\begin{equation}
\begin{array}{ccccc}
\phi: & U_{a} & \rightarrow & U_{b},\\
 & \mathbf{x} & \mapsto & \mathbf{x}' & =\phi\left(\mathbf{x}\right)
\end{array}\label{eq:iso}
\end{equation}
where $U_{a},U_{b}\subset\Sigma$ are (local) charts on $\Sigma$,
diffeomophic to $\mathbb{R}^{n}.$ Let us consider a scalar function
defined on the target patch:
\begin{equation}
\begin{array}{cccc}
f: & U_{b} & \rightarrow & \mathbb{R},\\
 & \mathbf{x}' & \mapsto & f_{\left[\mathbf{x'}\right]}.
\end{array}\label{eq:iso-1}
\end{equation}
The pull-back of $f$ by $\phi$, denoted $\phi^{*}$ is a function
on the domain $U_{a}$ defined by composition:
\begin{equation}
\begin{array}{ccccc}
\phi^{*}\left[f\right]: & U_{a} & \rightarrow & \mathbb{R}\\
 & \mathbf{x} & \mapsto & \phi^{*}\left[f\right]_{\left(\mathbf{x}\right)} & :=f_{\left[\phi\left(\mathbf{x}\right)\right]}=\left(f\circ\phi\right)_{\left(\mathbf{x}\right).}
\end{array}\label{eq:pullback}
\end{equation}
This pullback operation can be interpreted as a linear operator acting
on the Hilbert space $\mathcal{L}^{2}\left[U_{b}\right]\sim\mathcal{L}^{2}\left[U_{a}\right]:=\mathcal{H}_{x}$
(the isomorphism between $\mathcal{L}^{2}\left[U_{b}\right]\sim\mathcal{L}^{2}\left[U_{a}\right]$
stems from the fact that both $U_{a},U_{b}$ are diffeomorphic to
$\mathbb{R}^{n}$):
\begin{equation}
\begin{array}{cccc}
\phi^{*}: & \mathcal{H}_{x} & \rightarrow & \mathcal{H}_{x},\\
 & f_{\left[\mathbf{x}\right]} & \mapsto & \phi^{*}\left[f\right]_{\left(\mathbf{x}\right)},
\end{array}\label{eq:unitarytrans}
\end{equation}
and hence from relation (\ref{eq:pullback}) $f_{\left[\phi\left(\mathbf{x}\right)\right]}=\phi^{*}\left[f\right]_{\left(\mathbf{x}\right)}$,
the metric-preserving $\phi$ on $\Sigma$ will induce a transformation
$\phi^{*}$ on the function over $\Sigma$. 

Crucially, because $\phi$ is metric-preserving, it preserves the
Riemannian volume form $d\mu_{\Sigma}$ on $\Sigma$ (i.e., the Jacobian
of the transformation is unity). Therefore, the inner product is preserved:
\begin{equation}
\left\langle \phi^{*}f,\phi^{*}g\right\rangle =\intop_{U_{a}}f_{\left[\phi\left(\mathbf{x}\right)\right]}^{*}g_{\left[\phi\left(\mathbf{x}\right)\right]}d\mu_{\left[\mathbf{x}\right]}=\intop_{U_{b}}f_{\left[\mathbf{x}'\right]}^{*}g_{\left[\mathbf{x}'\right]}d\mu_{\left[\mathbf{x}'\right]}=\left\langle f,g\right\rangle .\label{eq:proof1}
\end{equation}
Thus, the metric-preserving $\phi$ on the manifold induces a unitary
transformation $\hat{U}_{\phi}=\phi^{*}$ on the Hilbert space $\mathcal{H}_{x}$. 

\subsection{The Role of Coordinate in GFT}

We established at the previous sections that the abstract GFT and
the Helmholtz equation are coordinate invariant. As illustrated in
Table I, a specific choice of coordinates often leads to a specific
eigenfunction family, and hence to a canonical choice of orthonormal
basis, which dictates the structure of the Fourier kernel.

\subsubsection{Coordinate Transformations, Isometries, and MASA (Representation)
Changes}

In the presence of spectral degeneracy, choosing an orthonormal eigenbasis
of the Laplace-Beltrami operator is equivalent to fixing a maximal
Abelian set of commuting self-adjoint operators (MASA) within the
commutant of $\triangle$. This choice is not unique; different MASAs
resolve the same spectral degeneracy in inequivalent ways, and any
two resulting bases are related by a unitary map acting within each
degenerate eigenspace.

It is crucial to distinguish three notions that are often conflated:
\begin{description}
\item [{(i)}] \textbf{Passive chart relabelling/coordinate transformation}:
\textcolor{blue}{The map $\varphi_{b}\circ\varphi_{a}^{-1}:\mathbf{x}\mapsto\mathbf{x}'_{\left[\mathbf{x}\right]}$}
are mere reparametrizations of the same geometric objects on $\Sigma$.
\textcolor{blue}{It does not move points in $\Sigma$} nor define
a new transform; it merely rewrites the same operators and eigenfunctions
in a different coordinate representation. If we force to view this
in the active perspective, this type of transform will fall to the
diffeomorphism class of $\mathrm{id}\left(\Sigma\right)$. These transformations
leave the function (or a quantum state vector in quantum mechanics)
invariant, inducing only a trivial unitary on $\mathcal{L}^{2}\left[\Sigma\right].$
\item [{(ii)}] \textbf{Active isometries:} These are metric-preserving
diffeomorphism $g\in I(\Sigma)$ satisfying (\ref{eq:iso}). They
induce a nontrivial unitary pullback (\ref{eq:unitarytrans}) and
satisfy $\phi^{*}\triangle=\triangle\phi^{*}$. Consequently, an isometry
acts as a symmetry operator that rotates eigenbases within each degenerate
eigenspace while preserving the joint spectrum (and hence the topology
of the associated $k$-space) up to the natural identification. 
\item [{(iii)}] \textbf{MASA representation changes (basis selection)}:
These are choices of different commuting families used to resolve
degeneracy. In practice, this is often realized via \textit{coordinate-adapted
separation schemes}. These choice can alter the degeneracy labelling
$\alpha$ and the resulting topology of the label space $\mathcal{F}_{\lambda}$
(and hence the apparent topology of $k$-space), while remaining unitarily
equivalent in $\mathcal{L}(\Sigma)$. This is a choice of representation/gauge,
physically distinct from a mere coordinate relabelling.
\end{description}

\subsubsection{Solve-Then-Transform vs Transform-Then-Solve Problem}

We distinguish these types of transforms because of an apparent non-commutativity
between two mathematical operations: solving for the eigenfunctions
of a PDE and changing variables. In 1-dimension (non-degenerate) problems,
these operations commute. In higher-dimensional (degenerate) PDEs,
they do not.

A classic example of this is the Helmholtz problem in $\mathbb{R}^{3}$,
viewed in Cartesian vs. spherical charts. The natural separation of
variables in the Cartesian chart leads to plane waves ($e^{i\mathbf{k}\cdot\mathbf{x}}$)
as its eigenfunction, while in spherical chart, it leads to spherical
waves (spherical harmonics). Simply coordinate-transforming a single
plane wave does not produce a single spherical wave (and vice-versa);
rather, one plane wave transforms into an infinite sum of spherical
waves (via the Rayleigh expansion). They are related by a unitary
transformation in $\mathcal{L}^{2}\left[\mathbb{R}^{3}\right],$ see
Section VII A. 

This non-commutativity arises because in concrete PDE problems, one
often performs a representation-dependent step before solving, e.g.,
choosing a coordinate-adapted separation ansatz or a preferred set
of commuting operators (a CSCO/MASA) to resolve degeneracy. This ``transform-then-solve''
route is therefore not solely a coordinate change: the choice of coordinate
chart to write the PDE implicitly dictates a separation of variable
natural to the chart; this is equivalent with fixing an orthonormal
basis in the space of solution, which in turns, fix the commuting
operators set (MASA). This ``transform-then-solve'' route is what
we label as \textit{coordinate-adapted separation schemes}, which
is sometimes get blurred with passive coordinate transformation.

This is also the case for our GFT. Different coordinate-adapted constructions
lead to different realizations of the label space $\mathcal{F}$ (Case
iii), whereas a passive rewriting of already-chosen eigenfunctions
leaves the underlying GFT data unchanged (Case i). In the next subsection,
we will discuss how the coordinate-adapted separation schemes and
isometry affect GFT.

\subsubsection{Non-Degenerate GFT vs Degenerate GFT}

Let us consider first the non-degenerate case in Section II.D.1. The
simplest example is the 1-dimensional manifold. On a 1D chart diffeomorphic
to a compact interval with separated self-adjoint boundary conditions
(e.g., Dirichlet or Neumann), the Laplace-Beltrami operator is guaranteed
to have a simple (non-degenerate) spectrum by Sturm-Liouville theory
\cite{ODE}. It is necessary to acknowledge that this theorem fade
for periodic boundary conditions (e.g., the ``ring''), where the
topology introduces degeneracy via rotation symmetry. Degeneracies
only appear if the manifold admits non-trivial continuous isometries
(e.g., the periodic boundary conditions of a circle allowing rotation
symmetry). Therefore, any invertible variable transformation (coordinate
change) leaves the uniqueness of the spectral basis invariant; the
basis functions are unique up to a phase factor.

For the $n$-dimensional non-degenerate cases (i.e., irregular drums
\cite{drum1,drum2}), their only isometry is the identity map. Lacking
a group action to force degeneracies, their spectra are generically
simple. Consequently, the non-degenerate GFT does not possess a degenerate
sector. All aspects of the GFT are invariant under invariant under
metric-preserving diffeomorphism (both coordinate transformation and
isometry) in the sense that the basis set is unique; only its functional
representation changes. See Table II.

In the degenerate GFT cases, the non-degenerate sector behaves exactly
as above. In general, the degenerate sector is sensitive to the specific
choice of representation. As illustrated in Table II, the structure
of the basis depends on whether the transformation is a pure isometry
I$\left(\Sigma\right)$ (Case ii) or a coordinate-adapted separation
scheme (Case iii).

\begin{table}[h]
\begin{centering}
\begin{tabular}{|l|c|c||c|c|c|}
\hline 
\multirow{3}{*}{$\qquad$\textbf{Non-degenerate Sector}} & \multicolumn{2}{c||}{Invariant under:} & \multirow{3}{*}{\textbf{Degenerate Sectors}} & \multicolumn{2}{c|}{Invariant under:}\tabularnewline
\cline{2-3}\cline{5-6}
 & \multirow{2}{*}{Diff$\left(\Sigma\right)$?} & Metric- &  & \multirow{2}{*}{Case iii?} & \multirow{2}{*}{Case ii?}\tabularnewline
 &  & preserving? &  &  & \tabularnewline
\hline 
\multirow{2}{*}{$\qquad$The geometrical objects in $\Sigma$} &  &  & The local chart representation &  & \tabularnewline
 &  &  & of the objects: &  & \tabularnewline
\hline 
\hline 
$\bullet$ Metric $\mathbf{q}$ & no & yes & $q_{ij}$ & no & yes\tabularnewline
$\bullet$ Measure $d\mu_{\Sigma}$ & no & yes & $\sqrt{q_{\left[\mathbf{x}\right]}}\textrm{d}^{n}\mathbf{x},$ & no & yes\tabularnewline
$\bullet$ Laplace-Beltrami operator $\triangle$ & no & yes & $\frac{1}{\sqrt{q_{\left[\mathbf{x}\right]}}}\frac{\partial}{\partial x^{i}}\left(\sqrt{q_{\left[\mathbf{x}\right]}}q^{ij}\frac{\partial}{\partial x^{j}}\right)$ & no & yes\tabularnewline
\hline 
Helmholtz equation $\left(\triangle+\lambda\right)\psi=0,$ & no & yes & $\frac{1}{\sqrt{q_{\left[\mathbf{x}\right]}}}\frac{\partial}{\partial x^{i}}\left(\sqrt{q_{\left[\mathbf{x}\right]}}q^{ij}\frac{\partial}{\partial x^{j}}\right)\psi=-\lambda\psi$ & no & yes\tabularnewline
\hline 
\multirow{2}{*}{Full solution (wave form) $\psi_{\left[\mathbf{x}\right]},$} & \multirow{2}{*}{no} & \multirow{2}{*}{yes} & $\bullet$ Eigensolution (GFT kernel)\footnote{``Eigensolution'' here is the eigenfunction obtained directly from
solving the Helmholtz equation in a chart. Of course a mere relabeling
would not affect the mathematical object.} $\hat{f}_{\left[\mathbf{x};\lambda,\alpha\right]}$ & no & no\tabularnewline
 &  &  & $\bullet$ Fourier pair $\phi_{\left[\lambda,\alpha\right]}$/$\psi_{\left[\mathbf{x}\right]}$'s
component & no & no\tabularnewline
\hline 
\multirow{2}{*}{Spectrum $\lambda$} & \multirow{2}{*}{no} & \multirow{2}{*}{yes} & $\bullet$ Degeneracy label $\alpha$ & no & no\tabularnewline
 &  &  & $\bullet$ Dual space $\mathcal{F}\ni\alpha$ & no & yes\tabularnewline
\hline 
$\bullet$ Full Hilbert space $\mathcal{H}_{x}\sim\mathcal{L}^{2}\left[\Sigma\right]$ & yes & yes &  &  & \tabularnewline
$\bullet$ The partition structure (\ref{eq:partition}) & no & yes &  &  & \tabularnewline
$\bullet$ Hilbert subspace $\left.\mathcal{H}_{x}\right|_{\lambda}=\mathcal{L}^{2}\left[\left.\mathcal{F}\right|_{\lambda}\right]$ & no & yes &  &  & \tabularnewline
$\bullet$ The spectral projection $\mathcal{\hat{P}}$ related to
$\triangle$ & no & yes &  &  & \tabularnewline
\hline 
\end{tabular}
\par\end{centering}
\caption{The non-degenerate and degenerate sector of the GFT. Each sector contains
objects and their invariance properties under diffeomorphism.}

\end{table}

While the non-degenerate spectrum stays invariant, the degenerate
sector in GFT is sensitive to coordinate-adapted transformations (Case
iii), this fact explains why different choice of coordinate chart
in $\Sigma$ canonically results in different sets of orthonormal
basis in $\mathcal{H}_{x},$ although they are related by a unitary
transformation (\ref{eq:unitarytrans}). This sensitivity reflects
the non-canonicity of the chosen representation inside degenerate
sectors, even though the underlying spectral decomposition of the
operator $\triangle$ remains invariant.

\subsubsection{Formalization: Invariance vs Representation-Dependence of $k$-Space.}

The definition of the momentum domain (\ref{eq:kspace}) assumes our
MASA includes the Laplace-Beltrami operator $\triangle$. In many
practical cases, requiring $\triangle$ to be explicitly in the generator
set would cause conflicts: either the subalgebra is redundant, not
maximal, or the sets containing non-local operators. 

A standard example is GFT in flat $\mathbb{R}^{3}$ using Cartesian
coordinates. Requiring $\triangle$ to belong to the subalgebra generally
leads to a tension with maximality. In flat $\mathbb{R}^{3}$, the
Cartesian MASA $\left\{ \hat{p}_{x},\hat{p}_{y},\hat{p}_{z}\right\} $
is already maximal. Replacing one momentum operator by $\triangle$
yields a local commuting set such as $\left\{ \triangle,\hat{p}_{x},\hat{p}_{y}\right\} $,
but this set is not maximal, since the Laplacian $\triangle=p_{x}^{2}+p_{y}^{2}+p_{z}^{2}$
is functionally dependent on the MASA generators, and $\hat{p}_{z}$
(which still commutes with the set) is not locally generated by it.
Restoring maximality while retaining $\triangle$ in the MASA set
$\left\{ \triangle,\hat{\mathcal{O}}_{1},\hat{\mathcal{O}}_{2}\right\} $
will result in either $\hat{\mathcal{O}}_{1},\hat{\mathcal{O}}_{2}$
are non-local (such as square-root functions of $\triangle-\hat{p}_{x}^{2}-\hat{p}_{y}^{2}$),
or local $\hat{\mathcal{O}}_{1},\hat{\mathcal{O}}_{2}$ but not separable
in the Cartesian coordinate. 

The definition of the momentum domain (\ref{eq:kspace}) is valid
for the case where our MASA includes the Laplace-Beltrami operator
$\triangle$. To accommodate cases where $\triangle$ is not a generator
but a function of the generators, we generalize the definition of
$k$-space as follows:
\begin{lyxlist}{00.00.0000}
\item [{\textbf{Definition.}}] ($k$-space as joint spectrum/label space).
\textit{Let $\mathcal{D}=\left\{ \mathcal{\hat{O}}_{1},...,\hat{\mathcal{O}}_{m}\right\} $,
$m\leq n,$ be a commuting family of (essentially) self-adjoint operators
on $\mathcal{H}$ that satisfies: (1) Each $\hat{\mathcal{O}}_{i}$
commutes with $\triangle$; (2) $\triangle$ is polynomial (or more
generally, functional in the sense of functional calculus/symbol relation)
of operators in $\mathcal{D}:$ $\triangle=\mathrm{Poly}\left(\hat{\mathcal{O}}_{1},...,\hat{\mathcal{O}}_{m}\right)$
(possibly of degree 0 if we include $\triangle$ itself), then, the
$k$-space associated to $\Sigma$ and $\mathcal{D}$ is the space
of joint spectrum:
\begin{equation}
\mathcal{F}:=\left\{ \left(k_{1},...,k_{m}\right)|k_{i}\in\mathcal{S}_{i}\subseteq\mathbb{R}\right\} ,\label{eq:kspacegen}
\end{equation}
where $k_{i}=\sigma\left(\hat{\mathcal{O}}_{i}\right)$ is the spectrum
of $\hat{\mathcal{O}}_{i}$. Equivalently, one may view $\mathcal{F}$
as a disjoint union of fibers as in (\ref{eq:kspace}), where $\mathcal{F}_{\lambda}$
encodes the degeneracy labels (joint eigenvalues inside $\mathcal{H}_{\lambda}$)
for a fixed energy} $\lambda=\mathrm{Poly}(k_{1},...,k_{m})$\textit{.}
\end{lyxlist}
$\triangle$ could or could not be a member of the MASA. The Laplace-Beltrami
eigenvalue $\lambda$ is recovered as a polynomial/functional in those
generators $\lambda=\mathrm{Poly}(k_{1},...,k_{m})$; inclusion of
$\triangle$ as a separate element is required only when that polynomial/functional
is non-trivial. Consequently, the Cartesian MASA $\left\{ \hat{p}_{x},\hat{p}_{y},\hat{p}_{z}\right\} $
produces $\mathcal{F}=\mathbb{R}^{3}$, whereas cylindrical and spherical
MASA\textquoteright s yield $\mathcal{F}=\mathbb{R}^{+}\times\mathbb{Z}\times\mathbb{R}$
and $\mathcal{F}=\mathbb{R}^{+}\times\mathbb{Z}^{2}$, ($\mathbb{Z}$
integers) respectively.

We emphasize that while the Hilbert spaces $\mathcal{L}^{2}\left[\mathcal{F}\right]$
(or $\mathcal{L}^{2}\left[\Sigma\right]$) are unitarily equivalent
(the unitarity acts on Hilbert spaces), the topological structures
of the chosen label spaces $\mathcal{F}$ (or $\Sigma$) are generally
distinct (hence in general different $\mathcal{F}$'s are not homeomorphic
to each other, not necessarily a diffeomorphism). This is the mathematical
manifestation of the coordinate-adapted separation scheme.
\begin{lyxlist}{00.00.0000}
\item [{\textbf{Theorem.}}] (Invariance of $k$-space under isometry/Case
ii). \textit{Let $\phi\in I\left(\Sigma\right)$ be an isometry satisfying
(\ref{eq:iso}) and let $\hat{U}_{\phi}=\phi^{*}$ be the pullback
operator on $\mathcal{H}$ defined by (\ref{eq:pullback}). Then:
(1) $\hat{U}_{\phi}=\phi^{*}$ is unitary on $\mathcal{L}^{2}\left(\Sigma,d\mu_{\Sigma}\right)$,
(2) $\hat{U}_{\phi}=\phi^{*}$ commute with the Laplace-Beltrami operator:
$\phi^{*}\triangle=\triangle\phi^{*}$. (3) If $\mathcal{D}=\left\{ \hat{\mathcal{O}}_{i}\right\} _{i=1}^{m}$
is a MASA for the commutant $\triangle$ with joint spectrum $\mathcal{F}$,
then the transformed family $\mathcal{D}'=\left\{ \hat{U}_{\phi}\hat{\mathcal{O}}_{i}\hat{U}_{\phi}^{\dagger}\right\} _{i=1}^{m}$
also commute with $\triangle$, and their joint spectrum is identical:
\[
\sigma\left(\triangle,\hat{\mathcal{O}}_{1},...,\hat{\mathcal{O}}_{m}\right)=\sigma\left(\triangle,\hat{\mathcal{O}}'_{1},...,\hat{\mathcal{O}}'_{m}\right).
\]
In particular, isometries preserve the $k$-space topology and act
as unitary rotations within degenerate fibers.}
\end{lyxlist}

\paragraph{Proof sketch. }

(1) Unitarity: Consider the inner product of transformed functions.
By definition:
\[
\left\langle \hat{U}_{\phi}f,\hat{U}_{\phi}g\right\rangle =\intop_{\Sigma}f_{\left[\phi\left(\mathbf{x}\right)\right]}^{*}g_{\left[\phi\left(\mathbf{x}\right)\right]}d\mu_{\left[\mathbf{x}\right]}.
\]
Perform a change of variables $\mathbf{x}'=\phi\left(\mathbf{x}\right)$.
Since $\phi$ is an isometry, the pullback of the metric is the metric
itself ($\phi^{*}g=g$) implying that the measure is preserved. Thus
$d\mu_{\left[\mathbf{x}\right]}=d\mu_{\left[\mathbf{x}'\right]}$,
and the integral becomes:
\[
\intop_{\Sigma}f_{\left[\mathbf{x}'\right]}^{*}g_{\left[\mathbf{x}'\right]}d\mu_{\left[\mathbf{x}'\right]}=\left\langle f,g\right\rangle .
\]
Hence, $\hat{U}_{\phi}$ is unitary. $\blacksquare$

(2) Commutativity: The Laplace--Beltrami operator $\triangle$ is
defined intrinsically by the metric tensor $\boldsymbol{q}$. Since
$\phi$ preserves $\boldsymbol{q}$, the operator is invariant under
the pullback: $\phi^{*}\circ\triangle\circ\phi^{*,-1}=\triangle,$
hence commutes with pullback by isometries. $\blacksquare$

(3) Spectral invariance: The spectrum of an operator is invariant
under unitary conjugation. Since$\hat{U}_{\phi}$ is unitary,\textit{
\[
\sigma\left(\phi^{*}\left(\hat{\mathcal{O}}_{1}\right),\dots,\phi^{*}\left(\hat{\mathcal{O}}_{m}\right)\right)=\sigma\left(\hat{U}_{\phi}\hat{\mathcal{O}}_{1}\hat{U}_{\phi}^{\dagger},\dots,\hat{U}_{\phi}\hat{\mathcal{O}}_{m}\hat{U}_{\phi}^{\dagger}\right)=\sigma\left(\hat{\mathcal{O}}_{1},\dots,\hat{\mathcal{O}}_{m}\right).
\]
}

Moreover, because conjugation is an algebra homomorphism, the joint
spectrum of the commuting tuple is preserved. $\blacksquare$
\begin{lyxlist}{00.00.0000}
\item [{\textbf{Theorem.}}] (Dependency of $k$-space under coordinate-adapted
MASA change/Case iii). \textit{Consider two distinct commuting families
$\mathcal{D}=\left\{ \hat{\mathcal{O}}_{i}\right\} _{i=1}^{m}$ and
$\mathcal{D}'=\left\{ \hat{\mathcal{O}'}_{j}\right\} _{j=1}^{m'}$,
each consisting of (essentially) self-adjoint operators commuting
with $\triangle$, and assume each family is used to resolve degeneracy
(i.e., fixes a fiberwise MASA). Let:
\[
\mathcal{F}:=\sigma\left(\triangle,\hat{\mathcal{O}}_{1},\dots,\hat{\mathcal{O}}_{m}\right),\qquad\mathcal{F}':=\sigma\left(\triangle,\hat{\mathcal{O}}'_{1},\dots,\hat{\mathcal{O}}'_{m}\right).
\]
Then the corresponding GFT kernels (joint eigenbases) are unitarily
equivalent on $\mathcal{H}$ in the sense that, for each $\lambda$,
the two bases are related by a unitary map acting within $\mathcal{H}_{\lambda}$.
In general, $\mathcal{F}$ and $\mathcal{F}'$ are topologically distinct.}
\end{lyxlist}

\paragraph{Proof sketch. }

Since both families commute with $\triangle$, they respect the spectral
decomposition of the Laplace-Beltrami operator:
\[
\mathcal{H}_{x}\cong\intop_{\mathcal{S}}^{\oplus}d\mu_{\mathcal{S}}\left.\mathcal{H}_{x}\right|_{\lambda}.
\]
The subspaces $\left.\mathcal{H}_{x}\right|_{\lambda}$ (the eigenspaces
of $\triangle$) are fixed geometric objects, independent of the choice
of MASA. The choice of $\mathcal{D}$ defines a basis $\left\{ \left.\hat{f}_{\left[\mathbf{x};\alpha\right]}\right|_{\lambda}\right\} $
for $\left.\mathcal{H}_{x}\right|_{\lambda}$, where $\alpha\in\left.\mathcal{F}\right|_{\lambda}$.
Similarly, the choice of $\mathcal{D}'$ defines a different basis
$\left\{ \left.\hat{f}'_{\left[\mathbf{x};\alpha'\right]}\right|_{\lambda}\right\} $
for $\left.\mathcal{H}_{x}\right|_{\lambda}$, where $\alpha'\in\left.\mathcal{F}'\right|_{\lambda}$.
Both constructions diagonalize $\triangle$ and therefore decompose
$\mathcal{H}$ into the same spectral fibers $\mathcal{H}_{\lambda}$.
In the discrete-multiplicity case, both choices define orthonormal
bases of the same (finite-dimensional) subspace $\mathcal{H}_{\lambda}$,
related by a unitary matrix $W_{\lambda}$ such that:
\[
\left.\hat{f}'_{\left[\mathbf{x};\alpha'\right]}\right|_{\lambda}=\sum_{\alpha}\left[W_{\lambda}\right]_{\alpha'\alpha}\left.\hat{f}_{\left[\mathbf{x};\alpha\right]}\right|_{\lambda},
\]
hence the kernels are related by a unitary mixing in the degenerate
sector. Nevertheless, the eigenvalue labels $\alpha$ are drawn from
spectrum $\sigma\left(\hat{\mathcal{O}}_{1},\dots,\hat{\mathcal{O}}_{m}\right),$
with \textit{$\mathcal{D}=\left\{ \hat{\mathcal{O}}_{i}\right\} _{i=1}^{m}$},
while the labels $\alpha'$ are drawn from spectrum $\sigma\left(\hat{\mathcal{O}}'_{1},\dots,\hat{\mathcal{O}}'_{m}\right),$
with \textit{$\mathcal{D}'=\left\{ \hat{\mathcal{O}}'_{i}\right\} _{i=1}^{m}$}.
Hence, the resulting label spaces $\mathcal{F}$ and $\mathcal{F}'$
need not coincide (and may have different product structure/topological
type).

In particular, coordinate-adapted separation schemes may select different
commuting families (different MASAs), leading to different realizations
of $\mathcal{F}_{\lambda}$ while leaving the underlying spectral
decomposition of $\triangle$ unchanged. The distinction lies in the
additional commuting operators used to label degeneracy: different
commuting families define different joint spectra, hence different
label spaces $\mathcal{F}$ and $\mathcal{F}'$ in general.

In the continuous-multiplicity case, the same conclusion is expressed
by replacing the discrete sum over $\alpha$ with the corresponding
integral transform on the spectral fiber. $\blacksquare$

\paragraph*{Remark {[}The Physical Distinction{]}.}

This clarifies the confusion regarding coordinate roles in GFT. A
passive change of chart on $\Sigma$ (Case i) solely rewrites the
same geometric objects (operators and eigenfunctions) in different
coordinates. For a \emph{fixed} degeneracy-resolving MASA (fixed commuting
family), such a re-description does not alter the abstract joint spectrum
$\mathcal{F}$. The dependence described in Case (iii) arises when
one commits to a \emph{coordinate-adapted} construction (e.g.\ a
separation ansatz) that implicitly selects a different commuting family/MASA
before solving; this is a representation (gauge) choice rather than
a physical change.

\subsection{St$\ddot{\textrm{a}}$ckel and Non-St$\ddot{\textrm{a}}$ckel Cases}

We established in Section IV B that the existence of a complete MASA
in $\Sigma$ is related to a natural choice of coordinate that highlight
the symmetries in $\Sigma$. This is due to the fact that they are
both obtained from the maximal set of commuting Killing fields in
$\Sigma$. Although, not all coordinate systems are created equal.
A special class of coordinates, known as \textit{St$\ddot{\textrm{a}}$ckel
coordinates}, allows the Helmholtz equation to separate into decoupled
ordinary differential equations (ODEs).

\subsubsection{St$\ddot{\textrm{a}}$ckel Manifold and St$\ddot{\textrm{a}}$ckel
Coordinate}

A Riemannian manifold $\left(\Sigma,\boldsymbol{q}\right)$ is a St$\ddot{\textrm{a}}$ckel
manifold / St$\ddot{\textrm{a}}$ckel-separable if the Hamilton-Jacobi
equation for the geodesic flow admits a complete additive separation
for each of its variables \cite{Sep1,Torres,Stackel}. Equivalently,
there exists a local coordinate chart (the St$\ddot{\textrm{a}}$ckel
coordinate) in $\Sigma$ that allows the separability of the Helmholtz/
Laplace-Beltrami equation, possibly after factoring out a conformal
factor/modulation function \cite{Stackel3,Stackel4}. In principle,
one could list all the possible orthogonal St$\ddot{\textrm{a}}$ckel
coordinate charts in $\Sigma$, if it is St$\ddot{\textrm{a}}$ckel;
for example, in Euclidean $\mathbb{R}^{3},$ the Helmholtz equation
separates in 11 inequivalent families of orthogonal coordinates (Cartesian,
spherical, cylindrical, elliptic, etc.) \cite{Sep5,moon-spencer,vectorAnalysis}.

The existence of St$\ddot{\textrm{a}}$ckel coordinate charts in $\Sigma$
guarantees the metric to be written in a diagonal form determined
by the St$\ddot{\textrm{a}}$ckel matrix $\mathbf{S}$, an $n\times n$
non-singular matrix whose element $S_{ij}$ depends only on the coordinate.
This allows $\triangle$ to be variable-separated. For the Helmholtz
equation (\ref{eq:Helmholtz}) to be separable, one needs to do the
R-separation trick \cite{Sep5,Rsep1,Rsep2}, this will be discussed
in Section V D.

Since the maximal set of commuting geometric operators and St$\ddot{\textrm{a}}$ckel
coordinate in $\Sigma$ are defined from Killing fields, the problem
of finding a set of commuting geometric operators in $\Sigma$ is
equivalent to finding the corresponding St$\ddot{\textrm{a}}$ckel
coordinate in it. 

St$\ddot{\textrm{a}}$ckel separability implies the existence of a
complete family of mutually commuting second-order symmetry operators
(equivalently: commuting rank-2 Killing tensors, together with Killing
vectors for cyclic variables), which in our language provides a complete
commuting set of geometric operators (i.e., complete MASA). The converse
(a complete commuting set in our sense \ensuremath{\Rightarrow} St$\ddot{\textrm{a}}$ckel
separability) need not hold.

Once we obtain the St$\ddot{\textrm{a}}$ckel coordinate, we only
need to write the Helmholtz problem (\ref{eq:Helmholtz}) in this
chart and solve it to obtain the corresponding orthonormal basis.
The orthonormal basis is guaranteed to correspond to a maximal set
of commuting geometric operators $\left\{ \triangle,\mathcal{\hat{O}}_{1},\dots,\mathcal{\hat{O}}_{n-1}\right\} $
within the second-order (St$\ddot{\textrm{a}}$ckel) construction.
One could refer to Table I for a well-known example in $\mathbb{R}^{3}$. 

\subsubsection{A Constructive Algorithm for St$\ddot{\textrm{a}}$ckel Coordinate }

Let us return to the algorithm to find the MASA in Section IV B. Building
on the MASA algorithm, we outline the constructive procedure to find
St$\ddot{\textrm{a}}$ckel coordinates. After Step 2, we could obtain
the maximal commuting sets of Killing vectors $\left\{ \sigma_{a}\right\} $,
$a=1,...,m$ where $m<n$. The next steps are the following:
\begin{description}
\item [{Step$\;$3}] \textbf{Build the cyclic coordinates.} Find the maximal
set of commuting Killing vectors $\left\{ \hat{\sigma}_{a}\right\} _{a=1}^{r}$.
Integrate the Killing vectors to obtain their corresponding integral
curve $\gamma_{\left[t\right]}^{\left(a\right)}$ by (\ref{eq:flow}),
i.e., $\frac{d}{dt}\gamma_{\left[t\right]}^{\left(a\right)}=\hat{\sigma}_{a}.$
The affine parameter $u^{a}$ along $\gamma_{\left[t\right]}^{\left(a\right)}$
is the new coordinate such that $\hat{\sigma}_{a}=\partial_{u^{a}}$.
By condition (\ref{eq:Killingvector}), the metric $q_{ij}$ is independent
of $u^{a},$ hence $u^{a}$ is cyclic. Normally, to complete the coordinate
chart, one could construct the remaining non-cyclic ones by picking
coordinates $\left\{ v^{p}\right\} $, $p=1,...,\left(n-1-r\right)$,
lying orthogonal to all $\left\{ \hat{\sigma}_{a}\right\} $, i.e.,
by $\hat{\sigma}_{a}\left[v^{p}\right]=0$ $\forall\;a.$ However,
for a coordinate to be St$\ddot{\textrm{a}}$ckel, it must arise from
the second-order Killing tensor.
\item [{Step$\;$4}] \textbf{Collect the sufficient second order Killing
tensors.} If $r<n,$ then we need $\left(n-1-r\right)$ more operators,
this could be obtained from the Killing tensors $\overline{\boldsymbol{K}}$. 
\begin{enumerate}
\item Solve the rank-2 Killing-tensor equation (\ref{eq:Killingtensor})
to obtain the symmetric tensors $\overline{\boldsymbol{K}}=\beta^{p}\hat{\kappa}_{p}$,
$\beta^{p}\in\mathbb{R}$, and $\hat{\kappa}_{p}$ is the basis/generator
that construct the space of solution $\mathcal{K}$. 
\item Construct a subspace inside $\mathcal{K}_{0}\subset\mathcal{K}$ where
all its elements commute with the Killing vectors $\left\{ \hat{\sigma}_{a}\right\} $.
This could be done by solving $\mathcal{L}_{\boldsymbol{K}}\overline{\boldsymbol{K}}=0,$
$\boldsymbol{K}$ is the Killing vector, or simply by checking $\mathcal{L}_{\sigma_{a}}\hat{\kappa}_{p}=0.$
\item Find a maximal Abelian set in $\mathcal{K}_{0},$ i.e., $\left\{ \hat{\kappa}_{a}\right\} $
where every element inside are in involution $\left[\hat{\kappa}_{a},\hat{\kappa}_{b}\right]_{\textrm{NS}}=0$.
If the rank of this space is lower than $\left(n-1-r\right)$, the
manifold does not possess a St$\ddot{\textrm{a}}$ckel coordinate
and hence a non-St$\ddot{\textrm{a}}$ckel manifold.
\end{enumerate}
\item [{Step$\;$5}] \textbf{Obtain the remaining non-cyclic coordinate
from the second order Killing tensors. }Choose $\left(n-1-r\right)$
functionally independent elements from the maximal Abelian set, let
us called it as $\left\{ \hat{\kappa}_{p}\right\} $. 
\begin{enumerate}
\item Notice that $\left\{ \hat{\kappa}_{p}\right\} $ is a set of rank
(2,0) tensors. As written, they would not give a standard eigenproblem
unless $\hat{\kappa}_{p}=\kappa_{p}^{ij}\partial_{i}\otimes\partial_{j}$
is an endomorphism. So let us define define an endomorphism $\hat{\kappa}_{p}^{*}=\left(\kappa_{p}\right)_{\:j}^{i}\partial_{i}\otimes dx^{j},$
where its components are obtained by metric contraction as follows:
$\left(\kappa_{p}\right)_{\:j}^{i}:=\kappa_{p}^{ik}q_{jk}$. 
\item Diagonalize each $\hat{\kappa}_{p}^{*}$ to obtain the eigenfunction
1-form (eigenline) $e_{\lambda}^{\left(p\right)}$, i.e.:
\begin{equation}
\hat{\kappa}_{p}^{*}e_{\lambda}^{\left(p\right)}=\lambda^{\left(p\right)}e_{\lambda}^{\left(p\right)},\label{eq:eigenline}
\end{equation}
with $\lambda^{\left(p\right)}$ is its corresponding eigenvalue.
\item Check the Frobenius integrability condition of the eigenfunction $e_{\lambda}^{\left(p\right)}$,
namely \cite{Frob1,Frob2}:
\begin{equation}
e_{\lambda}^{\left(p\right)}\wedge\mathrm{d}e_{\lambda}^{\left(p\right)}=0.\label{eq:Frobenius}
\end{equation}
If this is satisfied, $e_{\lambda}^{\left(p\right)}$ is integrable
by the Frobenius theorem and could be written as $e_{\lambda}^{\left(p\right)}=g_{\left[\mathbf{x}\right]}\mathrm{d}q_{\lambda\left[\mathbf{x}\right]}^{\left(p\right)};$
otherwise, find a linear combination of all possible eigenfunction
$e_{\lambda}^{\left(p\right)}$ inside the degenerate eigenspace $\left.\mathcal{H}_{x}\right|_{\lambda}\subset\mathcal{H}_{x}$.
If no such choice yields integrable eigenforms, the manifold is not
St$\ddot{\textrm{a}}$ckel- separable.
\item $q_{\lambda\left[\mathbf{x}\right]}^{\left(p\right)}:v^{p}$ is the
remaining $\left(n-1-r\right)$ coordinate, which, together with $\left\{ u^{a}\right\} $
gives $\left(n-1\right)$ coordinate variables. The last variable
$\chi$ is obtained by integrating \textit{any} scalar function whose
gradient annihilates all symmetry directions, namely:
\[
\alpha^{\left(a\right)i}\partial_{i}\chi=0,\qquad\beta^{\left(p\right)ij}\partial_{j}\chi=0.
\]
Locally, a solution exists provided the annihilator distribution is
integrable; the solution is unique up to reparametrization.
\item The St$\ddot{\textrm{a}}$ckel coordinate is $\left\{ u^{a},v^{p},\chi\right\} $,
with $a=1,...,r,$ $p=r+1,...,n-1$.
\end{enumerate}
\end{description}
One could verify that $\bar{\mathbf{x}}=\left\{ u^{a}:=\bar{x}^{a},v^{p}:=\bar{x}^{p},\chi:=\bar{x}^{n}\right\} $
is St$\ddot{\textrm{a}}$ckel by showing that the Laplace-Beltrami
equation and the Hamilton-Jacobi equation are separable in this coordinate.

\subsubsection{The R-Separation Trick}

In St$\ddot{\textrm{a}}$ckel coordinate, the metric $\mathbf{q}$
of a St$\ddot{\textrm{a}}$ckel manifold $\Sigma$ could always be
written in a diagonal form (with diagonal components $q_{i\left[\bar{x}^{i}\right]}$):
\begin{equation}
ds^{2}=\sum_{i=1}^{n}q_{i\left[\bar{x}^{i}\right]}\left(d\bar{x}^{i}\right)^{2}=\sum_{a=1}^{r}q_{a}\left(d\bar{x}^{a}\right)^{2}+\sum_{b=r+1}^{n}q_{b\left[\bar{x}^{b}\right]}\left(d\bar{x}^{b}\right)^{2},\label{eq:Stackelmetric}
\end{equation}
where $q_{a}=c_{a}$ is a constant (because $\bar{x}^{a}$ for $a=1,...,r$
are cyclic) and its $q_{b\left[\bar{x}^{b}\right]}$ is only a function
of the $\bar{x}^{b}$ (the non-cyclic coordinates $v^{b}$). This
comes from the theorem by Benenti \cite{separa2}, that the components
of the metric $\mathbf{q}$ in St$\ddot{\textrm{a}}$ckel coordinate
could always be written as:
\[
q_{\left[\bar{\mathbf{x}}\right]}^{ii}=\left[\mathbf{S}^{-1}\right]_{1i},
\]
with $\mathbf{S}$ is the St$\ddot{\textrm{a}}$ckel matrix, where
its components are:
\[
S_{ip\left[\bar{\mathbf{x}}\right]}:=\bar{\kappa}_{p\left[\bar{\mathbf{x}}\right]}^{ii},
\]
obtained from the Killing tensor $\hat{\kappa}_{p}=\kappa_{p}^{ij}\partial_{i}\otimes\partial_{j},$
but written in the new St$\ddot{\textrm{a}}$ckel coordinate, i.e.:
\[
\bar{\kappa}_{p\left[\bar{\mathbf{x}}\right]}^{ij}=\frac{\partial\bar{x}^{i}}{\partial x^{k}}\frac{\partial\bar{x}^{j}}{\partial x^{l}}\kappa_{p\left[\mathbf{x}\right]}^{kl}=\bar{\kappa}_{p\left[\bar{\mathbf{x}}\right]}^{ii}.
\]
The Jacobian guarantees the Killing tensors are diagonal in the new
coordinate $\bar{\mathbf{x}}.$ See \cite{Rsep3,Rsep4}.

Even if the metric diagonalizes, the Helmholtz equation (\ref{eq:Helmholtz}),
written as $\left(\triangle+\lambda\right)\psi=0,$ may require a
modulation factor to separate fully. This is the Robertson condition/R-separation
trick as follows \cite{Rsep1,Rsep2}.

We seek solutions of the form $\psi_{\left[\bar{\mathbf{x}}\right]}=R\bar{\psi}_{\bar{\mathbf{x}}}$,
where the modulation factor $R$ is typically related to the metric
determinant:
\[
R=R_{\left(\bar{\mathbf{x}}\right)}=q^{-1/4}=\Bigl[q_{1}q_{2}\dots q_{n}\Bigr]^{-1/4}\prod_{i=1}^{n}\;q_{i\left[\bar{x}^{i}\right]}^{-\nicefrac{1}{4}},
\]
and $\bar{\psi}_{\bar{\mathbf{x}}}$ is the product of separable individual
ODE's solutions:
\[
\bar{\psi}_{\bar{\mathbf{x}}}=\prod_{i=1}^{n}\bar{\psi}_{i\left[\bar{x}^{i}\right]}.
\]

Substituting this ansatz into the Helmholtz equation yields $n$-decoupled
ODEs if and only if the Robertson condition holds \cite{Rsep1}:
\[
\partial_{i}\partial_{j}\left[\ln\sqrt{\left|\boldsymbol{q}\right|}q^{ii}\right]=0,\qquad i\neq j.
\]
Inserting this ansatz $\psi_{\left[\bar{\mathbf{x}}\right]}=R\bar{\psi}_{\bar{\mathbf{x}}}$
to the Helmholtz equation and using the separable metric (\ref{eq:Stackelmetric})
allows the Helmholtz PDE to be written as:
\begin{align}
\sum_{i=1}^{n}\left(\frac{1}{q_{i}}\partial_{i}^{2}\bar{\psi}-\left(\frac{1}{q_{i}^{2}}\partial_{i}q_{i}\right)\partial_{i}\bar{\psi}+V_{i\left[\bar{\mathbf{x}}\right]}\bar{\psi}\right)+\lambda\bar{\psi} & =0,\label{eq:Helmholtzseparable}\\
V_{i\left[\bar{\mathbf{x}}\right]}=\frac{1}{4}\frac{1}{q_{i}}\left(\frac{1}{2}\Bigl(\partial_{i}\ln q_{i}\Bigr)^{2}-\partial_{i}^{2}\ln q_{i}\right),\nonumber 
\end{align}
where each individual ODE's gives a separation constant $\alpha_{i}$
(degeneracies of the eigenvalue $\lambda$). Since $q_{i}$ is only
a function of a single variable $\bar{x}^{i}$, (\ref{eq:Helmholtzseparable})
is additively separable, with a separable solution:
\[
\psi_{\left[\bar{\mathbf{x}}\right]}=\prod_{i=1}^{n}\;q_{i\left[\bar{x}^{i}\right]}^{-\nicefrac{1}{4}}\;\bar{\psi}_{i\left[\bar{x}^{i}\right]}.
\]

To summarize this section, only sets of orthonormal basis that correspond
to sets of commuting, local, differential operators originating from
the Killing field will gives orthogonal coordinates in $\Sigma.$
Furthermore, if one requires the Helmholtz equation to be solvable
via separation of variables, the choice of orthonormal basis needs
to be St$\ddot{\textrm{a}}$ckel. Not all manifold $\Sigma$ admits
St$\ddot{\textrm{a}}$ckel coordinate, although its existence guarantee
the existence of the complete maximal Abelian sets of geometric operators
and hence  the existence of a complete label of the momentum domain
$\mathcal{F}$. This will be important for the classification of GFT
in the next section.

In the end, the freedom to choose the orthonormal basis in Hilbert
subspace affect the degenerate sector of GFT. \textbf{The abstract
GFT stay invariant, but the choice of coordinate will naturally lead
to a specific commuting set of geometric operators. This will affect
the topology of the $k$-space $\mathcal{F}$ for the case of degenerate
GFT.}

\section{GFT Classifications}

The construction in Section IV and V allow us to propose classifications
for GFTs on Riemannian manifolds. Such classifications are vital for
highlighting the possible topological and geometrical structures of
the spectral domain. While existing literature often isolates specific
cases, here we propose a unified framework that categorizes GFTs based
on their algebraic tractability. The classification could also serve
as a practical guidance for predicting the possible GFT structure
given limited information, such as the topology of $\Sigma$, its
symmetry class, and algebraic completeness.

\subsection{First Classification: MASA Completeness and St$\ddot{\textrm{a}}$ckel
Existence}

\subsubsection{Semi-Algorithm and Completeness at Rank $m$}

The operator construction procedure in Section IV B is a semi-algorithm
in the sense that it can certify completeness of the MASA, but not
its incompleteness. We say that the MASA is complete at Killing-tensor
rank $m$ if by rank $m$ one succesfully identifies a maximal set
of $\left(n-1\right)$ mutually commuting operators in addition to
$\triangle$ (or $n$ if excluding $\triangle$). If the maximal
set is not filled, the procedure may be extended to higher ranks,
and the status remains inconclusive.

To obtain a decidable classification at finite order, we impose an
upper bound $m=m_{0}$. If the algorithm terminates successfully before
or at rank $m_{0}$, the MASA (and hence the GFT) is certified complete
(at rank $m_{0}$), otherwise it is certified incomplete at rank $m_{0}$.
In this work, we fix $m_{0}=2$ as the upper bound; beyond second
order, the rank-raising search is treated as a semi-algorithm.
\begin{lyxlist}{00.00.0000}
\item [{\textbf{Definition.}}] (Certified Rank-2 Completeness). \textit{A
GFT system is certified complete at rank $m_{0}$ if there exists
a MASA $\mathcal{D}=\left\{ \hat{\mathcal{O}}_{i}\right\} _{i=1}^{n-1}$
(in addition to $\triangle$) where all generators $\hat{\mathcal{O}}_{i}$
are local differential operators of order $\leq m_{0}$, and their
principal symbols are functionally independent.}\textbf{\textit{ }}\textit{When
$m_{0}=2$, the existence of an orthogonal St$\ddot{a}$ckel structure
provides a standard geometric certificate for a complete commuting
family in the second-order (rank-2) class. Failure at $m=2$ implies
incompleteness at rank 2 (hence the manifold is non-St$\ddot{a}$ckel),
forcing us to rely on either higher-order (non-geometric) symmetries
or numerical diagonalization.}
\end{lyxlist}

\subsubsection{Why Rank 2? The St$\ddot{\textrm{a}}$ckel Certificate}

The choice $m_{0}=2$ is not arbitrary. Rank-2 Killing tensors admit
several distinguished physical and mathematical features that make
them the natural cutoff for a standard GFT classification:
\begin{description}
\item [{(1)}] \textbf{Natural companions to the Laplace-Beltrami operator}:
$\triangle$ is second order and canonically built from the metric;
the next simplest local commuting symmetry class beyond Killing vectors
consists of second-order operators generated by rank-2 Killing tensors.
\item [{(2)}] \textbf{Direct classical correspondence (quadratic integrals)}:
Rank-2 Killing tensors correspond to integrals of the geodesic flow
that are quadratic in the momenta, generalizing linear momentum and
angular momentum.
\item [{(3)}] \textbf{A canonical certificate package (St$\ddot{\textrm{a}}$ckel
theory)}: At rank 2 there exists an operational framework relating
separability of the Hamilton-Jacobi/Helmholtz problems to geometric
data (St$\ddot{\textrm{a}}$ckel matrices, orthogonal coordinate webs,
and Frobenius-type integrability tests).
\item [{(4)}] \textbf{Controlled quantization and reduced ambiguity}: Second-order
symmetry operators admit comparatively standard quantization prescriptions,
whereas higher-order operators typically introduce more ordering ambiguity
and potential commutator anomalies.
\item [{(5)}] \textbf{Compatibility with separation of variables practice}:
The separation-of-variables literature for Laplace (or Helmholtz)
equations is largely organized around orthogonal rank-2 structures
(e.g., hydrogen atoms, harmonic oscillator), making this class interpretable
and computationally tractable in low dimensions.
\item [{(6)}] \textbf{Minimal hidden symmetry beyond isometries}: Killing
vectors encode isometries; rank-2 Killing tensors are the minimal
setting in which genuinely new (non-group) hidden symmetries appear.
\end{description}
Accordingly, we treat rank-2 completeness as the primary threshold
in our taxonomy (the St$\ddot{\textrm{a}}$ckel certificate). Systems
failing this test are classified as \textquotedblleft algebraically
incomplete\textquotedblright{} within the geometric framework. Completeness
beyond rank 2 may be certified by the rank-raising construction when
successful, but the procedure is not guaranteed to terminate in general.

\subsubsection{Completeness Classification at Rank $m=2$}

Here, we classify the GFT systems based on their MASA completeness/
incompleteness at rank $2$ and the existence of St$\ddot{\textrm{a}}$ckel
coordinate. It must be emphasized that this classification is restricted
to the cutoff $m=2.$ The completeness certificate is absolute: once
the $\left(n-1\right)$-quota is achieved at a finite order $m$,
higher-rank commuting operators (if they exist) are redundant for
labeling the regular set. By contrast, failure at a given cutoff $m$
only certifies $m$-incompleteness and does not exclude completion
at higher rank. Therefore, the label ``incomplete'' should be interpreted
as ``\textit{the number of MASA does not reach the quota $\left(n-1\right)$
within the class of Killing tensors of order $\leq2$; whether it
reaches the quota on higher ranks remains inconclusive}\textquotedbl .

The classification defined as follows:
\begin{itemize}
\item \textbf{Type I (Geometrically separable):} The manifold $\Sigma$
admits a complete MASA (quota $n-1$ is satisfied in addition to $\triangle$)
at $m=2$ and the eigenforms satisfy the Frobenius integrability condition
(\ref{eq:Frobenius}). This implies the existence of St$\ddot{\textrm{a}}$ckel
coordinate, hence $\Sigma$ is St$\ddot{\textrm{a}}$ckel and the
Helmholtz equation is separable in an (orthogonal) St$\ddot{\textrm{a}}$ckel
chart of $\Sigma$. Examples of this type are the well-known cases:
GFT in $\mathbb{R}^{n}$, $\mathbb{S}^{n}$, $\mathbb{H}^{n}$, and
flat tori.
\item \textbf{Type II (Algebraically complete/geometrically non-separable)}:
The manifold $\Sigma$ admits a complete MASA at $m=2$ but fails
the Frobenius integrability condition (\ref{eq:Frobenius}). This
implies that $\Sigma$ does not admit a St$\ddot{\textrm{a}}$ckel
coordinate. The second order Killing tensors are sufficient to label
the states algebraically (to complete the MASA), but at least one
of them does not satisfies the Frobenius integrability (\ref{eq:Frobenius})
(notice that the Frobenius integrability (\ref{eq:Frobenius}) is
defined exclusively for the $2^{\mathrm{nd}}$-order Killing tensor).
Hence their eigenvectors do not form a holonomic coordinate grid.
The system is MASA-complete but the Helmholtz equation in $\Sigma$
is non-separable at least for 2 variables. Type II arises exclusively
in the degenerate-branch of the classification.
\item \textbf{Type III (Rank-2 incomplete)}: If the degenerate GFT lives
in $\Sigma$ does not admit a complete MASA (rank $\mathfrak{h}=r_{\mathrm{max}}<$
$n-1$) at $m=2$, this is classified as Type III. This implies that
$\Sigma$ does not admit a St$\ddot{\textrm{a}}$ckel coordinate.
Hence $\Sigma$ is non-St$\ddot{\textrm{a}}$ckel and the Helmholtz
equation in $\Sigma$ is only separable via the St$\ddot{\textrm{a}}$ckel
method for a maximal $r_{\mathrm{max}}$ variable or lower. If we
insist on stopping the algorithm at $m=2$, then any remaining degeneracies
in this case must be resolved by operators that are either higher-order
or non-local; in the quantum mechanical case, this will lead to a
non-complete CSCO. Examples for this type may include ergodic or even
chaotic geodesic flows. On the other hand, the non-degenerate GFT/
simple $\triangle$ is always of this type except for 1-dimensional
cases, which is MASA complete and St$\ddot{\textrm{a}}$ckel (Type
I), since the MASA for these cases are simply $\left\{ \triangle\right\} $.
Here, incompleteness refers to whether the semi-algorithm produce
full quota of $\left(n-1\right)$ independent commuting symmetry operators
of Killing-tensor order $\leq2$, not to maximality of an Abelian
(von Neumann) algebra.
\end{itemize}
We provide a flowchart to classify the GFT into Type I, II, and III,
see FIG. 2.

\begin{figure}[h]
\centering{}\includegraphics[scale=0.7]{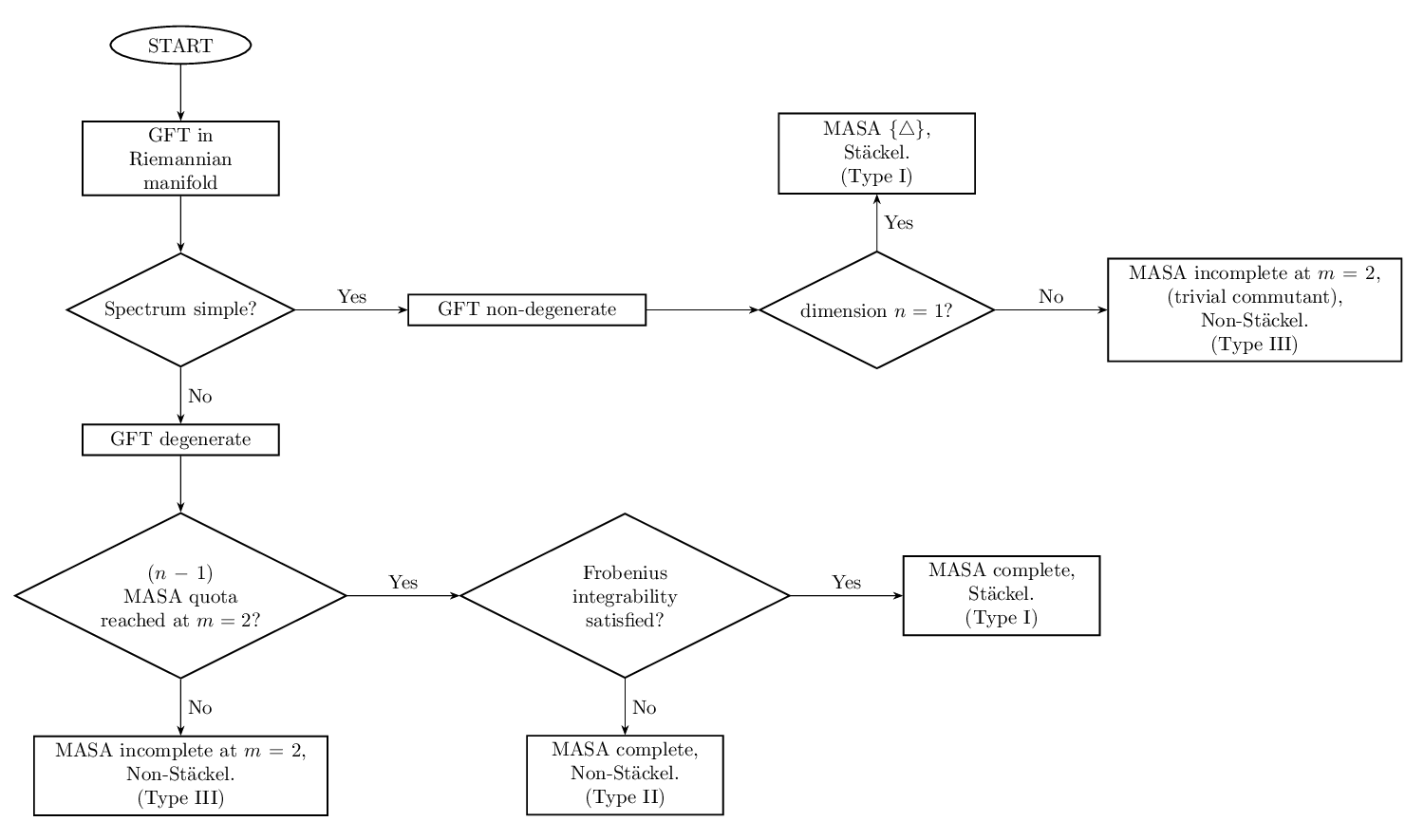}\caption{The GFT classification, based on MASA completeness at $m=2$ and existence
of St$\ddot{\textrm{a}}$ckel coordinate.}
\end{figure}

\paragraph*{Remark {[}On non-degenerate spectra{]}.}

For the non-degenerate branch, the spectrum $\lambda$ is simple,
in this case the eigenfunctions $\hat{f}_{\left[\mathbf{\mathbf{x};\lambda}\right]}$
are uniquely labeled by the spectral parameter $\lambda$ (up to phase).
Consequently, any symmetry operator commuting with $\triangle$ must
preserve each one-dimensional spectral subspace and hence acts as
multiplication by a scalar on $\hat{f}_{\left[\mathbf{\mathbf{x};\lambda}\right]};$
equivalently it belongs to the functional calculus $f\left(\triangle\right).$
Within our semi-algorithm restricted to Killing-tensor data of order
$\leq2$ (i.e. $m=2$), this yields no additional independent commuting
operators beyond the trivial commutant, and the $\left(n-1\right)$-quota
is not reached for $n\geq2$. We therefore classify this branch as
MASA-incomplete at $m=2$ (Type III) because the semi-algorithm yields
no spatial separation structure. However, unlike chaotic Type III
cases, here the \textquotedblleft incompleteness\textquotedblright{}
refers to the absence of independent commuting auxiliary symmetry
operators, not an inability to uniquely label the basis.

The first classification we introduce above is geometrical: for the
degenerate sector, the classification depends on the intrinsic symmetries
of $\Sigma$, independent of the specific basis choice within the
fiber.

\subsection{Second Classification: Topology of the Spectral Domain $\mathcal{F}$}

We propose a second classification based on the topology of the Fourier
dual domain ($k$-space) $\mathcal{F}$. Unlike the first classification,
which was intrinsic to the manifold geometry, this classification
is spectral and may depend on the choice of MASA in the degenerate
sector, see Table II.

\subsubsection{Topological Classification}

We classify the GFT into three categories based on the structure of
the joint spectrum $\sigma\left(\hat{\mathcal{O}}_{1},\dots,\hat{\mathcal{O}}_{m}\right)$
(the topology of the $k$-space $\mathcal{F}$):
\begin{itemize}
\item \textbf{Continuous (C)}: Occurs when the spectrum is purely continuous.
This arises in non-compact, unbounded manifolds (e.g., Euclidean plane
$\mathbb{R}^{2}$ and hyperbolic plane $\mathbb{H}^{2}$ with Cartesian
MASA). The spectral domain $\mathcal{F}$ is a continuous manifold
(typically homeomorphic to $\mathbb{R}^{n}$).
\item \textbf{Discrete (D)}: Occurs when the spectrum of the Laplace-Beltrami
operator (and all commuting operators) consists purely of point eigenvalues.
This is characteristic of compact manifolds (e.g., sphere $\mathbb{S}^{2}$
and torus $\mathbb{T}^{2}$). The spectral domain $\mathcal{F}$ forms
a lattice or discrete grid.
\item \textbf{Semi-Discrete (SD)}: A hybrid case where the spectrum is discrete
in some dimensions and continuous in others (at least one band is
continuous). This occurs in manifolds with mixed boundary conditions
or cylindrical topologies (e.g., the infinite cylinder $\mathbb{R}\times\mathbb{S}^{1}$),
or non-compact, unbounded manifold with compact choice of MASA (e.g.,
$\mathbb{R}^{3}$ with cylindrical or spherical MASA).
\end{itemize}
See the corresponding flowchart as follows.

\begin{figure}[h]
\centering{}\includegraphics[scale=0.77]{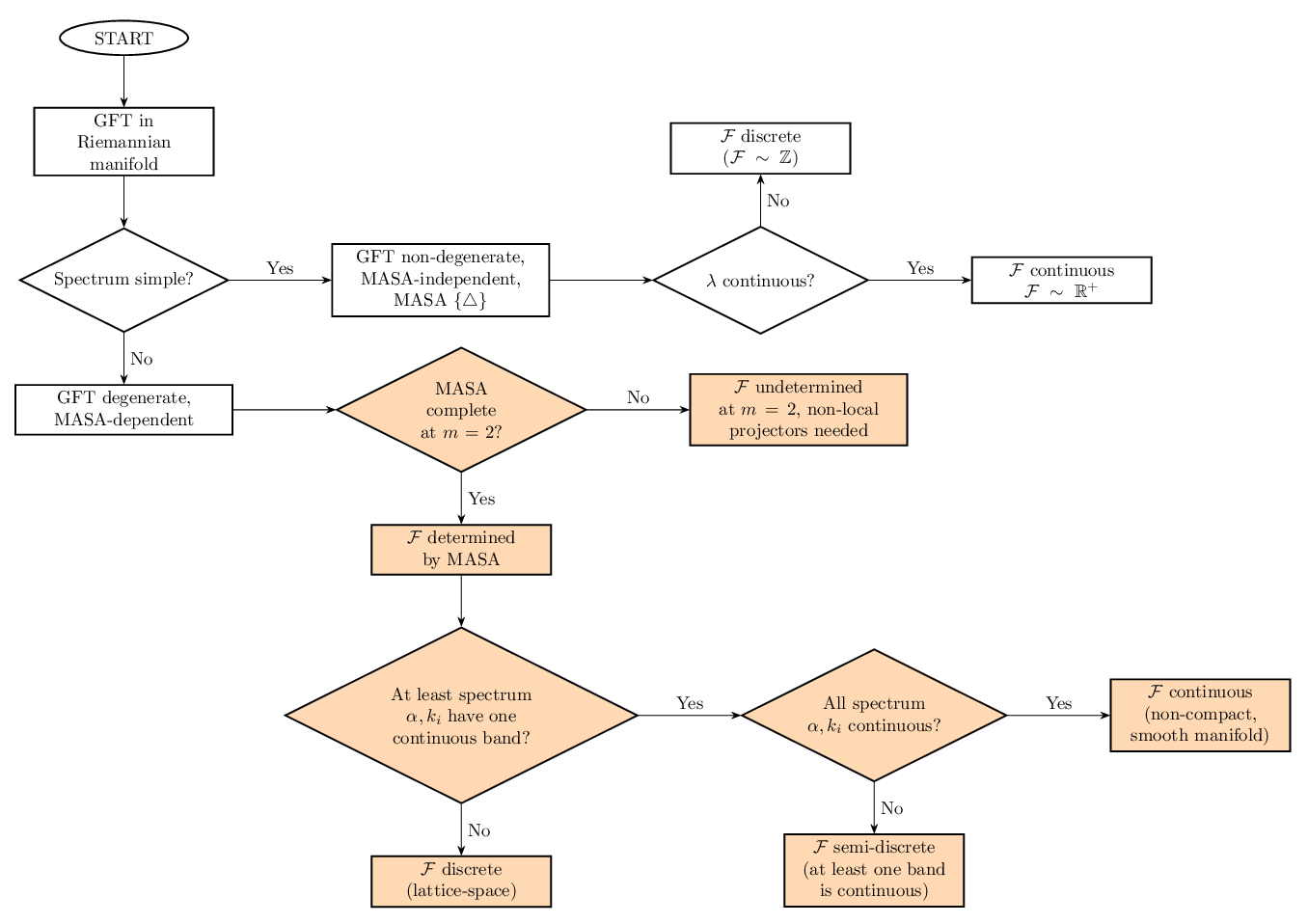}\caption{Classification of GFT based on the topology of $\mathcal{F}.$ Here
the upper-bound for the rank of Killing tensor is $m=2$.}
\end{figure}

\paragraph*{Remark {[}The role of degeneracy{]}.}

As discussed in Section V, the degenerate sector of the GFT is sensitive
to the choice of orthonormal basis (the choice of MASA). Consequently,
for degenerate eigenvalues, the specific topology of the $k$-space
depends on which commuting operators are chosen to lift the degeneracy.
This dependency is illustrated in the \textquotedbl degenerate Sector\textquotedblright{}
(orange path) of the flowchart above.

\subsubsection{The $3\times3$ Classification Chart}

Together with the topological classification, we could construct a
unified taxonomy of GFTs on Riemannian manifolds, based on their algebraic
completeness (MASA completeness in $m=2$ and Helmholtz separability):
Type I-III from Section VI A, and spectral (Fourier dual-space) topology:
Class D, C, SD from Section VI B.

\paragraph{The non-degenerate branch.}

First, we address the non-degenerate sector (simple spectrum). Here,
the classification simplifies significantly into 2 categories: \textbf{(a)
The 1D cases ($n=1$)}. These are always Type I. Since for this case
$n-1=0$ (in addition to $\triangle$), the Laplace-Beltrami alone
constitutes a complete set. The spectrum can be discrete (topologically
equivalent to $\mathbb{Z}$), e.g., circle $\mathbb{S}^{1}$, or continuous
(homeomorphic to $\mathbb{R}^{+}$) e.g., positive line. \textbf{(b)
Higher Dimensions ($n\geq2$)}: If the spectrum is simple, it implies
an absence of auxiliary symmetries \textcolor{blue}{providing} new
spectral labels. Consequently, these cases generically fall into Type
III. While the eigenfunctions are unique (up to phase), the system
admits no auxiliary spatial operators to form a coordinate grid, rendering
it geometrically non-separable.

As both cases have the possibility of the $k$-space to be discrete
or continuous, the discreteness is only guaranteed if $\Sigma$ is
compact\footnote{The reason for this is due to the fact that $\triangle$ is an elliptic
operator, which, in a compact support, will cause the resolvent to
be compact and hence posses a discrete spectrum.}, otherwise for non-compact $\Sigma$, it falls either into discrete
or continuous class. The branch do not admit degenerate sector, hence
the only element of its MASA is only $\triangle$, which is complete
for 1-dimensional cases, but incomplete for higher dimensions.

\paragraph{The degenerate branch.}

The degenerate sector allows for rich geometric structure. Crossing
the three algebraic types with the three topological classes yields
9 distinct categories (see Table II). 

\begin{table}[h]
\begin{centering}
\includegraphics[scale=1]{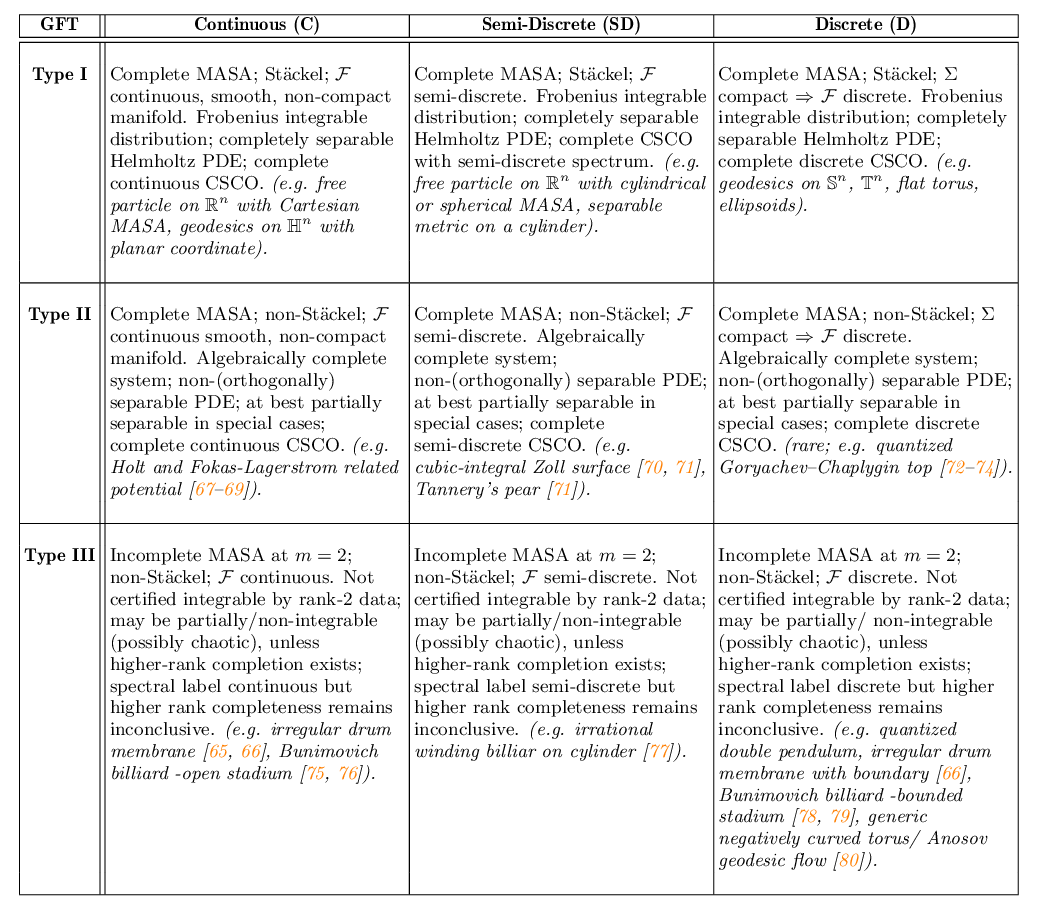}
\par\end{centering}
\caption{The $3\times3$ GFT classification chart.}
\end{table}

All GFTs on Riemannian manifolds fall into one of these cell. Type
III, where $\mathcal{F}$ is practically undetermined without the
use of non-local operators, is divided into 3 cases by the topology
of its incomplete MASA spectrum. Type IC (the ``standard'' class
where they are St$\ddot{\textrm{a}}$ckel-separable and continuous)
is the class where we could confidently state the spectral domain
$\mathcal{F}$ inherits a smooth $n$-dimensional manifold structure
admitting a globally defined (Lebesgue) measure (\ref{eq:measurek}).
The remaining eight cells may contains obstructions that are either
algebraic ($k$-space dimension is lower than $n$, i.e., $\mathcal{F}$
is labeled only by $m<n$ tuple of scalars), geometric (failure of
Frobenius integrability), and/or topologic ($\mathcal{F}$ is not
continuous and smooth).

\paragraph{Remarks {[}On separability and joint-spectrum topology (scope){]}.}

Our classification grid (MASA-completeness Type I--III versus spectral-topology
classes: discrete/continuous/mixed) is intended as an organizing principle
supported by physical examples. A general necessary-and-sufficient
criterion linking \textbf{St$\ddot{\textrm{a}}$ckel} separability
(i.e. the existence of complete commuting symmetry operators generated
by Killing data) to the topology of the \emph{joint} spectrum is an
interesting problem in quantum integrability and would require substantial
additional development, beyond the scope of the present work.

\section{Subtleties and Examples}

In this section, we examine two critical subtleties of the GFT framework
through well-known examples. The first subtlety concerns the analytical
status of the basis functions: the eigenfunction of $\triangle$ often
lies outside of the Hilbert space $\mathcal{H}_{x}\sim\mathcal{L}^{2}\left[\Sigma\right]$.
This problem could be solved with the standard rigged Hilbert space
formalism, discussed in Section VII A.

The second subtlety emphasizes the influence of the choice of MASA
on the degenerate sector of the GFT. We provide two illustrative examples:
(a) Coordinate-adapted MASA transformation (Case iii) in Subsection
VII B, where we take 2 different MASA related by a coordinate-adapted
MASA transformation in $\Sigma=\mathbb{R}^{3}$. This case clearly
highlights the topological difference on the corresponding $k$-space;
and (b) Isometry and unitary transformation (Case ii) in Subsection
VII C, where we apply the 2 different MASA related by an isometry
on the (universal cover of the) 2-dimensional torus $\mathbb{T}^{2}.$
For this case, the sets of the orthonormal basis are related by a
unitary transformation, yet topology of the $k$-spaces remains invariant.

\subsection{Isomorphism of the Rigged Hilbert Spaces under GFT}

We now step back from classification schemes to address functional-analytic
details of the GFT on the rigged Hilbert spaces. For many cases, particularly
for non-compact $\Sigma$, the eigensolution to the Helmholtz equation
(\ref{eq:Helmholtz}) lies \textit{outside} of the square-integrable
space $\mathcal{L}^{2}\left[\Sigma\right]$. These Helmholtz eigenfunctions
are generalized eigenfunctions, $\delta$-normalized, and lives in
the distributional space rather than $\mathcal{L}^{2}\left[\Sigma\right]$;
their continuous spectrum is handled via spectral theorem (direct
integral decomposition). A standard example is the Helmholtz equation
in $\mathbb{R}^{n},$ where its eigensolution $e^{\mathbf{i}\mathbf{k\cdot x}}$
that serve as an orthonormal basis in $\mathcal{L}^{2}\left[\mathbb{R}^{n}\right]$,
is not actually square-integrable in the domain $\mathbb{R}^{n}$,
hence $e^{\mathbf{i}\mathbf{k\cdot x}}$ lives in a larger non-separable
Hilbert space.

To formalize the solution to this problem, one could construct a rigged
Hilbert space (or Gelfand triple):
\begin{equation}
\Phi\left[\Sigma\right]\subset\mathcal{L}^{2}\left[\Sigma\right]\subset\Phi^{*}\left[\Sigma\right],\label{eq:Gelfand}
\end{equation}
with $\Phi\left[\Sigma\right]$ is a dense subspace of test functions
in $\mathcal{L}^{2}\left[\Sigma\right]$ (usually a Schwartz space
$\mathcal{S}\left[\Sigma\right]$ if $\Sigma$ is non-compact or $C^{\infty}\left[\Sigma\right]$
if $\Sigma$ is compact), and $\Phi^{*}\left[\Sigma\right]$ is the
(usually non-separable) continuous dual space to $\Phi\left[\Sigma\right],$
see \cite{rigged}. The dual space $\Phi^{*}\left[\Sigma\right]$
is guaranteed to contain $\mathcal{L}^{2}\left[\Sigma\right]$ via
the Riesz map, the delta distribution (\ref{eq:ddeltaq}), and the
generalized eigenfunctions $\hat{f}_{\left[\mathbf{x};\lambda,\alpha\right]}$
of the Laplace-Beltrami operators (which sometimes lies outside $\mathcal{L}^{2}\left[\Sigma\right]$)
\cite{Spectral1}. Consequently, while the basis elements $\hat{f}_{\left[\mathbf{x};\lambda,\alpha\right]}\in\Phi^{*}\left[\Sigma\right]$,
they effectively span $\mathcal{L}^{2}\left[\Sigma\right]$.

The GFT (\ref{eq:map-1}) could be applied to the rigged Hilbert space
(\ref{eq:Gelfand}) over $\Sigma$ to construct a rigged Hilbert space
over $\mathcal{F}$ as follows. The space $\mathcal{L}^{2}\left[\Sigma\right]$
is isomorphic to $\mathcal{L}^{2}\left[\mathcal{F}\right]$ via Plancherel-Parseval
theorem. We construct the spectral test space $\Phi\left[\mathcal{F}\right]$
by applying the GFT (\ref{eq:map-1}) directly to the spatial test
functions $\Phi\left[\Sigma\right]$ such that $\mathcal{U}\left(\Phi\left[\Sigma\right]\right)\sim\Phi\left[\mathcal{F}\right].$
$\Phi^{*}\left[\mathcal{F}\right]$ could be obtained by taking the
continuous dual to $\Phi\left[\mathcal{F}\right];$ $\mathcal{L}^{2}\left[\mathcal{F}\right]$
is automatically included in $\Phi^{*}\left[\mathcal{F}\right]$,
together with the delta distribution (\ref{eq:ddeltak}) and $\hat{f}_{\left[\mathbf{x};\lambda,\alpha\right]}$.
Hence, the rigged Hilbert space over $\mathcal{F}$:
\begin{equation}
\Phi\left[\mathcal{F}\right]\subset\mathcal{L}^{2}\left[\mathcal{F}\right]\subset\Phi^{*}\left[\mathcal{F}\right],\label{eq:Gelfandk}
\end{equation}
is isomorphic to (\ref{eq:Gelfand}).

\paragraph{Remark {[}General construction of the Gelfand triple for $\mathcal{F}${]}.}

One might attempt to construct the Gelfand triple (\ref{eq:Gelfandk})
not via the map (\ref{eq:map-1}) of GFT, but by independently assuming
$\Phi\left[\mathcal{F}\right]\sim\mathcal{S}\left[\mathcal{F}\right]$.
This is possible, but in general, $\mathcal{S}\left[\mathcal{F}\right]$
does not necessarily coincide with $\mathcal{U}\left(\Phi\left[\Sigma\right]\right)$
and so does their duals, hence the isomorphism between (\ref{eq:Gelfandk})
and (\ref{eq:Gelfand}) is not guaranteed by this construction.

\subsection{$\mathbb{R}^{3}$ with different Maximal Abelian Sets}

In this subsection, we demonstrate how the choice of (coordinate-adapted)
MASA in a fixed manifold $\Sigma$ fundamentally alters the GFT classification
and the topology of the resulting $k$-space. Let $\Sigma$ be a 3-dimensional
Euclidean space $\mathbb{R}^{3}$(non-compact, degenerate), equipped
with a Euclidean metric $\boldsymbol{\delta}$. Using the metric,
one could define a canonical measure on $\mathbb{R}^{3}$ as the volume
form $d\mu=\mathrm{d}^{3}\mathbf{x}$. $\mathbb{R}^{3}$ is St$\ddot{\textrm{a}}$ckel,
it possesses 11 orthogonal St$\ddot{\textrm{a}}$ckel coordinates
where each of them are related to a distinct set of MASA \cite{vectorAnalysis}.
Here, we will compare 2 different sets of MASA, the set $\left\{ \hat{p}_{x},\hat{p}_{y},\hat{p}_{z}\right\} $
and $\left\{ \triangle,\left|\hat{L}\right|^{2},\hat{L}_{z}\right\} $
which we will call respectively as the Cartesian and the spherical
MASA. The properties of these sets are listed on Table I.

The Cartesian and spherical MASA will canonically lead to two of the
11 St$\ddot{\textrm{a}}$ckel webs, namely, the Cartesian $\left(x,y,z\right)$
and the spherical coordinate $\left(r,\theta,\phi\right)$, this could
be done by applying the algorithm in Subsection V D. One could easily
check that the Helmholtz PDE are separable in these coordinates, leading
to two different expansions of the full waveform $\psi$:
\begin{align*}
\psi_{\left[x,y,z\right]} & =\intop_{\mathcal{F}}dk_{x}dk_{y}dk_{z}\,\phi_{\left[k_{x},k_{y},k_{z}\right]}\hat{f}_{\left[\mathbf{x,k}\right]},\qquad\hat{f}_{\left[\mathbf{x,k}\right]}=\frac{1}{^{\left(2\pi\right)^{\nicefrac{3}{2}}}}e^{\mathbf{i}\mathbf{k\cdot x}},\quad\mathbf{k\cdot x}=k_{x}x+k_{y}y+k_{z}z,
\end{align*}
and:
\begin{align*}
\psi_{\left[r,\theta,\phi\right]} & =\intop_{\bar{\mathcal{F}}}dk\sum_{\ell,m}\phi_{\ell m\left[k\right]}\hat{f}_{\left[\mathbf{r,k}\right]},\qquad\hat{f}_{\left[\mathbf{r,k}\right]}=R_{\ell\left[kr\right]}Y_{\ell m\left[\theta,\phi\right]},
\end{align*}
where $R_{\ell\left[kr\right]}$ and $Y_{\ell m\left[\theta,\phi\right]}$
are, respectively, the radial Bessel function and the spherical harmonics:
\begin{align}
R_{\ell\left[kr\right]} & =\frac{1}{r}\left(c_{1}j_{\ell\left[kr\right]}+c_{2}y_{\ell\left[kr\right]}\right),\qquad c_{1},c_{2}\;\textrm{constants;}\quad k\in\mathbb{R}^{+},\nonumber \\
Y_{\ell m\left[\theta,\phi\right]} & =\sqrt{\frac{\left(2\ell+1\right)}{4\pi}\frac{\left(\ell-m\right)!}{\left(\ell+m\right)!}}P_{\ell m\left[\cos\theta\right]}e^{\mathbf{i}m\phi},\qquad\left(\ell,m\right)\in\mathcal{D}\sim\mathbb{Z}^{2}.\label{eq:spherical}
\end{align}
$\hat{f}_{\left[\mathbf{x,k}\right]}$ and $\hat{f}_{\left[\mathbf{r,k}\right]}$
are respectively, the p-wave and s-wave orthonormal bases that span
$\mathcal{H}\sim\mathcal{L}^{2}\left[\mathbb{R}^{3}\right]$, but
they live in the larger nonseparable space of $\Phi^{*}\left[\mathbb{R}^{3}\right].$ 

The Cartesian $\left(x,y,z\right)$ and the spherical coordinate $\left(r,\theta,\phi\right)$
are related by a coordinate transformation
\begin{equation}
x=r\sin\theta\cos\phi,\quad y=r\sin\theta\sin\phi,\quad z=r\cos\theta,\label{eq:7}
\end{equation}
with $0\leq r<\infty$, $0\leq\theta\leq\pi$, $0\leq\phi<2\pi$,
which belongs to the subgroup $\mathrm{id}\left(\Sigma\right)$ of
the local metric-preserving diffeomorphism. 

\textcolor{blue}{One could easily check that a mere coordinate transformation
(\ref{eq:7}) does not convert a p-wave to the s-wave; rather, the
two bases are related by a unitary transformation $U_{\ell m\left[k;k_{x},k_{y},k_{z}\right]}$}:
\begin{align}
e^{\mathbf{i}\mathbf{k}\cdot\mathbf{r}} & =\intop_{\mathbb{R}^{+}}dk\sum_{\ell,m}U_{\ell m\left[k;k_{x},k_{y},k_{z}\right]}R_{\ell\left[kr\right]}Y_{\ell m\left[\theta,\phi\right]},\label{eq:unitary}
\end{align}
where $\mathbf{k}\cdot\mathbf{r}$ is an abbreviation of:
\[
\mathbf{k}\cdot\mathbf{r}=r\left(\left(k_{x}\cos\phi+k_{y}\sin\phi\right)\sin\theta+k_{z}\cos\theta\right)=\mathbf{k}\cdot\mathbf{x}.
\]

Let us construct the $k$-spaces related to these two different sets
of MASA in $\mathbb{R}^{3}.$ Since the GFT is degenerate, they admit
degenerate sectors, hence the $k$-space depend on the choice of MASA.
Using the definition of the $k$-space as the Fourier dual (\ref{eq:kspacegen})
(for Cartesian MASA) or (\ref{eq:kspace}) (for spherical MASA), the
$k$-space of $\mathbb{R}^{3}$ with Cartesian MASA is $\mathcal{F}_{\mathrm{c}}\sim\mathbb{R}\times\mathbb{R}\times\mathbb{R},$
labeled by $\left(k_{x},k_{y},k_{z}\right)$. Hence $\mathcal{F}_{\mathrm{c}}\sim\mathbb{R}^{3}$
is continuous (Type I-C). Meanwhile, the $k$-space of $\mathbb{R}^{3}$
with spherical MASA is $\mathcal{F}_{\mathrm{s}}\sim\mathbb{R}^{+}\times\mathbb{Z}^{2}$,
a semi-discrete, topological space where the index $\left(\lambda,\ell,m\right)$
labeling the orthonormal basis lives, hence Type I-SD. This space
is continuous in the $\mathbb{R}^{+}$-part, which is Fourier-dual
to itself and labeled by the spectrum $\lambda,$ but discrete in
the $\mathbb{Z}^{2}$ subspace, labeled by the spherical harmonics
index $\left(\ell,m\right)$. $\mathcal{F}_{\mathrm{s}}$ is not homeomorphic
to $\mathcal{F}_{\mathrm{c}}.$ This is the subtlety that we would
like to highlight in our work: If we want to consistently define the
Fourier dual $k$-space as the abstract space of indices (or equivalently,
the topological space of the eigenvalues), then the topology of this
$k$-space depends on how we select the MASA in $\Sigma.$ This choice
 will be reflected in the eigensolution to the Helmholtz equation
(\ref{eq:Helmholtz}), and hence, affect the $k$-space via the spectrum
of the MASA.

\textcolor{blue}{From a physical point of view, a choice of MASA may
be interpreted as a choice of maximal compatible commuting observables,
and hence as a choice of spectral or measurement context in the quantum-mechanical
framework. Different MASAs correspond to different symmetry-adapted
decompositions of the same Hilbert space. In particular, for our case
in $\mathbb{R}^{3}$, the Cartesian MASA $\left\{ \hat{p}_{x},\hat{p}_{y},\hat{p}_{z}\right\} $
is adapted to a linear-momentum description, while the spherical MASA
$\left\{ \triangle,\left|\hat{L}\right|^{2},\hat{L}_{z}\right\} $
is adapted to central symmetry and angular-momentum observables. The
passage from one MASA to the other is therefore not merely a coordinate
relabeling, but a change in the commuting observables used to spectrally
resolve the state.}

It needs to be kept in mind that the geometry of $\Sigma,$ $\mathcal{F}_{\mathrm{c}}$,
$\mathcal{\mathcal{F}_{\mathrm{s}}}$, the Laplace-Beltrami operator
$\triangle$, and all geometrical objects involved are independent
from the choice of coordinate. However, there always exists a coordinate
system that naturally captures the symmetry of $\Sigma,$ this choice
of coordinate \textit{dictates} the form of the Laplace-Beltrami operator
and select the \textcolor{blue}{coordinate-adapted} MASA, which provides
a \textquotedbl natural\textquotedbl{} spectral representation that
aligns with the separation of variables written in this coordinate.
Our example in this subsection clearly demonstrates this fact.

\subsection{Flat Torus with Rational/Irrational Slope}

The flat torus is defined by the space $\mathbb{R}^{2}/2\pi\mathbb{Z}^{2},$
i.e., by identifying a rectangle on its adjacent sides. For simplicity,
let each side of the rectangle's length be $L$. Let us use a Cartesian
coordinate $\left(x,y\right)$ to label points on the rectangle, their
coordinate lines are parallel to the sides of the square. To define
the flat torus from the square, we impose a periodicity condition
as follows:
\begin{align}
x & =x+mL\;\left(\mathrm{mod}L\right),\quad m\in\mathbb{Z},\quad x\in\mathbb{R},\label{eq:period}\\
y & =y+nL\;\left(\mathrm{mod}L\right),\quad n\in\mathbb{Z},\quad y\in\mathbb{R},\nonumber 
\end{align}
i.e., the coordinate $\left(x,y\right)$ lives on the quotient space
$\mathbb{R}^{2}/L\mathbb{Z}^{2}$. With this, the coordinate curves
defined by $x=\mathrm{constant}$ and $y=\mathrm{constant}$ are two
orthogonal families of closed geodesics on $\mathbb{T}^{2}.$

Let us define another coordinate $\left(x',y'\right)$ on $\mathbb{T}^{2}$
related to $\left(x,y\right)$ by an $SO(2)$ (local) frame transformation
as follows:
\begin{equation}
\left[\begin{array}{c}
x'\\
y'
\end{array}\right]=\left[\begin{array}{cc}
\cos\varepsilon & -\sin\varepsilon\\
\sin\varepsilon & \cos\varepsilon
\end{array}\right]\left[\begin{array}{c}
x\\
y
\end{array}\right],\qquad0\leq\varepsilon\leq\pi.\label{eq:isomet}
\end{equation}
Notice that (\ref{eq:isomet}) is an isometry on $\mathbb{R}^{2},$
the universal cover of $\mathbb{T}^{2}$ (before quotienting the space
with $\mathbb{Z}^{2}$). Now we need to apply the periodicity condition
of the torus (\ref{eq:period}) to our new coordinate. Let us write:
\begin{align*}
x & =x_{0}+\Delta x,\\
y & =y_{0}+\Delta y,
\end{align*}
where $\left(x_{0},y_{0}\right)$ is the origin of both coordinates.
Let us focus mainly on the family of curves defined by $y'=\mathrm{constant}$,
labeled with $\left.x'\right|_{y'}:=\left.x'_{\left[t\right]}\right|_{y'}$
parameterized by its length $t=\left|x'\right|$, that could be written
in the old coordinate as $\left.x'\right|_{y'}=\left(\left|x'\right|\cos\varepsilon,\left|x'\right|\sin\varepsilon\right),$
or:
\begin{align}
\left.x'\right|_{y'} & =\left(x,y\right)=\left(x_{0}+\Delta x,y_{0}+\Delta y\right).\label{eq:huff}
\end{align}
The slope of the coordinate line $\left.x'\right|_{y'}$ is then $\frac{\Delta y}{\Delta x}=\tan\varepsilon$.
Now let the $x$-component of $\left.x'\right|_{y'}$ complete $m$
circles around the torus and returns to the same point $x_{0}$, hence
$\Delta x=mL$ and (\ref{eq:huff}) could be written as:
\begin{align}
\left.x'\right|_{y'}=\left(x,y\right) & =\left(x_{0}+mL,y_{0}+mL\tan\varepsilon\right),\nonumber \\
 & =\left(x_{0},y_{0}+mL\tan\varepsilon\right).\label{eq:hiks}
\end{align}
 Notice that only the $x$-component of $\left.x'\right|_{y'}$ returns
to its original position after $m$ rotation, while the $y$-component
has shifted from its original position $y_{0}$ by $L\tan\varepsilon$. 

\textit{If we require $\left.x'\right|_{y'}$ to return to its original
position} $\left(x_{0},y_{0}\right)$, we need to impose the periodicity
condition (\ref{eq:period}) for the $y$-component, and hence, we
obtain:
\begin{equation}
\tan\varepsilon\equiv\frac{n}{m}\in\mathbb{Q},\quad n,m\in\mathbb{Z},\label{eq:req}
\end{equation}
$\mathbb{Q}$ is the set of rational numbers. 

Inserting this condition to (\ref{eq:hiks}) returns $\left.x'\right|_{y'}$
to its original position $\left(x_{0},y_{0}\right)$. (\ref{eq:req})
is a constraint on the space of $\mathbb{Z}^{2}\ni\left(n,m\right);$
it picks only the possible value of $\left(n,m\right)$ $-$\textcolor{blue}{the
winding pair of the closed orbit around the two fundamental cycles
of $\mathbb{T}^{2},$} such that $\left.x'\right|_{y'}$, with slope
$\tan\varepsilon$, returns to its original position (example: Given
the slope $\tan\varepsilon=\nicefrac{3}{4},$ then the possible sets
of $\left(n,m\right)$ is $\left\{ \left(3,4\right),\left(6,8\right),\left(9,12\right),..\right\} $).
If (\ref{eq:req}) is satisfied, namely, if the slope $\tan\varepsilon$
is rational, the coordinate line $\left(x',\mathrm{constant}\right)$
forms a closed loop in $\mathbb{T}^{2},$ hence \textit{periodic},
see FIG. 4.

It needs to be kept in mind that in general, \textit{it is not necessary
for $\left.x'\right|_{y'}$ to return to its original position, hence,
the curve }$\left.x'\right|_{y'}$ \textit{is not necessarily a closed
loop. }This case occurs when condition (\ref{eq:req}) is violated,
i.e., the slope $\tan\varepsilon$ is irrational: $\tan\varepsilon\in\mathbb{I}$.
In this case, the $x$ and $y$ components of $\left.x'\right|_{y'}$
will individually return to their original position, but they will
never reach their original point \textit{simultaneously}, hence the
coordinate line $\left.x'\right|_{y'}$ never return to its original
position (\textit{aperiodic}). The coordinate line is homeomorphic
to $\mathbb{R}^{+}$ and is dense in $\mathbb{T}^{2}$. Hence, \textcolor{blue}{an
irrational slope implies incommensurate frequencies leading to a dense
orbit that explores the entire phase space}, see FIG. 4. A similar
treatment could be done for the remaining coordinate lines $y'$ (defined
by the family of curves $x'=\mathrm{constant}$, labeled with $\left.y'\right|_{x'}:=\left.y'_{\left[t\right]}\right|_{x'}$).
\begin{figure}[h]
\centering{}\includegraphics[scale=0.65]{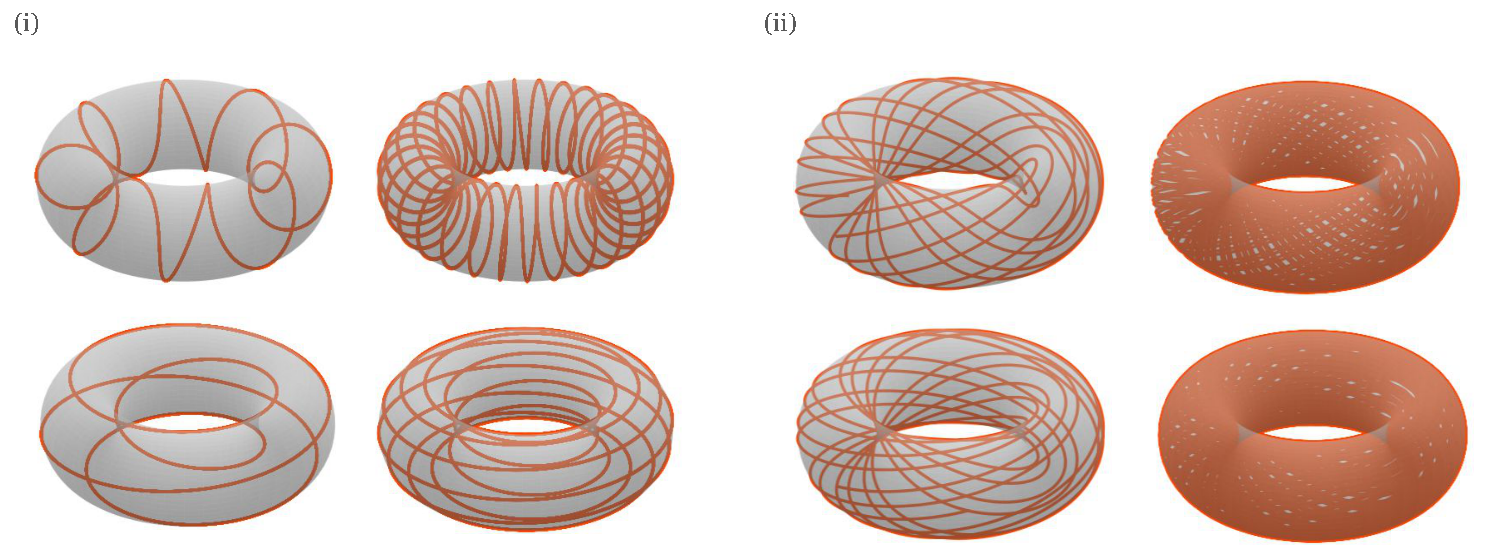}\caption{Killing flows on $\mathbb{T}^{2}$ with varying slopes. (i) Rational
slope; Top, from left to right: $\tan\varepsilon=\left\{ \nicefrac{1}{5},\nicefrac{1}{10}\right\} ;$
Bottom, from left to right $\tan\varepsilon=\left\{ 5,10\right\} .$
(ii) Irrational slope; Top: $\tan\varepsilon=\nicefrac{1}{\sqrt{2}};$
from left to right: 1000 time steps, 4000 time steps. Bottom: $\tan\varepsilon=\sqrt{2};$
from left to right: 1000 time steps, 4000 time steps. Notice that
for the irrational slopes, the Killing integral curve covers the surface
of the torus.}
\end{figure}

The next part is to apply GFT on $\mathbb{T}^{2}.$ Let us use the
(global) angle version of the coordinate $\left(x,y\right)$, namely:
\begin{equation}
\theta:=\frac{2\pi x}{L},\quad\phi:=\frac{2\pi y}{L}.\label{eq:globalangle}
\end{equation}
One could calculate the Killing vectors in $\mathbb{T}^{2},$ which
are $\partial_{\theta}$ and $\partial_{\phi}$ (they are functionally
independent), and construct the \textcolor{blue}{coordinate-adapted}
MASA as $\left\{ -\mathbf{i}\partial_{\theta},-\mathbf{i}\partial_{\phi}\right\} .$
Diagonalizing each member of the MASA gives the set of spectrum and
eigenfunctions $\left\{ \left(m,e^{\mathbf{i}m\theta}\right),\left(n,e^{\mathbf{i}n\phi}\right)\right\} $,
with $\left(m,n\right)\in\mathbb{Z}^{2}$. Defining a flat metric
on $\mathbb{T}^{2}$ as $ds^{2}=d\theta^{2}+d\phi^{2},$ one could
calculate its Laplace-Beltrami operator: $\triangle=\partial_{\theta}^{2}+\partial_{\phi}^{2}$,
which clearly commute with the MASA, hence sharing the same eigenfunctions
$e^{\mathbf{i}m\theta}e^{\mathbf{i}n\phi}$, with a discrete spectrum
$\lambda_{mn}=m^{2}+n^{2}\in\mathbb{Z}$. Therefore, the GFT on $\mathbb{T}^{2}$
with MASA choice $\left\{ -\mathbf{i}\partial_{\theta},-\mathbf{i}\partial_{\phi}\right\} $
has $e^{\mathbf{i}m\theta}e^{\mathbf{i}n\phi}$ as its kernel, and
$\mathcal{F}\sim\mathbb{Z}^{2}\ni\left(m,n\right)$ as its $k$-space;
it belongs to Type I-D.

Let us define a new ``oblique'' angle coordinate on $\mathbb{T}^{2}$
as follows:
\begin{equation}
\theta':=\frac{2\pi x'}{L},\quad\phi':=\frac{2\pi y'}{L}.\label{eq:that}
\end{equation}
$\left(\theta',\phi'\right)$ are the local ``oblique'' angles related
to the new coordinate $\left(x',y'\right)$. Notice that (\ref{eq:that})
is only valid locally in a chart, in contrast with the global angles
(\ref{eq:globalangle}), since (\ref{eq:that}) does not share a same
periodicity condition of $\mathbb{T}^{2}$ as in (\ref{eq:globalangle}).

The local coordinate transformation between angles $\left(\theta',\phi'\right)$
and $\left(\theta,\phi\right)$ in a chart in $\mathbb{T}^{2}$ is
then:
\begin{equation}
\left[\begin{array}{c}
\theta'\\
\phi'
\end{array}\right]=\left[\begin{array}{cc}
\cos\varepsilon & -\sin\varepsilon\\
\sin\varepsilon & \cos\varepsilon
\end{array}\right]\left[\begin{array}{c}
\theta\\
\phi
\end{array}\right],\qquad0\leq\varepsilon\leq\pi.\label{eq:localcoord}
\end{equation}
generated by the MASA/generator transformation (the infinitesimal
covector transformation):
\begin{equation}
\left[\begin{array}{c}
\partial'_{\theta}\\
\partial'_{\phi}
\end{array}\right],=\underset{R_{\varepsilon}}{\underbrace{\left[\begin{array}{cc}
\cos\varepsilon & \sin\varepsilon\\
-\sin\varepsilon & \cos\varepsilon
\end{array}\right]}}\left[\begin{array}{c}
\partial_{\theta}\\
\partial_{\phi}
\end{array}\right],\qquad0\leq\varepsilon\leq\pi.\label{eq:unitar}
\end{equation}
Since $\left(\begin{array}{c}
\partial'_{\theta},\end{array}\partial'_{\phi}\right)$ is a linear combination of $\left(\begin{array}{c}
\partial{}_{\theta},\end{array}\partial{}_{\phi}\right)$, then the set $\left\{ -\mathbf{i}\partial'_{\theta},-\mathbf{i}\partial'_{\phi}\right\} $
also construct a valid MASA on $\mathbb{T}^{2}.$ Both MASA are related
by $R_{\varepsilon},$ an element of an isometry group $I\left(\mathbb{R}^{2}\right),$
where $\mathbb{R}^{2}$ is the universal cover to $\mathbb{T}^{2}$.
Notice that the Killing vectors $\left(\begin{array}{c}
\partial{}_{\theta},\end{array}\partial{}_{\phi}\right)$ and $\left(\begin{array}{c}
\partial'_{\theta},\end{array}\partial'_{\phi}\right)$ are well-defined \textit{globally} on $\mathbb{T}^{2},$ even if
their integral flows (\ref{eq:localcoord}) are not. One could show
that the isometry $R_{\varepsilon}$ preserves the form of metric
and the Laplace-Beltrami operator in this new coordinate: $\triangle=\partial_{\theta}^{'2}+\partial_{\phi}^{'2}$.

At this point, we need to consider the 2 cases we discused previously:
\begin{itemize}
\item First, the rational case where $\tan\varepsilon\in\mathbb{Q}.$ Diagonalizing
each member of the MASA gives the set of spectrum and eigenfunctions
$\left\{ \left(m',e^{\mathbf{i}m'\theta'}\right),\left(n',e^{\mathbf{i}n'\phi'}\right)\right\} $,
where:
\begin{align}
m' & =m\cos\varepsilon+n\sin\varepsilon,\label{eq:condd}\\
n' & =-m\sin\varepsilon+n\cos\varepsilon.\nonumber 
\end{align}
Hence $\left(m',n'\right)$ is an element of a countable set homeomorphic
to $\mathbb{Z}^{2},$ i.e. $R_{\varepsilon}\left[\mathbb{Z}^{2}\right]$.
The eigenfunctions are truly elements of $\mathcal{L}^{2}\left[\mathbb{T}^{2}\right]$
since the domain is $\left[0,2\pi\right)$ and countable. The spectrum
of $\triangle$ for this case is $\lambda_{m'n'}=m'{}^{2}+n'{}^{2}=m^{2}+n^{2}\in\mathbb{Z},$
supporting the fact that $\lambda$ is invariance under coordinate
transformation and the choice of MASA. The GFT on $\mathbb{T}^{2}$
with MASA choice $\left\{ -\mathbf{i}\partial'_{\theta},-\mathbf{i}\partial'_{\phi}\right\} $
with $\tan\varepsilon\in\mathbb{Q}$ has $e^{\mathbf{i}m'\theta'}e^{\mathbf{i}n'\phi'}$
as its kernel, and $\mathcal{F}'\sim R_{\varepsilon}\left[\mathbb{Z}^{2}\right]\ni\left(m',n'\right)$
as its $k$-space; it belongs to Type I-D, exactly as our previous
choice of MASA.
\item Second, the irrational case where $\tan\varepsilon\in\mathbb{I}.$
Diagonalizing each member of the MASA still gives the set $\left\{ \left(m',e^{\mathbf{i}m'\theta'}\right),\left(n',e^{\mathbf{i}n'\phi'}\right)\right\} $
satisfying (\ref{eq:condd}), $\left(m',n'\right)\in\mathbb{Z}^{2}$.
The irrationality of the slope does not affect the discreteness of
the spectrum since the sine and cosine terms cancels, hence $\lambda_{m'n'}=m'{}^{2}+n'{}^{2}=m^{2}+n^{2}\in\mathbb{Z},$
even for the irrational cases. Hence, for $\tan\varepsilon\in\mathbb{I}$,
GFT on $\mathbb{T}^{2}$ with MASA choice $\left\{ -\mathbf{i}\partial'_{\theta},-\mathbf{i}\partial'_{\phi}\right\} $
has exactly similar properties with the rational case except for their
choice of MASA (Killing vectors) and hence their corresponding Killing
integral curves (or coordinate lines): The first is periodic, closed
(compact) and sparse in $\mathbb{T}^{2},$ while the later is aperiodic,
non-compact, and dense in $\mathbb{T}^{2}.$ 
\end{itemize}

\textcolor{blue}{From a physical perspective, this distinction may
be read as the difference between commensurate and incommensurate
flows on a compact manifold. Rational slopes correspond to periodic
closed trajectories, while irrational slopes generate quasi-periodic
trajectories that densely explore the torus (ergodicity). Similar
geometric distinctions are familiar in several physical settings,
including Bloch-type lattice dynamics and torus-based spectral models
such as the Harper-Hofstadter problem; we mention this only as a qualitative
analogy and do not pursue those applications here.}

In contrast with the previous $\mathbb{R}^{3}$ example, different
choices of MASA in the flat-torus case do not change the GFT classification
or the topology of the spectral space; they only change the kernel
and the associated Killing flows. The reason is that the transformation
(\ref{eq:unitar}) is induced by an isometry of the universal cover
$\mathbb{R}^{2}$ of $\mathbb{T}^{2}$. \textcolor{blue}{Thus, although
the rational and irrational cases differ qualitatively at the level
of periodic versus quasi-periodic flow, both lead to the same Type
I-D spectral classification. }This example supports the general observation
summarized in Table II.

In this paper, we only consider examples for Type I. For a more exotic
examples inside each grid of the $3\times3$ classification chart,
one could consult the references listed in Table III.

\section{Discussions and Summary}

\subsection{Discussions}

\subsubsection{Coordinate Transformations, Isometries, and Gauge Freedom}

Choosing a basis within each degenerate eigenspace of the Laplace-Beltrami
operator is equivalent to choosing additional commuting self-adjoint
operators (preferably local) whose joint eigenfunctions resolve the
degeneracy, i.e. a maximal commuting choice within each degenerate
spectral fiber (a fiberwise MASA). While many choice of MASAs exists,
only a subset is geometrically distinguished. In particular, commuting
families generated by Killing fields (and their associated conserved
quantities) are intrinsic to $\left(\Sigma,\boldsymbol{q}\right)$
and encode its isometry structure. For this reason, we prioritize
Killing data when constructing degeneracy-resolving algebras: unlike
an arbitrary MASA chosen abstractly, symmetry-generated MASAs are
tied to the manifold\textquoteright s intrinsic geometry and therefore
admit a clear physical/geometric interpretation. Since Killing generators
need not commute with each other (isometry groups are generally non-Abelian),
different inequivalent commuting subalgebras may be selected, and
any two orthonormal bases that diagonalize these fiberwise-MASAs are
related by a unitary transformation on $\mathcal{L}^{2}\left[\Sigma\right]$.

\textcolor{blue}{As discussed in Section V.C, in PDEs, particularly
within the GFT framework, three transformations are frequently conflated:
(1) passive coordinate transformations (Case i), (2) active isometries
(Case ii), and (3) changes of MASA representation (Case iii). While
Case i is merely a rewriting of the field representation without any
geometrical change, the latter two cases induce distinct classes of
unitary transformations on $\mathcal{L}^{2}\left[\Sigma\right]$.}

If the unitary transformation stems from an isometry/Case ii, the
geometric and topological structure of the GFT is invariant; the basis
is merely rotated. If instead one changes the coordinate-adapted degeneracy-resolution/separation
scheme used to construct the transform (case iii), the kernel can
change more profoundly: the induced label space ($k$-space)/spectral
domain $\mathcal{F}$ may acquire a fundamentally different product/topological
structure (see Table II).

\paragraph{The Order of Operations Subtlety.}

A potential source of confusion is the \textcolor{blue}{apparent}
\textquotedblleft coordinate dependence\textquotedblright{} of the
label space $\mathcal{F}$: \textcolor{blue}{selecting a coordinate
system on $\Sigma$ may appear to affect the topological structure
of $\mathcal{F}$. As explained in Section V.C, Cases i and iii are
frequently conflated. The \textquotedblleft coordinate dependence\textquotedblright{}
of $\mathcal{F}$ is in fact a consequence of Case iii, where one
implicitly selects a MASA through a coordinate-adapted separation
scheme, rather than performing a pure coordinate transformation as
in Case i. This happens because}, in the PDE setting, two different
operations, i.e., solving the eigenvalue problem and changing variables
(pure coordinate transformation, depend on the order in which they
are performed. If one first solves the spectral problem for $-\triangle$
(or the relevant PDE) and only then constructs the transform, a change
of coordinates on $\Sigma$ merely rewrites the same eigenfunctions
and does not alter the underlying label space (the \textit{solve-then-transform}
order). \textcolor{blue}{This is Case i}. In contrast, if one first
adopts a coordinate-adapted transform/separation scheme and only then
solves the resulting ODEs in that representation, the induced degeneracy
labeling (and hence the resulting $\mathcal{F}_{\lambda}$) is determined
by that prior choice (the \textit{transform-then-solve} order). \textcolor{blue}{This
is Case iii} and reflects a representation choice rather than the
intrinsic physics.

\paragraph{Physical Interpretation (Observer Dependence).}

From a physical standpoint, the choice of MASA is not merely a mathematical
artifact. In the 3+1 point of view such as quantum mechanics, the
choice of MASA correspond to the choice of CSCO we choose to measure
the system. Different CSCOs correspond to different observable we
choose to measure, although the physical system stays invariant.

In a covariant field theory context like QFT, the choice of MASA is
often dictated by the choice of coordinate chart, where coordinates
are attached to a specific reference frames/observers. This will illustrate
the different perspectives experienced by different observers on different
reference frames. In this case, the isometries (Case ii) behave like
Lorentz boosts: a plane-wave solution remains a plane wave, its four-momentum
simply Lorentz-transformed. 

However, a coordinate transform associated with changing observer
(the coordinate-adapted separation schemes/Case iii) can have drastic
effects. The classic example is the transition from inertial (Minkowski)
to Rindler coordinates. This transformation maps Minkowski plane waves
into Rindler mode superpositions, producing the Unruh effect: an inertial
observer\textquoteright s sharp momentum eigenstate appears (thermally)
spread to a uniformly accelerated observer.

Thus, especially in the degenerate case, a GFT is not uniquely defined
until the ``gauge freedom'' is fixed by choosing one definite MASA.
Only then does a single, well-posed space-frequency duality emerge.

\subsubsection{Toward a Unified Definition of Momentum}

At present the word ``momentum'' in physics terminology has several
context-dependent meanings:
\begin{enumerate}
\item Canonical/Noether perspective (local): Momentum is a cotangent vector,
the conjugate variable to position that generates spatial translations;
that is, the quantity $\mathbf{p}=\frac{\partial L}{\partial\dot{\mathbf{x}}}$
$\in$ $T_{\mathbf{x}}^{*}\Sigma$ with $\dot{\mathbf{x}}=\frac{d\mathbf{x}}{d\lambda}$
($\lambda$ a real parameter) that is well-defined if we provide a
scalar functional $L_{\left[\mathbf{x},\dot{\mathbf{x}},\lambda\right]}$
on $\Sigma$. The Special and General Relativity perspective of momentum
as part of the covariant four-vector $\left(E,p_{i}\right)$, a local
densitized version could be packaged in the stress--energy tensor
$T_{\mu\nu}$, is included in this category.
\item The GFT/spectral perspective (global): Momentum is a spectral label
belonging to the eigenvalue set of a self-adjoint operator (MASA)
on $\Sigma.$ These joint eigen-labels form the familiar $k$-space.
In our framework, the momentum space is the topological space of real
spectral parameters and its degeneracy $\mathcal{F},$ that label
the irreducible representations of the symmetry group.
\end{enumerate}

\paragraph{The Euclidean Coincidence.}

In the flat case where $\Sigma=\mathbb{R}^{n}$ ($n=3$ in particular),
these two definitions of momentum coincide: the canonical momentum
$\mathbf{p}$ $\in$ $T_{\mathbf{x}}^{*}\Sigma$ where the cotangent
fiber $T_{\mathbf{x}}^{*}\Sigma$ is $T_{\mathbf{x}}^{*}\mathbb{R}^{n}\sim\mathbb{R}^{n,*}$,
while from the spectral momentum perspective, the GFT in $\Sigma=\mathbb{R}^{n}$
gives $\mathcal{F}$ isomorphic to $\mathbb{R}^{n,*}.$ The (plane)
wave vector $\boldsymbol{k}$ lives in $\mathbb{R}^{n,*},$ hence
allow us to identify the spectral label $\boldsymbol{k}$ directly
with the cotangent vector $\mathbf{p}$. This nice coincidence is
the reason standard quantum mechanics works so intuitively in flat
space. For a general curved manifold $\Sigma$, however, these two
different spaces, i.e., the local cotangent space $T_{\mathbf{x}}^{*}\Sigma$
and global spectral domain $\mathcal{F}$, are not isomorphic.

\paragraph{The Symplectic Unification (The Moment Map).}

Our work is a part of a larger project with an attempt to supply a
single mathematical framework for those context-dependent definitions.
It is possible to package both canonical and spectral momentum as
two faces of one single object. The object is simply the tautological
1-form $\theta$ on the cotangent bundle / phase space $T^{*}\Sigma$
defined as:
\[
\theta_{\left[\mathbf{x},\mathbf{p}\right]}\left(V\right):=\mathbf{p}\left(\pi_{*}V\right),\qquad\forall\;V\;\exists\;T_{\left(\mathbf{x,p}\right)}\left(T^{*}\Sigma\right),
\]
where $\pi:T^{*}\Sigma\rightarrow\Sigma$ is the projection from the
section of a cotangent bundle to its basespace.

Let us gives the sketch of the framework. The two perspectives are
simply \textquotedblleft reading off components\textquotedblright{}
of that same object: (1) The canonical momentum $\mathbf{p}\in T_{\mathbf{x}}^{*}\Sigma$
is simply the fiber on a fixed point in basespace $\mathbf{x}.$ (2)
The spectral momentum could be obtained from $\theta$ as follows.
Assume a Lie group $\mathcal{G}$ acting on $\Sigma$ (e.g. isometries).
For any generator $\boldsymbol{K}\in\mathfrak{g},$ there is an induced
vector field $\boldsymbol{K}_{\Sigma}$ on $\Sigma$ and a lifted
vector field $\boldsymbol{K}_{T^{*}\Sigma}$ on phase space. Then
define the momentum map:
\[
J:T^{*}\Sigma\rightarrow\mathfrak{g}^{*},
\]
by a single pairing rule:
\begin{equation}
\left\langle J\left(\mathbf{x},\mathbf{p}\right),\boldsymbol{K}\right\rangle :=\mathbf{p}\left(\boldsymbol{K}_{\Sigma\left[\mathbf{x}\right]}\right)=\iota_{\boldsymbol{K}_{T^{*}\Sigma}}\theta,\label{eq:mommap}
\end{equation}
 with $\iota$ is the interior product on $T^{*}\Sigma$.

From this unified framework, the MASA (which comes from the Killing
field $\boldsymbol{K}$) that select the GFT and the corresponding
spectral parameter space $\mathcal{F}$ are related to the canonical
momentum $\mathbf{p}$ through the momentum map $J$ defined in (\ref{eq:mommap}).
The covector $\mathbf{p}$ is the geometrical object, while the Noether
charge for symmetry direction $\boldsymbol{K}$ is simply \textquotedblleft $\mathbf{p}$
evaluated on the symmetry vector field $\boldsymbol{K}_{\Sigma}$\textquotedblright .
Hence, here lies the connection: the \textquotedblleft canonical momentum\textquotedblright{}
is the object $\mathbf{p}$, while the \textquotedblleft spectral
momentum\textquotedblright{} (eigenvalues) corresponds to the quantized
level sets of the moment map function $J$.

On $\Sigma=\mathbb{R}^{n}$, translations symmetry give $\boldsymbol{K}=\partial_{i}$,
then (\ref{eq:mommap}) gives:
\[
\left\langle J\left(\mathbf{x},\mathbf{p}\right),\partial_{i}\right\rangle =\mathbf{p}\left(\partial_{i}\right)=p_{i},
\]
and simultaneously, the plane waves diagonalize $-\triangle$ with
$\lambda=\left|\boldsymbol{k}\right|^{2}$; everything collapses into
one familiar vector. The full construction of unified framework for
momentum in curved space $\Sigma$ will be studied elsewhere.

Finally, there are at least two ways to view the joint spectrum \textit{$\mathcal{F}$}.
We list them on the Table IV: 

\begin{table}
\begin{centering}
\includegraphics[scale=1]{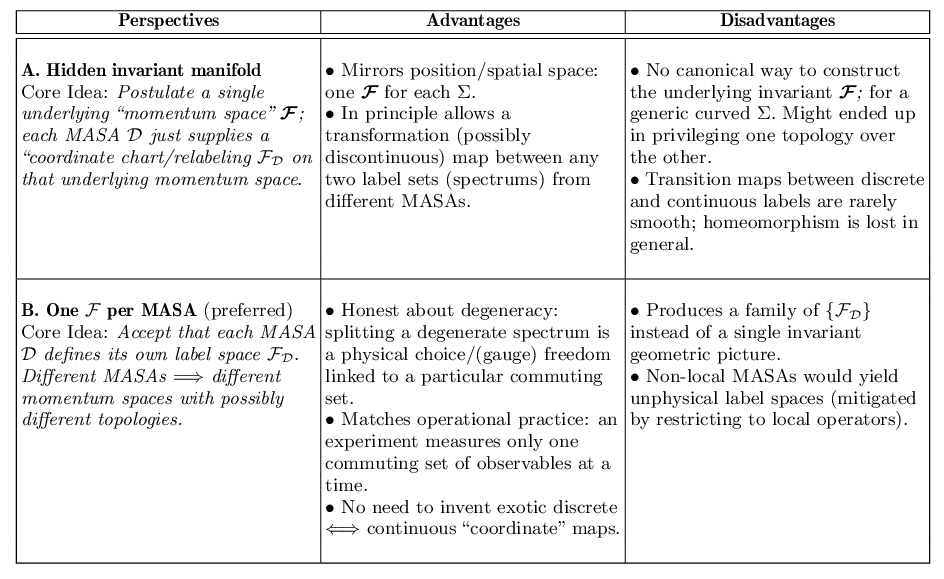}
\par\end{centering}
\caption{Two possible perspectives on the \textquotedbl momentum space''
\textit{$\mathcal{F}.$}}
\end{table}

We adopt Perspective B: momentum space is not a universal invariant
but MASA-dependent; locality selects the physically meaningful MASAs.
This viewpoint recovers the standard flat-space definitions when $\Sigma=\mathbb{R}^{n}$
(Cartesian MASA), reproduces angular-momentum labels on spheres-like
fibers on $\Sigma=\mathbb{R}^{n}$ (spherical MASA), and extends naturally
to the curved/operator-based momenta \textcolor{blue}{for general
$\Sigma$ (this could possibly} includes generalized momentum re-definition,
curved $k$-space, physical examples like Unruh effect, Aharonov--Bohm
effect, etc.).

\subsubsection{The Remaining Freedom: Geometrizing the $k$-Space}

We noted in Section II that the GFT is not unique. Our framework reveals
three inherent levels of freedom in defining the spectral domain:
\begin{itemize}
\item \emph{Basis rotation inside degenerate fibers.}\textbf{ This is related
to MASA (orthonormal basis) freedom}: The choice of MASA will affect
the orthonormal basis, consequently, fixes the Fourier kernel (up
to a normalization constant) and the topology of the dual momentum
space $\mathcal{F}$. This has been the primary focus of this paper.
\item \emph{Spectral measure / normalization convention.}\textbf{ This is
related to geometrization freedom}: In the continuous case where $\mathcal{F}$
is treated as a smooth manifold (in particular, Type I-C), the spectral
measure is fixed only up to equivalent representations (i.e. up to
changes that preserve the Plancherel/isometry property). One may rescale
the generalized eigenmodes by a positive measurable weight function
$w_{\left[\lambda,\alpha\right]}$:
\[
\hat{f}_{\left[\mathbf{x};\lambda,\alpha\right]}\mapsto\hat{f}'_{\left[\mathbf{x};\lambda,\alpha\right]}=\sqrt{w_{\left[\lambda,\alpha\right]}}\hat{f}_{\left[\mathbf{x};\lambda,\alpha\right]}
\]
provided the associated weight/measure is transformed inversely: $\rho_{\left[\lambda,\alpha\right]}\mapsto\rho'_{\left[\lambda,\alpha\right]}=\rho_{\left[\lambda,\alpha\right]}/w_{\left[\lambda,\alpha\right]}$.
This guarantees that the completeness relation and Plancherel theorem
remain intact. This freedom allows us to \textquotedblleft geometrize\textquotedblright{}
$\mathcal{F}$ by identifying the measure $d\mu_{\mathcal{F}}=\rho_{\left[\lambda,\alpha\right]}d\lambda d\alpha$
with the Riemannian volume form $d\mathrm{vol}_{\mathcal{F}}$ of
a chosen metric on the momentum space. Although not mathematically
necessary, imposing an appropriate metric on $k$-space is physically
motivated, particularly for models requiring curved momentum space.
\item \emph{Coordinate (label) freedom on $\mathcal{F}$: }\textbf{This
is related to coordinate transformation in }$\mathcal{F}$.\textbf{
}\textit{\emph{Once a smooth manifold }}\textit{$\mathcal{F}$}\textit{\emph{
is endowed }}with a chosen metric/volume form, the spectral labels
\textit{\emph{$\left(\lambda,\alpha\right)$}} may be interpreted
as coordinates on a (possibly local) chart of \textit{$\mathcal{F}$.}
Coordinate changes on\textit{ $\mathcal{F}$} then correspond to reparameterizations
(diffeomorphism) of the same geometrized label space and do not alter
the underlying physics of the transform.
\end{itemize}
One of the motivations for geometrizing $k$-space and endowing it
with a metric is to allow for a well-defined notion of curvature in
momentum space. A curved $k$-space introduces richer structures and
offers new perspectives in both mathematical physics, particularly
in quantum gravity contexts \cite{curvedmomemtum1,curvedmomentum2,curvedmomentum3}.
This remains a fertile ground for future investigation

\subsection{Summary}

In this work, we develop a systematic framework for constructing and
classifying Generalized Fourier Transforms (GFTs) on Riemannian manifolds.
Our derivation relies on three minimal axiomatic requirements: (a)
\textbf{invertibility}, i.e. the integral transform defines a bijection
(isomorphism) between the relevant function spaces (i.e., the physical
function space and the spectral domain); (b) \textbf{isometricity,}
i.e. the transform and its inverse preserve the $\mathcal{L}^{2}$
inner-product norm (unitarity); and (c) \textbf{spectral diagonalization},
i.e. the kernel diagonalizes the Laplace--Beltrami operator. Within
this setting, we established a generalized Parseval-Plancherel theorem
for curved Riemannian manifolds.

A central feature of our construction is the rigorous treatment of
spectral degeneracy, which introduces an intrinsic freedom in choosing
orthonormal basis functions within degenerate spectral sectors. To
control this non-uniqueness, we emphasize the role of local differential
operators and argue that geometrically/physically meaningful bases
are naturally associated with commuting families of local, symmetry-respecting
operators; most notably those constructed from intrinsic symmetry
data such as Killing fields and St$\ddot{\textrm{a}}$ckel webs (when
available). We provide an explicit algorithm for constructing such
operators and illustrate it on canonical examples in $\mathbb{R}^{3}.$

We explicitly examine how symmetries and coordinate-adapted choices
affect the structure of the GFT. Isometries provide a canonical class
of metric-preserving maps that act unitarily on $\mathcal{L}^{2}\left[\Sigma\right]$
and preserve the Laplace--Beltrami spectrum. By contrast, when the
transform is constructed through a coordinate-adapted separation/labeling
scheme, changes of coordinates can lead to different induced $k$-space
labelings in the degenerate sector. In particular, the existence of
St$\ddot{\textrm{a}}$ckel coordinates on $\Sigma$ plays a special
role in enabling complete separability of the Helmholtz problem, hence
constraining the availability and completeness of degeneracy-resolving
commuting structures and the resulting momentum-space label space
$\mathcal{F}.$

To organize these phenomena, we introduce a double classification
scheme: (i) \textbf{algebraic type}: based on the completeness of
the (fiberwise) MASA/ degeneracy-resolving structure and the existence
of St$\ddot{\textrm{a}}$ckel (Types I--III), and (ii) \textbf{topological
type}: based on the topological nature of the dual momentum space
(discrete (D), continuous (C), semi-discrete (SD)). Finally, we also
highlight subtleties of the framework and provide representative examples
illustrating how isometries and coordinate-adapted constructions affect
the degenerate sector.

Together, these results offer a principled and physically-motivated
framework for performing harmonic analysis on curved manifolds, i.e.,
one that connects local geometry (metrics and Killing vectors) and
local operator structure, to the global spectral topology of function
spaces. This framework may serve as a foundation for further studies
in spectral geometry and manifold-based representations, and, when
coupled to additional dynamical/observer structure, provides the necessary
tools for defining \textquotedblleft momentum\textquotedblright{}
in curved-space physical models.

\section*{Acknowledgement}
The authors are grateful to the anonymous reviewer for a very careful reading of the manuscript and for numerous constructive suggestions that substantially improved both the clarity and the presentation of this work.

%\appendix

%\section*{Declarations}
%Funding information - not applicable.


\begin{thebibliography}{10}

\bibitem{Spectral1}M. Reed, B. Simon. \textit{Functional Analysis}
(Methods of Modern Mathematical Physics Volume \textbf{I}). Academic
Press. 1980.

\bibitem{Folland}G. B. Folland. \textit{Fourier Analysis and its
Applications}. Wadsworth and Brooks. California. 1992.

\bibitem{Grigoryan}A. Grigor'yan. \textit{Heat Kernel and Analysis
on Manifolds}. American Mathematical Soc. 2009.

\bibitem{Spectral2}P. Berard. \textit{Spectral Geometry: Direct and
Inverse Problems}. Lecture Notes in Mathematics \textbf{1207}. 1986. 

\bibitem{Pontryagin1}S. A. Morris. \textit{Pontryagin Duality and
the Structure of Locally Compact Abelian Groups}. London Math. Soc.
Lecture Notes \textbf{29}. Cambridge U. Press. 1977.

\bibitem{Pontryagin2}H. Reiter, J. D. Stegeman. \textit{Classical
Harmonic Analysis and Locally Compact Groups}. 2nd ed. Clarendon Press.
Oxford. 2000.

\bibitem{Pontryagin3}W. Rudin. \textit{Fourier Analysis on Groups}.
D. van Nostrand Co. 1962.

\bibitem{Plancherel1}S. Helgason. \textit{Geometric Analysis on Symmetric
Spaces}. Am. Math. Soc. Providence .1994.

\bibitem{Plancherel2}S. Helgason. \textit{Groups and Geometric Analysis}:
425-444. Academic Press. New York. 1984.

\bibitem{Helgason1}S. Helgason. \textit{The Fourier transform on
symmetric spaces}. $\mathrm{\acute{E}}$lie Cartan et les math$\mathrm{\acute{e}}$matiques
d'aujourd'hui no. \textbf{131}: 151-164. 1984. 

\bibitem{Helgason2}P. Mohanty, S. K. Ray, R. P. Sarkar, A. Sitaram.
\textit{The Helgason--Fourier Transform for Symmetric Spaces II}.
Journal of Lie Theory Volume \textbf{14}: 227--242. Heldermann-Verlag.
2004.

\bibitem{Helgason3}M. Boujeddaine, M. E. Kassimi, S. Fahlaoui.\textit{
Helgason--Gabor--Fourier transform and uncertainty principles}.
International Journal of Wavelets, Multiresolution, and Information
Processing \textbf{19} (01): 2050056. 2021.

\bibitem{Helgason4}S. Helgasson. \textit{Wave equations on homogeneous
spaces.} Lie Group Representations III. Lecture Notes in Math. \textbf{1077}:
254-287. Springer-Verlag, New York. 1984.

\bibitem{Helgason5}A. Terras. \textit{Harmonic Analysis on Symmetric
Spaces and Applications I}. 1985. Springer-Verlag.

\bibitem{Gelfand1}I. M. Gelfand, M. A. Naimark. \textit{On the imbedding
of normed rings into the ring of operators on a Hilbert space}. Mat.
Sbornik. \textbf{12} (2): 197--217. 1943.

\bibitem{Gelfand2}E. Hewitt, K. A. Ross. \textit{Abstract Harmonic
Analysis. Volume I: Structure of Topological Groups Integration Theory
Group Representations.} Springer-Verlag. 1979.

\bibitem{NoncommuteHarmonic1}M. E. Taylor. \textit{Noncommutative
Harmonic Analysis}. American Mathematical Society. 1986.

\bibitem{NoncommuteHarmonic2}J. Carmona, P. Delorme, M. Vergne. \textit{Noncommutative
harmonic analysis: in honor of Jacques Carmona}. Springer. 2004.

\bibitem{NoncommuteHarmonic3}K. I. Gross. \textit{On the evolution
of noncommutative harmonic analysis}. Amer. Math. Monthly. \textbf{85}
(7): 525--548. 1978.

\bibitem{FIO1}L. H$\mathrm{\ddot{o}}$rmander. \textit{Fourier integral
operators I}. Acta Mathematica \textbf{127}: 79--183. Springer Netherlands.
1970.

\bibitem{FIO2}J. J. Duistermaat. \textit{Fourier Integral Operators}.
Progress in Mathematics. Birkhäuser. 1995.

\bibitem{curvedmomemtum1}J. Kowalski-Glikman. \textit{Living in Curved
Momentum Space.} Int. J. Mod. Phys. A \textbf{28} (12): 1330014. 2013.
\url{https://arxiv.org/abs/1303.0195}.

\bibitem{curvedmomentum2}S. A. Franchino-Vi$\tilde{\mathrm{n}}$as,
S. Mignemi, J. J. Relancio. \textit{The beauty of curved momentum
space.} Proceedings of the Corfu Summer Institute School and Workshops
on Elementary Particle Physics and Gravity. 2022. \url{https://arxiv.org/abs/2303.08220v1}.

\bibitem{curvedmomentum3}N. Jafari. \textit{Evolution of the concept
of the curvature in the momentum space}. \url{https://arxiv.org/abs/2404.08553}.

\bibitem{Hopf}H. Hopf. \textit{$\ddot{U}$ber die Abbildungen der
dreidimensionalen Sph$\ddot{a}$re auf die Kugelfl$\ddot{a}$che}.
Mathematische Annalen \textbf{104}: 637 - 665. 1931.

\bibitem{CNYang}C. N. Yang. \textit{Generalization of Dirac’s monopole
to SU2 gauge fields}. J. Math. Phys. \textbf{19}: 320--328. 1978.

\bibitem{Unruh}W. G. Unruh. \textit{Notes on black-hole evaporation.}
Phys. Rev. D \textbf{14} (4): 870--892. 1976. 

\bibitem{Fulling}S. A. Fulling. \textit{Nonuniqueness of Canonical
Field Quantization in Riemannian Space-Time}. Phys. Rev. D \textbf{7}
(10): 2850--2862. 1973. 

\bibitem{Davies}P. C. W. Davies. \textit{Scalar production in Schwarzschild
and Rindler metrics}. J. Phys. A \textbf{8} (4): 609--616. 1975.

\bibitem{ABeffect}Y. Aharonov, D. Bohm. \textit{Significance of Electromagnetic
Potentials in the Quantum Theory}. Phys. Rev. \textbf{115}: 485. 1959.

\bibitem{Sep1}L. P. Eisenhart. \textit{Separable systems of St$\ddot{a}$ckel}.
Ann. of Math. \textbf{35}: 284--305. 1934.

\bibitem{Sep2}E. G. Kalnins. \textit{On the separation of variables
for the Laplace equation $\Delta\Psi+K^{2}\Psi=0$ in two and three-dimensional
Minkowski space}. SIAM J. Math. Anal. \textbf{6}: 340--374. 1975.

\bibitem{Sep3}S. Benenti. \textit{Intrinsic characterization of the
variable separation in the Hamilton-Jacobi equation}. J. Math. Phys.
\textbf{38}: 6578--6602. 1997.

\bibitem{Sep4}S. Benenti. \textit{Separability in Riemannian manifolds}.
SIGMA \textbf{12}: 013. 2016. \url{https://arxiv.org/abs/1512.07833}.

\bibitem{Sep5}E. G. Kalnins, J. M. Kress, W. Miller, Jr. \textit{Separation
of Variables and Superintegrability: The symmetry of solvable systems}.
IOP Publishing Ltd. 2018.

\bibitem{Sep6}K. Rajaratnam, R. G. McLenaghan, C. Valero. \textit{Orthogonal
Separation of the Hamilton--Jacobi Equation on Spaces of Constant
Curvature.} SIGMA \textbf{12}: 117. 2016. \url{https://arxiv.org/pdf/1607.00712}.

\bibitem{Sep7}C. M. Chanu, G. Rastelli. \textit{Block-Separation
of Variables: a Form of Partial Separation for Natural Hamiltonians.}
SIGMA \textbf{15}: 013. 2019. \url{https://arxiv.org/abs/1808.01889}.

\bibitem{Peetre}J. Peetre. \textit{Une caractérisation abstraite
des opérateurs différentiels.} Mathematica Scandinavica \textbf{7}:
211--218. (1959).

\bibitem{prolongation}A. R. Gover, T. Leistner. \textit{Invariant
prolongation of the Killing tensor equation.} Annali di Matematica
Pura ed Applicata (1923 -) \textbf{198}: 307--334. 2019.

\bibitem{prolongation1a}M. Eastwood. \textit{Higher symmetries of
the Laplacian}. Annals of Mathematics, \textbf{161}: 1645--1665.
2005.

\bibitem{prolongation2}M. Eastwood, T. Leistner. \textit{Higher Symmetries
of the Square of the Laplacian.} IMA Vol. Math. Appl. \textbf{144}:
319-338. Springer, New York. 2008. \url{https://arxiv.org/abs/math/0610610}.

\bibitem{obstruct1}O. P. Santillan. \textit{Killing-Yano tensors
and some applications}. J. Math. Phys. \textbf{53}: 043509. 2012.
\url{https://arxiv.org/pdf/1108.0149}.

\bibitem{obstruct2}G. Clemente. \textit{A curvature obstruction to
integrability}. Math. Commun. \textbf{28}: 29--48. 2023. \url{https://arxiv.org/abs/2108.03376}

\bibitem{uhlenbeck}K. Uhlenbeck. \textit{Generic properties of eigenfunctions}.
Amer. J. Math. \textbf{98} (4): 1059-1078. 1976.

\bibitem{separa}B. Carter. \textit{Hamilton-Jacobi and Schr$\ddot{o}$dinger
Separable Solutions of Einstein’s Equations}. Commun. Math. Phys.
\textbf{10}: 280--310. 1968.

\bibitem{separa1}E. G. Kalnins, W. Miller, Jr. \textit{Killing Tensors
and Variable Separation for Hamilton-Jacobi and Helmholtz Equations}.
SIAM J. Math. Anal. \textbf{11}: 6. 1980.

\bibitem{separa2}S. Benenti. \textit{Orthogonal Separable Dynamical
Systems.} Differential Geometry and Its Applications Proc. Conf. Opava
(Czechoslovakia): 163-184. 1993.

\bibitem{ODE}P. Hartman. \textit{Ordinary Differential Equations}:
Second Edition. Vol. \textbf{38} of Classics in Applied Mathematics.
SIAM e-books. 1982.

\bibitem{Nijenhuis}A. Nijenhuis. \textit{Jacobi-type identities for
bilinear differential concomitants of certain tensor fields. II. }Indagationes
Mathematicae (Proceedings) \textbf{58}: 398-403. 1955.

\bibitem{Schouten}J. A. Schouten. \textit{Ricci-Calculus: An Introduction
to Tensor Analysis and Its Geometrical Applications.} Grundlehren
der mathematischen Wissenschaften, 2nd ed. Springer. 1954. 

\bibitem{Operatorbook}I. Kol$\mathrm{\acute{a}\check{r}}$, P. W.
Michor, J. Slovak. \textit{Natural operations in differential geometry.}
Springer. 1993.

\bibitem{nonlocalproject}P. Busch, P. J. Lahti, P. Mittelstaedt.
\textit{The Quantum Theory of Measurement.} Lecture Notes in Physics
Monographs. 2nd ed. Springer. 2013.

\bibitem{Torres}G. F. Torres del Castillo. \textit{The St$\ddot{a}$ckel
theorem in the Lagrangian formalism and the use of local times}. Revista
Mexicana de Fisica \textbf{67} (3): 44-451. 2021

\bibitem{Stackel}P. St$\ddot{\textrm{a}}$ckel. \textit{Uber die
Integration der Hamilton-Jacobischen Differential Gleichung mittelst
Separation der Variabeln}. Habilitationsschrift. Halle, 1891.

\bibitem{Stackel3}P. St$\ddot{\textrm{a}}$ckel. \textit{Uber Die
Bewegung Eines Punktes In Einer N-Fachen Mannigfaltigkeit}. Math.
Ann. \textbf{42}: 537--563. 1893. 

\bibitem{Stackel4}A. V. Tsiganov. \textit{The St$\ddot{a}$ckel systems
and algebraic curves}. J. Math. Phys. \textbf{40}: 279--298. 1999.

\bibitem{moon-spencer}P. Moon, D. E. Spencer. \textit{Eleven Coordinate
Systems.} Chapter \textbf{1}: 1-48. Field Theory Handbook. Springer-Verlag.
1988.

\bibitem{vectorAnalysis}M. R. Spiegel, S. Lipschutz, D. Spellman.
\textit{Vector Analysis.} Schaum’s Outlines 2nd ed. McGraw Hill. 2009.

\bibitem{Rsep1}C. P. Boyer, E. G. Kalnins, W. Miller, Jr. \textsl{R-Separable
Coordinates for Three-Dimensional Complex Riemannian Spaces.} Trans.
Am. Math. Soc. \textbf{242}: 355-376. 1978.

\bibitem{Rsep2}E. G. Kalnins, W. Miller, Jr. \textit{The Wave Equation
and Separation of Variables on the Complex Sphere $S^{4}$}. Jour.
Math. Anal. App. \textbf{83}: 449-469. 1981.

\bibitem{Frob1}F. W. Warner. \textit{Foundations of Differentiable
Manifolds and Lie Groups.} Series Title Graduate Texts in Mathematics.
Springer. New York. 1983.

\bibitem{Frob2}J. M. Lee . \textit{Introduction to Smooth Manifolds}.
Graduate Texts in Mathematics. Springer. New York. 2012.

\bibitem{Rsep3}P. Moon, D. E. Spencer. \textit{Field Theory Handbook,
Including Coordinate Systems, Differential Equations, and Their Solutions}.
2nd ed. Springer-Verlag, New York. 1988.

\bibitem{Rsep4}P. M. Morse, H. Feshbach. \textit{Tables of Separable
Coordinates in Three Dimensions.} Methods of Theoretical Physics,
Part I. McGraw-Hill. New York. 1953.

\bibitem{drum1}M. Kac. \textit{Can One Hear the Shape of a Drum?}
Am. Math. Monthly. \textbf{73} (4): 1-23. Part 2: Papers in Analysis.
1966.

\bibitem{drum2}P. Amore. \textit{Solving the Helmholtz equation for
membranes of arbitrary shape: numerical results.} J. Phys. A: Math.
Theor. \textbf{41}: 265206. 2008. \url{https://iopscience.iop.org/article/10.1088/1751-8113/41/26/265206}

\bibitem{II-Ca}C. R. Holt. \textit{Construction of new integrable
Hamiltonians in two degrees of freedom}. J. Math. Phys. \textbf{23}:
1037--1046. 1982.

\bibitem{II-Cb}A. S. Fokas, P. A. Lagerstrom. \textit{Quadratic and
cubic invariants in classical mechanics}. J. Math. Anal. Appl. \textbf{74}:
325--341. 1980.

\bibitem{II-Cc}R. Campoamor-Stursberg, J. F. Cari$\mathrm{\tilde{n}}$ena,
M. F. Ra$\mathrm{\tilde{n}}$ada. \textit{Higher-order superintegrability
of a Holt related potential}. \url{https://arxiv.org/abs/1303.0195}.

\bibitem{IIMa}V. S. Matveev, V. V. Shevchishin. \textit{Two-dimensional
superintegrable metrics with one linear and one cubic integral}. Jour.
Geom. Phys. \textbf{61} (8): 1353-1377. 2011. \url{https://arxiv.org/abs/1010.4699}.

\bibitem{IImb}G. Valent. \textit{Zoll and Tannery Metrics from a
Superintegrable Geodesic Flow}. Lett. Math. Phys. \textbf{104}: 1121--1135.
2014. \url{https://arxiv.org/abs/1404.1793}.

\bibitem{IIDa}I. V. Komarov. \textit{Goryachev-Chaplygin top in quantum
mechanics}. Theor. Math. Phys. \textbf{50}: 265--270. 1982.

\bibitem{IIDb}E. K. Sklyanin. \textit{Goryachev-Chaplygin top and
the inverse scattering method}. J. Math. Sci. \textbf{31}: 3417--3431.
1985.

\bibitem{IIDc}I. V. Komarov, E. I. Novikov. \textit{Spectral surface
for the quantum Goryachev-Chaplygin top.} Phys. Lett. A \textbf{186}:
396--402. 1994.

\bibitem{IIICa}L. A. Bunimovich. \textit{On the Ergodic Properties
of Nowhere Dispersing Billiards}. Commun. Math. Phys. \textbf{65}:
295--312. 1979.

\bibitem{IIICb}H. Ishio. \textit{Quantum transport and classical
dynamics in open billiards}. J. Stat. Phys. \textbf{83}: 203. 1996.

\bibitem{IIISD}T. Gilbert, D. P. Sanders. \textit{Stable and Unstable
Regimes in Higher-Dimensional Convex Billiards with Cylindrical Shape}.
N. Jour. Phys. \textbf{23}: 043012. 2021. \url{https://arxiv.org/pdf/1009.0337}.

\bibitem{IIIDa}O. Bohigas, M. J. Giannoni, C. Schmit. \textit{Characterization
of Chaotic Quantum Spectra and Universality of Level-Fluctuation Laws}.
Phys. Rev. Lett. \textbf{52}: 1--4. 1984.

\bibitem{IIIDb}E. J. Heller. \textit{Bound-State Eigenfunctions of
Classically Chaotic Hamiltonian Systems.} Phys. Rev. Lett. \textbf{53}:
1515--1518. 1984.

\bibitem{IIIDc}D. V. Anosov. \textit{Geodesic flows on closed Riemannian
manifolds of negative curvature}. Proc. Steklov Inst.\textbf{ 90}
.1967.

\bibitem{rigged}A. B$\mathrm{\ddot{o}}$hm, J. D. Dollard. \textit{The
Rigged Hilbert Space and Quantum Mechanics}. Lectures in Mathematical
Physics at the University of Texas. Part of the book series: LNP \textbf{78}.
Springer Berlin. 2005.





\end{thebibliography}
\end{document}